Shteryo Nozharov

# SOCIAL COSTS OF CIRCULAR ECONOMY IN EUROPEAN UNION

**Shteryo Nozharov**

# SOCIAL COSTS OF CIRCULAR ECONOMY IN EUROPEAN UNION

# PREFACE

Two fundamental issues are incorporated in the present monograph: the issue related to the quantification of the social costs and the issue, related to the defining of the circular economy concept as a theoretical model. The analysis is based on the methodology of the new institutional economics, which fact distinguishes it from the many other circular economy analysis based on the neo-classical methodological apparatus.

The originality of the ideas presented in the monograph is highlighted by their approbation to the *United Nations Economic Commission for Europe* (UNECE). In relation to invitation on behalf of UNECE, Assoc. prof. Dr. Shteryo Nozharov presented a paper on topic: „Circular economy concept implementation in the inland waterway transport" on 4$^{th}$ of November 2021 (UNECE, 2021,p.13). This invitation by UNECE is based on the author's interest in the topic.

Assoc. prof. Dr. Shteryo Nozharov is a full faculty member at UNWE since 2013. His publications are cited in two UN monographs, devoted to the analysis of the global public sector (United Nations 2021,p.iv; United Nations 2018,p.iv). His publications are also cited in official documents of EBRD and of other international institutions.

Assoc. prof. Dr. Shteryo Nozharov is also a deputy-chair of the Union of Scientists in Bulgaria – Department "Economic sciences" since February 2023.

# TABLE OF CONTENTS

## LIST OF FIGURES

# INTRODUCTION

## Relevance of the studied problems.

The global economy is entering a new phase of its development. After 2015 more and more scientific publications go beyond the basic fundament of general and conceptual analysis of the problems of sustainable development, and they put the accent on two main aspects. The first aspect is the establishment of the digital economy, which is studied as software and hardware infrastructures, organized in technical and technological processes where humans and artificial intellect fully interact to one another (Neeraj, 2019; Szalavetz, 2019; Nambisan, Wright and Feldman, 2019). The second aspect is the implementation of the circular economy, which upgrades the vision for low-carbon economy. The scarcity of resources, negative impact on the environment and at the same time the desire to increase economic benefits and growth need new economic paradigm (Ritzén and Sandström, 2017).

According to Winans et al. (2017) there is ample evidence that the consumption and the market for reused and recycled products are increasing in the last decades. The circular economy is examined as an economic concept which main purpose is waste flows to be reduced without leading to a reduced satisfaction of the economic agent's needs. (Svensson and Funck, 2019; Gusmerotti, Testa, Corsini, Pretner and Iraldo, 2019; Figge and Thorpe, 2019). This economy should be realized through closing the material cycles and resource-efficient increase so as the highest utility and value of resources to be maintained over time. In the circular economy, the economic growth does not depend on the amount of the scarce nature resources. In this way, the systematic risks for the nature capital are minimized, incl. those by pollution, and at the same time the scarce nature stocks are preserved and also positive effects for the social capital emerge (Moreau, Sahakian, Van Griethuysen, & Vuille, 2017; Barrett and Kathleen, 2019).

The circular economy is both a theoretical model and an official public policy. It is a model with a regenerative design, based on the usage of two material types. The first types are the renewable (nature) materials, which could be returned to the biosphere after their final economic usage. The second types are the non-renewable (technical) materials, which could be recycled many times without reducing their value or quality. The principles of the circular economy concept upgrade the 3Rs rules (reduce, reuse, recycle) and bring it down to 6Rs rules (reuse, recycle, redesign, reproduce, reduce, restore) The perception in this economic type is that the main value is created through the management of markets of resources but not through the production processes. (Barrett and Kathleen, 2019; Winans et al. 2017).

The circular economy is also an official government policy of many countries in the world. In People's Republic of China, it was legally implemented in 2009. The legislative package for circular economy in EU was adopted in 2015 (Moreau, Sahakian, Van Griethuysen and Vuille, 2017). In 2020, EU has adopted a legal definition for circular economy, which definition covers the abovementioned definitions for circular economy in the scientific literature (Regulation (EU) 2020/852). The EU model for circular economy has the purpose to achieve not only sustainable economic growth, but also a global competitiveness (Rathore & Sarmah, 2020).

That is why, every scientific research in both fields (digital and circular economy) could be perceived as relevant. On the other hand, the social costs analysis is a fundamental issue for the economic science, which has been always relevant to the existing economic problems. There is a Nobel Prize in economics in this field for the research of Coase, titled "The problem of social cost" (1960). At the same time, the studies in this area are continuing to be of interest for researchers and they are still searching for an answer to this question. (Davidson and Potts, 2016; Singer, 2018; Garnache, Mérel, Lee and Six, 2017).

Kapp (1970) says: „*Let me add, that there will be a considerable need of quantitative determination and accurate treating of the problems, concerning the environmental disruption and social costs* ". The quantification of this relationship is still being sought and it is bound up to controversial definition of these economic categories. There is a wide range of scientific research that has evaluated the value of human life, human health, the value of various environmental services, the effects of the global warming, the public concern about the location of waste treatment facilities in the vicinity of residential areas (Miranda & Hale, 1997; Porter, 2010).

According to Moreau et al., (2017) two main aspects in the scientific literature, related to the circular economy, are missing: „*the first one is a comprehensive view of the biophysical dimensions and introduction of the institutional dimensions and the second one is the social aspects* ". This fact outlines the necessity the institutional and social aspect in the context of circular economy to be more substantially studied.

The concept, concerning externalities, also contains some ambiguities. Berta & Bertrand (2014) have defined externalities as a "residual concept", which refers to the non-evaluated effects of the economic activities. But then, the definition, concerning external effect and its potential internalization are put in dependance on the definition of the relevant market to which these effects are applied. This in practice changes the evaluation of the relationship between transaction costs when internalizing the external effects and the potential benefits of this internalization. This in turns brings risk of wrong choice amongst market or non-market approach for internalization of the negative external effects. In this regard, the following question is arising, whether the externalities are exogenous to the market. Berta & Bertrand (2014) citing Arrow, (1969) indicate that it is surprising that nowhere in the scientific literature there is an accurate common definition of the term "externalities". The existing definitions are either general and difficult to use for analysis, or they are too specific and could be applied for special cases. The social costs of circular economy could not be examined without the use of the externalities' theory, which fact in the case of existing ambiguities in the theory is to some extent a scientific challenge.

Having in mind the abovementioned, the relevance of the current research is underlined by its subject area – the circular economy, as well as by its attempt to develop the understanding for quantification of the social costs.

## Purpose and content of the research

**The purpose of the research** is to bring out an alternative concept for social costs specifics as a structure, share and effectiveness in the process of introduction of circular economy in EU.

Thus, the purpose of the study defines its **object** – social costs are examined as a theoretical concept.

**A subject of the research** are models for social costs measurement in the context of their possible usage in the field of circular economy.

**Hypothesis of the research**: only a model, based on transaction costs, could be used for derivation of the social costs.

The presented hypothesis above is related to the assumption that the main definition in the economic theory, according to which Social Cost equal Private Costs plus External Costs is not universal (Pigou, 1954; Kapp, 1953; Berger, 2017) and could not be applied to the circular economy concept in the European Union.

In relation to the purpose of the research, the following **tasks** will be solved:

1. Critical analysis of the scientific literature in field of the purpose of the research and delineation of the limits of scientific achievements on the subject with a view to the possibility of their future upgrading;
2. Examination of the possibilities social costs to be derived by transaction costs.
3. Identification of the institutional failures in the circular economy concept in EU.

## Research methods and methodology

The present study makes a distinction between the term "methodology" and term "method". According to Thorn, (1980) methodology represents the logic of research, and it should be distinguished by the applied methods, which could be dialectical, econometric, statistical, or mathematical. Various methodologies could use the same methods. To avoid confusion between method and methodology, some authors are talking about "research programs", which apply one or more methods.

In this regard, the present study will perceive the term "methodology" as a concept which is close to content and nature to the term "research program", introduced by Thorn, (1980). According to the same author, the pragmatism requires the relevant theories to be evaluated not according to the realism of their structures, but to their contribution in solving the problems they analyze. Distinguishing the causes and effects, as well as the number of significant variables in relevant situation amongst the existing huge data and information, depends on the choice of the appropriate methodology of research.

In methodological aspect, the present research is based on the methodology of the **new institutional economics**. This methodology has certain features which will be presented in the context of analysis in accordance with the vision of Ménard, (2001):

**Theoretical basis.** Transactions and the costs, associated to them, are the basis of the theory of new institutional economics. The focus is put on studying the impact of various institutional environments on specific modes for transactions organization. According to this theory, transactions are the basis of labour division and the evolution of innovative technologies. Thus, transactions take precedence over the conditions of production which conflicts with the neoclassical view that underestimates the transaction costs influence.  These costs are function of the existing institutions, containing the ways for transactions organization and the allocation of property rights. In this context, the new institutional economics does not exclude the neoclassical analysis as an alternative theory, but it makes it more complicated and probably leads to its restructuring.

In this context, the present study uses the theoretical basis of the new institutional theory as it will be established on its propositions in the field of waste management and circular economy concept. The circular economy in the context of waste recovery organizations is examined as institutionalized imitating administrative market. This market is analyzed by identifying the level of transaction costs and the effectiveness of property rights allocation. It is also examined the interaction between the institutional environment and the management structure of this market in accordance with the vision of waste recovery organizations as special hierarchies.

The new institutional economics has not yet represented large enough set of concepts for studying issues through the interpretation of many facts and the correlations amongst these facts. That is why, every new research, which uses such set of methodology, has contribution to clarifying its theoretical conception, even though this research relates to specific issue such as the social costs of circular economy.

Modelling. In the context of new institutional economics, the neoclassical microeconomic models could not be directly used for institutions analysis. Their characteristics are related to consumers' rational behavior, the repeating organization of market transactions and these characteristics conflict to the theoretical principles of the new institutional economics.

The present research is trying to establish its own model, based on the traditions and theory of the new institutional economics. By establishing this model, the study is trying also to challenge the neoclassical model for social costs analysis and to create its own, even though this model could only be applicable to specific circumstances, for example the waste management processes in the context of circular economy concept.

The established model in the present research could be expanded in the future by analyzing a wider range of economic problems, if it is protected in accordance with the principles of the intellectual property. For the establishment of the model, the author has applied empirical statistical tests, which are published as articles in order the conclusions of the present monograph to be approbated (Nozharov 2018a, 2018b).

Testing. In the context of the present analysis, the process of collecting statistical data for testing the model is the main problem, insofar testing is perceived as a measurement that aims to establish whether the facts correspond to the predictions. Firstly, the circular economy concept is brand new, and it was introduced to the European legislation in 2015. That is why, the presented statistical data in the analysis cover a 5-year period, which fact does not allow a full-scale statistical testing with the appropriate reliability to be made. On second place, the statistical information needed could only be obtained through the Bulgarian public institutions, but most of them do not maintain open-access and public information registries. This fact complicates the testing of the proposed model, but we hope that soon there will be accumulated enough statistical information this model to be confirmed or rejected. At this stage, the testing of the model is indicative, and it is based on the fact that the model is logic and responds to the standard scientific infrastructure, as well as it sufficiently explains the economic reality of the case for which purposes it is created.

According to Ménard, (2001): „*Science requires measurement. However, the measurement itself is not a science: there are infinite examples for measurement in the history of science without using theoretical basis.* " If we try to paraphrase another citation of Ménard: *Counting of planets,…, does not prove that extraterrestrials exist!*

Consequently, empirical tests are important, but they don't have to be taken for granted.

Thorn, (1980) also opposes absolutization of testing by saying: *„It should be mentioned that speculative hypothesis have put foundations for many dramatical achievements in the field of natural sciences. The general theory of Keynes abounds speculative hypothesis and this could explain its durability as a starting point for new productive studies. "*

In accordance with the accepted practice in the Bulgarian scientific literature, there will be presented some methodological remarks, concerning the current research from the viewpoint of the methodological apparatus of the neo-classical theory:

In methodological aspect, it is confirmed the understanding that circular economy is a complex phenomenon, which includes ecological, economic, and political characteristics and its territorial aspect could not be limited only to the borders of the national economies. Having in mind that Bulgaria as an EU member-state participates in a complex interaction in the single European market, the analysis of the social costs in the context of circular economy concept further complicates the choice of appropriate methodological apparatus.

An approach, based on the convergence method where objects are studied simultaneously in several directions and sections (Lulanski, 2005). In the present research, regarding the organizational-typological aspect, the understanding for social costs in the context of the neoclassical theory is compared to the understanding of social costs in the context of environmental institutional economics. The structural-morphological aspect of social costs is analyzed through their form and constituent elements (e.g., their construction of private costs and negative externalities). From the functional aspect viewpoint, it is made an analysis of the social costs' behavior in the circular economy context. The essential aspect of social costs in the context of circular economy is focused on the rejection of possibility for negative environmental externalities creation. The genetic aspect of social costs is examined in the context of circular economy concept of EU creation in 2015 and subsequently the development stages of this process. The divergence of social costs is examined as variation of the general case of the neoclassical model of market in the context of an imitating administrative market, established in accordance with the circular economy concept in EU.

The social costs are also examined through the method of science abstraction, as in the context of neoclassical theory and respectively of the ecological institutional economics (which is also used for their analysis in the context of circular economy) some factors are confirmed as significant ones while others are ignored.

They are also used the logical and historical methods. The social costs in the context of circular economy are examined through their constituent elements in their logical interaction. And in the viewpoint of the historical approach and novelty of the circular economy concept in EU, the social costs for waste management are examined.

Analysis and synthesis. Through the methods of analysis, the complex construction of social costs in the context of circular economy is divided into constituent elements, as the focus is put on the functioning of waste recovery organizations. As a result of in-depth analysis of the waste recovery organizations, based on the methods of synthesis, a general model for explanation of the social costs is established.

Induction and deduction. Based on the induction method, some statements concerning social costs in the context of circular economy are made. Conversely, examining the social costs in their neoclassical understanding, some statements concerning social costs in the context of circular economy, are made.

Comparative analysis. In the present study, alternative theoretical views for the social costs in neoclassical context and in the context of environmental institutional theory, are examined. They have been analyzed the pros and cons of both theoretical views for social costs.

Descriptive analysis. The social costs in the context of circular economy are examined through the understanding for institutional ineffectiveness.

Quantitative analysis. It is established a vision for comprehensive statistical model for social costs analysis in the context of circular economy. This model is partially approbated because of the limited statistical data available publicly and the short existing period of the circular economy concept in EU.

## Structure of the research

The logic and consistency in structuring the present study is in accordance with its purpose, object, subject and methods of research. The content covers three chapters, conclusion, list of references and appendices.

**The first chapter** represents the theoretical basis of the understanding for social costs in the context of circular economy. It is presented the general view of neo-classical economic theory, as well as the view of environmental institutional economy for social costs. The main theoretical views and their criticisms in the scientific literature are examined in detail. In this way, some conclusions were drawn to serve as starting point for establishing a model for relationship analysis between social and transaction costs in the context of circular economy. This helps to be outlined the limitations of the scientific achievements in this thematic area. Such an approach implies relationship between circular economy and waste management – from general to private, which serves as basis for the analysis in the present study.

**The second chapter** represents a critical analysis of the scientific literature, devoted to the relationship between social costs and waste management by supplementing the theoretical foundations, made in the first chapter. The main elements are presented both individually and in correlation of the neo-classical model for social costs in the waste management process. Given the limited number of scientific publications devoted to the social costs of circular economy, this appears to be a logical transition from induction to deduction which has the purpose to deepen the theoretical focus of the research.

**The third chapter** represents the modeling of the correlation social costs and transaction costs in the context of circular economy concept. The analysis starts with the examination of the theory of institutional ineffectiveness as possible basis for establishment of analytical model for measuring the correlation social-transaction costs. In the second section of the third chapter, there have been proposed a model for investigation of the correlation social-transaction costs in the context of circular economy concept. There have been examined also the existing models in theoretical aspect in order to be distinguished the new model and to be outlined the pros and cons of these models. They are also outlined the main requirements that must be covered by a model of an imitating administrative market in the field of waste management with government intervention. A partial statistical testing of the proposed model is also made.

**The conclusion** summarizes the main implications of the research, and it represents some recommendations to future studies, which could upgrade the contributions of the present one.

# FIRST CHAPTER

# THEORETICAL BASIS OF THE UNDERSTANDING FOR SOCIAL COSTS IN THE CONTEXT OF CIRCULAR ECONOMY CONCEPT

The introduction of theoretical basis for the understanding for social costs in the context of circular economy concept will be made through the critical analysis of the scientific literature in this field. In this way, there will be made conclusions that will function as a starting point for the construction of a model for analysis of the correlation between social and transaction costs in the context of circular economy concept. There will also be outlined the boundaries of the scientific achievements in this field. Such an approach supposes a correlation between the circular economy concept and the waste management procedures – from general to private and it could facilitate the analysis.

## 1.1. Theoretical problems of social costs

The discussion, concerning the essence of social costs has been existing for more than 100 years in the economic science. In this regard, it could be mentioned the famous work of Knight, (1924): "Some fallacies in the interpretation of social cost". However, the defining and management of social costs is still a significant challenge for the economists around the world. The attempts for introducing policies based on taxes (for internalization of the external costs), subsidies (for stimulating the positive externalities) or shrinking the market mechanisms and replacing them with government interventions – do not give the desired effects. The choice between „market failure“and „government failure“ in most of the cases does not present the government as an efficient opportunity/alternative to the market (Cheung, 1978).

There exist two main economic theories which analyze this issue: the neo-classical economic theory and the institutional environmental economics.

The first view - of **the neo-classical economic theory** for social costs is a dominating approach (Cheung, 1978; Lichfield, 1996; Demsetz, 1996; Medema, 2011; Gruber, 2012; Pigou, 2013; Schlag, 2013; Mohrman, 2015). According to this theory, social costs are:

**Social costs = Private costs + Externalit (1)**

Private costs are those costs that the buyer pays to the seller.

The external factors (favorable or unfavorable) are presented as uncompensated externalities, in which the behavior of a person or a separate company affects the well-being of a third-party observer who has not given his consent for them (Lichfield, 1996; Cheung, 1978; Hu, 2013). The concept for externalities (in the neo-classical theory) refers to interdependencies which are external to the price system and therefore they are not considered in the market transactions (Demsetz, 1996).

According to McClure and Watts (2016), the externalities should be differentiated into two categories: monetary and technological. The „price externalities“are those in which the price policy of a company from a separate economic sector leads to changes in the industry price or the competitive conditions of the other companies in the sector. These actions could harm the financial interests of the other companies in the sector, but at the same time they could lead to a transfer of wealth to the customers. On the other hand, „technological externalities"are those that are accepted in the (neo-classical) economic theory as externalities – pollution, traffic jams, etc. The authors (McClure and Watts, 2016) consider that if the two types of externalities are not being distinguished this will lead to favoring of the government intervention in the market conditions.

The social costs are opportunity costs of the resources, used for the production of goods, various combinations of production factors and business arrangements based on the value of the output, assessed by the market (Coase, 1960; Mohrman, 2015). In the neo-classical economic theory, these costs are perceived as external (exogenous) for the economic system, which are rare and accidently happen in extremely conditions (Swaney, 2006; Schlag, 2013; Hu, 2013). Some kind of market failure is seen as a reason for them.

In a perfect competition condition, the private and social costs will be equal one to another (Stigler ,1966; Berta & Bertrand, 2014).

As possible solutions are perceived: Pigou's taxes, Coase theorem, regulations for control and management, emission trading permits (Cheung, 1978; Lichfield, 1996; Medema, 2011; Hu, 2013).

A brief classification of the possible solutions is proposed by Cheung (1978) and Lichfield, (1996):

- Through Pigou's input tax: restricting the owner's behavior by instruments, such as taxation or compulsory compensations for the affected parties.

- Through Viner’s output tax: balancing the production price and the social production costs based on marginal values.
- Through considering the Coase’s transaction costs: weighing the benefits for the one party against the losses for another, recognizing that the social costs are reciprocal and the implemented policy must aim to avoid more serious harms.
- Through Cheung’s aggregated balance: compensates the costs, performed by the contractor against the sum of the values of effects generated, regardless of whether they will be included in the organizer’s calculations: agreed or not.

Litigation is also an opportunity, but this approach has many disadvantages: it is too long, payment of legal fees, which could be higher than the damages for the victims, uncertain outcome of the dispute (Cheung, 1978). In many cases, courts are studied as cases not from the viewpoint of economic effectiveness, but from the viewpoint of fairness (Frischmann and Marciano, 2015). However, lawsuits could have a deterrent effect on generators of externalities. This leads to lower number of the cases filed, which in turn avoids the shortcomings of this instrument for overcoming the negative externalities. The victim could strategically sue without considering the amount of the legal costs, relying on the fact to reach quick settlement. All sorts of actions are possible in order to make this approach (of legal actions) still viable (Kaplow, 1986).

The most popular neo-classical approach is the Pigou’s input tax (Pigou, 2013). If this tax is studied in-depth, it could be determined that its main purpose is to equalize private costs to the social costs, related to the main activity of the company, that generates externalities. Thus, responsibility the owners of the company that generates a negative external effect will be realized towards the persons affected by it, and its production activities will be limited in areas where the damage is high. The marginal tax will be equal to the marginal damage, caused by the pollution (Cheung, 1978). This approach involves active government intervention. The government intervention is usually justified by the following motives. By stopping or limiting the unethical activities, however for the economic theory, these activities are subjective or institutional concepts (like „crimes “). Other motives are the inefficient allocation of incomes and resources. Presence of high transaction costs to achieve private bargaining for solving the problem. According to the scholarly criticism (which will be presented below), Pigou’s analysis is overly simplistic and contains implicit bias to constant government intervention. In this approach, it is missing a comparison of the total social product, obtained under opportunity/alternative social arrangements. However, the government intervention intended to remove the externalities in many cases creates other externalities (Cheung, 1978; Demsetz, 1996; Frischmann and Marciano, 2015).

There are many scholar critics of the Pigou’s view (Pigou, 2013) about the social costs:

**First,** social costs are reciprocal. According to Frischmann and Marciano, (2015) and contrary to the conventional economic analysis, the damages caused by the negative externalities are co-generated and caused by reciprocal, interdependent actions, and actors. The reciprocal nature of social costs requires not only the causer of the externalities to consider the costs imposed on the affected third party. It requires also the “affected third party” to consider the costs, imposed by the producer (i.e., the lost profits) of any reduction in its activities because of an imposed adjustment-tax. This means that the reciprocal nature of social costs requires that actions should be taken only if the marginal costs for controlling the damage are lower than the marginal profit for the damaged individuals (Coase, 1960; Regan,1972; Cheung, 1978; Medema, 2011; Schlag, 2013; Mohrman, 2015). On the next place, the reciprocal nature of the problem requires not only to be restrained the damaging party, but also to be restrained both the damaging and damaged parties in such a way that the gain to one party cannot be greater than the loss to the other, in total and on the margin. The use of taxes (subsidies) to correct the externalities while considering their reciprocal nature will require bilateral taxation, which is administratively too complex and expensive. Very few public policies could achieve such an effectiveness. Neither the government, nor the companies have free and complete information and the costs for enforcing the adjustments and they could not be determined to be negligibly low (Cheung, 1978; Schlag, 2013; McClure and Watts, 2016).

**Second,** „the existing of the problem social costs in itself indicates the presence of transaction costs„ (Cheung, 1978). Transaction costs are partially responsible for the market failures, and they really influence the functioning of the markets. Once the transaction costs are considered, then a realignment of the property rights will occur only when and to an extent, according to which the related profits of the production values exceed their costs. This calls into question not only the optimality of markets but also the availability of the markets to direct the resources to more valuable goals. In this regard the legal (institutional) regime, which define and apply the property rights, starts to be significant.  At the same time, the neo-classical model of Pigou (Pigou, 2013) has assumed that transaction costs equal zero and he has not taken them into account (Coase, 1960; Medema, 2011; Ménard, 2013). However, the perfect competition model requires publicly available complete information about prices and technologies. Thus, transaction costs could be perceived as a barrier to access such information, which will limit competition. Then the pricing system will not solve the problem with the resource allocation in the right way. In this way, transaction costs are obstacle to the existence of perfect competition, and they reject the assumptions, based on its existence (Regan,1972; Demsetz, 1996; Frischmann and Marciano, 2015; Mohrman, 2015). Still, it should be discussed the question whether the absence of transaction costs and the possibility property rights to be transferred, is analogous to the presence of perfect competition (Medema, 2011). Determining the level of transaction costs leads to various opportunities for solving the problems, related to social costs. They could be high, low, or indeterminate. When transaction costs are very high, this could justify the government intervention and vice versa (Cheung, 1978). Transaction costs, which are related to problems, concerning the environment, are the costs which exclude "free riders" from the resource consumption, which they have not paid for or whose consumption they are abusing. These costs occur when the property rights are not defined or they do not exist, as well as when they are incorrectly or unclearly defined. Natural resources are very often characterized as general resources. However, serious problems occur whenever property is defined as general rather than private (Cheung, 1978). It should be indicated that there also exist problems, related to the definition of transaction costs, not only with social costs and externalities. According to Dahlman, (1979) and Coase, (1988), transaction costs are: "*search and information costs, bargaining and decision costs, policing and enforcement costs*". But according to Allen, (2015) this is only a list of transaction costs and not a proper definition. He believes that there is a confusion between information and transaction costs. And that it is important to be made a distinction between economic property rights, legal property rights and transaction costs, as well as to be developed the understanding for the relationship amongst them. According to Allen, (2015) transaction costs are the costs for establishing and maintaining the economic property rights, which are defined as an opportunity to freely exercise choice. Property rights are complete when all the attributes of the thing are possessed, and they are perfect when actual choice can be fully manifested. Transaction costs are determined in terms of perfect property rights. When transaction costs are positive, the optimal degree of rights is not perfect, and the degree of wealth depends on the property rights allocation (Allen, 2015). This in turns influence the choice of an approach for coping with the problems, related to the transaction costs.

**Third,** lack of determination or incorrect determination of property rights in terms of externalities (Coase, 1960; Cheung, 1978; Lichfield, 1996; Demsetz, 1996; Frischmann and Marciano, 2015; Mohrman, 2015). According to this understanding, what is traded on the market is a set of rights (which gives the opportunities certain activities to be performed), but not physical goods (including also factors of production). Consequently, a key issue in the analysis of the definition, is the allocation of rights and the mechanism of their transfer (Ménard, 2013). However, the law is not an exogenous factor to the economic system, but an endogenous one. Then, if the transaction costs are high and they hinder bargaining, then the incorrect property allocation by the law will lead to inefficient resources and incomes allocation (Schlag, 2013). When delineating property rights, the legal system efficiently, but imperfectly balances the full set of benefits and costs as it need to also consider the reciprocal character of the problem (Frischmann and Marciano, 2015). In many cases, social costs occur from economic activity that leads to transformation of freed goods to scarce goods. It is also important to be considered the character of the external effect – if it is permanent, periodically occurring, or random. Because this affects the decision-making process as in case of accidental events, an insurance can be provided, which excludes the government intervention. Pigou does not specify property rights (for example in the case of roads), which casts doubt on his theory Property rights are a necessary constraint on every economic decision (Cheung, 1978). For persistent externalities, there are other possible solutions that exclude government intervention. According to „Coase's theorem"(Coase, 1960), if the property rights are well defined, the number of participants is low and transaction costs could be negligible, then the private bargaining will lead to an efficient outcome, regardless of the initial property right allocation. Under these circumstances, the government needs to limit its role in promoting bargaining amongst interested individuals or groups (Cheung, 1978). The expected degree of ineffectiveness, which will remain after bargaining, will decrease as the transaction costs decrease (Regan,1972). In addition, contracts as mutually beneficial transactions amongst two or more parties will always implicitly include public interests in order to regulate social costs by limiting the subject of the contract, the relationships of the contracting parties with the regulation authorities and the opportunity the Court to interpret and modify the contract (Hoffman & Hwang, 2021).

**Fourth,** uncertainty effect of externalities (Cheung, 1978). In many cases, there exist causation of external effect, doubt about its actual causer, doubt about the degree of its negative effect (and its compensation by the positive external effect, created by the same activity). Then it could hardly be regulated both by the government and a bargain that satisfies all the contracting parties. Examples about the negative external effects have been already given. An example for a compensating positive external effect in the contemporary economics is the creation of knowledge, which is readily available and diffuses rapidly. This stimulates competitiveness and allows the rest of the companies in the sector or those in other sectors to continuously develop their technological level, which fact will be of benefit also for the customers. If this effect is not considered, this will lead to loss of welfare. Solving these problems will require at least additional information (or scientific) costs (Cheung, 1978; McClure and Watts, 2016).

**Fifth,** the concept of so-called „government failure“ (Cheung, 1978; Lichfield, 1996). According to the public choice theory, idealizing government intervention is wrong. There exist the so-called „coordination costs“, which very often are significant in size and could make the government intervention ineffective market alternative (Medema, 2011; Allen, 2015; McClure and Watts, 2016). In most cases, governments are not guided by supreme social goals, but by asserting their power and influence. In this process the lobbying intervenes and influence on the government decisions is exerted by the party that have greater opportunities to influence the government (companies, causing the externalities or affected parties – voters). The government not always have the stimulus to be much more informed than the affected parties. It will make such efforts only if these actions will help the government to be re-elected. Just the opposite, the affected individuals (in the case of absence of the free-rider problem) have strong stimulus to be informed. The government makes decisions, based on a short-term horizon (a mandate that is limited by political elections), because it is influenced by its desire to be re-elected. The state agencies have an interest in increasing their budgets and inflating the public measures they are responsible for, which fact will increase their influence. After all, there are no guarantees that the government intervention will solve the problem with social costs in efficient way and will not lead to additional problems, related to the resources and incomes allocation (Coase, 1960; Mueller, 1976; Cheung, 1978; Lichfield, 1996; Frischmann and Marciano, 2015). The approach of using taxes for solving the problem with externalities has other disadvantages. This avoids critical examination of the net social product under one or another assignment. And the reciprocal character of the problem with social costs requires a tax levy on the injured in such amount that accounts the steps not taken in order the damage to be reduced. Also, the effort for receiving information in the imposing a tax on the externalities could not be financially evaluated without a clear idea of the future real improvement of the net social product (Mohrman, 2015). If the property rights are well defined and transaction costs are low, then the market decision is always more efficient than the government intervention. However, even Coase (Coase, 1960) does not completely exclude the government intervention. In contrast to the Pigouvian view, according to Coase the government is not perceived as benevolent and idealized agent who is trying to do its best in order to solve the problem with negative externalities. On the contrary, the government is seen also as a potential generator of transaction costs (Ménard, 2013). The government intervention even if it is efficient in the field of one economic sector, could disrupt competitiveness in the case of substituting products from other economic sectors or from abroad (Stabile, 1993). Ultimately, the use of government intervention for solving problems, related to social costs is not out of question but it could be of high risk, and it should be a last resort.

**Sixth,** the use of the concept „Pareto efficiency"in the problem solving process with social costs is not reliable enough (Cheung, 1978). This concept proposes a way for evaluation of one state of resource allocation against another. In this way, when the private costs and social costs are different, then the decision-making process in the private sector (which trends to ignore the externalities) will lead to inappropriate resource allocation (Lichfield, 1996). However, according to the "Pareto efficiency" concept, it is not clear what is the level of reaching a maximum value. Instead, it is analyzed whether one individual could gain profit without another individual loses it (Cheung, 1978). That is why, if we take into account the total product of both activities (activity, causing the externalities and affected activity), in some cases the non-liability regime will be preferable to the liability regime. Recognizing the reciprocal nature of the social costs needs to pursuit a wealth maximizing goal The equalization of the marginal private costs and social costs does not guarantee wealth maximization (Medema, 2011; Mohrman, 2015; McClure and Watts, 2016). Some clarifications, it is correct to talk about “distribution of welfare” rather than “distribution of wealth”, because the effects of the distribution affect not only the parties who directly participate in the process (Regan,1972). Another possibility, different from the “Pareto efficiency” principle when comparing alternative social arrangements is to correctly compare the total social product, obtained by these different regimes. This could be done on basis of the Kaldor-Hicks rule, which is different from the Pareto Efficiency principle. It only requires a theoretical possibility of compensation amongst parties without the necessity this compensation to be realized in relation to specific party if the total welfare is increasing. From this viewpoint, the comparison between private products and social products, realized by the government will not be efficient. Yet, efficient Kaldor-Hicks distributions do not necessarily lead to better distributions according to the Pareto efficiency principle (Coleman, 1980). The government intervention (implementing liability) will influence the level of production and costs not only to the damaging activity, but also to other related activities in the supply chain. That is why the accent on the comparison between private product and social product will lead to incorrect results in the economic viewpoint, including in cases of correction through the law (Coase, 1960; Schlag, 2013; Frischmann and Marciano, 2015).

**Seventh**, the effect of introducing the problems, concerning aesthetics, morality, and justice in the analysis of social costs. According to Mohrman (2015), who has cited Knight (1951), studying the problems of economics of welfare dissolves with the problems of aesthetics and morality. And finding an approach how to be evaluated these effects without market values is a difficult action. According to Knight himself (1951), the physical concept of welfare poses a dilemma. Welfare could hardly be studied only as a physical phenomenon. It must be subjective or a matter of values in an objective sense, but not a physical reality. However, on the other side, there must be causal or functional dependence of welfare on measurable objective quantities, otherwise, the term welfare will have no economic meaning. Such a measurable objective economic basis of welfare is the income. However, Knight (1951) reflects whether welfare is an issue, depending on the total income of the nation or it depends on the average incomes of households. According to him, it is wrong to accept income as homogeneous, consisting of a single final product, which fact also includes the issues, concerning its structure and allocation. Income consists of services consumed or of additional “capacity” to produce such services. It is disputable whether the welfare could be treated as a mathematical function of consumption qualities of measurable goods and services (including the utilities and inconveniences they create) What society members (consumers) want and what they must want are two different questions. Other authors except Knight (1951), for example Regan,(1972), question the rationality of individuals in the context of social costs determination. The rational individual behavior does not always guarantee group rational behavior. The freedom as a value is related both to the desire and ethical ideal for responsible behavior. An important element of this reasoning is the synchronization of the freedom with curiosity and desire to see what outcome will follow, as opposed to the single-minded pursuit of a previously desired goal. This fact changes the vision of predictability of the relationship for maximization of given goals by given means and hinders the possibility of scientifical definition of the welfare. In the real life the family (households) is the basic economic unit. There are “countless” other units:

political communities, voluntary groups and etc. – operating economically to various degree in different areas and they are not affected by the general abstract economic principles. The synchronization of group actions depends on the possibilities for reaching agreement on the normative values, introduced by the Law. The alternative to the open market system is the regulation by a central authority, which means a tyranny of the majority. Therefore, a key issue to the analysis of welfare is the freedom protection or its limitation in favor of another values Knight (1951).

The understanding of neo-classical approach, which studies the public preferences regarding the consumption of certain goods as a function of the sum of individual net utilities, is also contested by Söderholm and Sundqvist, (2000). According to them, this would mean that the calculation of benefits over costs determines behavior of individuals and society. However, the ecological evaluation of individuals in many cases is based on the deontological (or rights-based) approach in the decision-making process. The deontological approach involves judging whether the action being performed is right or wrong, regardless of its consequences. However, when the decision of consumers for protection of their rights includes a monetary payment (WTP) in many cases they refuse to pay. However, individuals could be seen in two roles: as consumers with private preferences (oriented towards what is best for them) and as citizens with public preferences (oriented towards what is best for the society). This makes public values important, because they define what is right or wrong, as well as the related to them behavior norms, regarding the public goods or goods with complex nature. This concept makes a connection with the dilemma for the price of the human life. How much of the public budget should be spent on saving every endangered human life and is it possible the human life loss to be monetary compensated. The answer to this question involves a combination of cost-benefit analysis and the public discourse (Söderholm and Sundqvist, 2000).

Litigations as an opportunity to solve problems, caused by negative externalities also includes the issue of justice. In many cases, Courts consider these cases from the economic effectiveness viewpoint, not from the viewpoint of justice (Frischmann and Marciano, 2015).

Of the abovementioned seven groups of criticizing theories in relation to Pigou's view (Pigou, 2013) for solving problems with social costs, the most popular critics are those of Coase (Coase, 1960). However, the Coase's theorem as an alternative to Pigou's approach also faces serious criticism. Many of the externalities include large number of stakeholders, high transaction costs and public goods (which are public property) such as air and water. In such situations, the private bargaining does not work as a remedy (Medema, 1994; Mäki,1998; Allen, 2015; Mohrman, 2015). If social costs are the result of the lack of certain property rights, then whether the right solution is to define such rights and bid for them. For example, if a system of "parent leave" permits is created that can be traded by workers with higher wages, then this must lead to higher economic effectiveness. However, at the same time, it will damage the sense of community and will erode the social capital. The market does not provide enough information about social costs of consumers to make socially responsible decisions (Stabile, 1993).

The neo-classical economic theory is not monolithic. It does not envisage a single solution to the problem of social costs. In the cases of negative externalities, there should be compared the possible "alternative social arrangements” (markets, companies, government, and their mixed structures) and to be evaluated their respective costs and benefits (or revenues). However, it is more correct to be compared the total public product, received by the various arrangements. All possible options involve the transfer of rights (of pollution and damage) with their accompanying transaction costs (Ménard, 2013).

There are many challenges in the neo-classical theory, which complicate the analysis of the social costs in the context of the circular economy. Equalization of private and social costs related to the activity of a company that generates an external effect is not enough to solve all problems in this area. But on the other hand, these challenges make it possible, through an analysis of the circular economy, to derive a new view of social costs in general.

The second major view is that of **the institutional environmental economics**.

The prototype of modern publications in this field could be found in Kapp’s studies. His publication „Environmental disruption and social costs: a challenge to economics“(Kapp, 1970) has raised the question of necessity of adequate economic theory, through which social costs could be analyzed.

The definition of this term is important because Kapp assumes that phenomena, related to environmental disruption are perceived as social costs, but they are far from exhausting the concept of social costs. He assumes social costs in a wider context, related to the disruption of the “human natural and social environment”. According to him these (social) costs are everywhere borne by economic and politically weaker elements of the society, such as seasonal workers and minorities. He believes that the accounting of these costs would limit the possibilities of racism, chauvinism and authoritarianism through higher standards, available to all people, concerning health care, educational and cultural systems.

The author advocates the internalization of social costs of the economic system. He has contributed to the measurement of the social costs. According to him, it is wrong to work with static process output pollution coefficients and linear correlations of national or regional output regarding environmental disruptions.

The publication of Kapp (1970) rises many questions, which are still not correctly answered. The first of them is related to the missing of universal definition of the social costs. Its presence is important from the viewpoint of the possibility for accurate measurement of social costs, as well as the measurement of the socially benefits of the certain public goods, which are related to concrete social costs.

On second place, he underlines the necessity of accurate measurement of the cumulative effects of the correlations amongst various social costs, which could generate non-linear impact and even new social costs.

On third place, beyond the narrow sample definitions for social costs (environmental disruption, violation of labor rights) he outlines some very broad limits of their possible definition. These limits, he relates to the human and social environment disruption, which is the basis for individual wellbeing. However, according to his views, they go as far as the concept of human "amenities". But he limits the notion of these "amenities" by equating them with the natural assets on which they are based. All these questions, posed by Kapp (1970) have been looking for solutions till now, as the present research also tries to answer these questions.

**The institutional environmental economics** studies social costs as:

**Social costs = Social opportunity costs - Privat (2)**

According to this view, the definition for social costs includes: all those harmful effects and damages that other persons or society suffer (directly or indirectly) because of production processes and for which private entrepreneurs are not easily held responsible (Kapp, 1971; Swaney and Evers, 1989; Stabile, 1993; Hu, 2013). And other authors study social costs in the broad context, such as Lewin (1982), who related them to the missed social benefits from the point of view of the concept of "opportunity costs". In general, in institutional environmental economics, “social costs” are a much broader term than “external costs”. They concern much more to spillover effects on third parties in a static partial institutional equilibrium. They are an unusually complex set of interdependent and delayed cumulative effects in which multiple, often distant individuals or groups of individuals, or entire societies, or human-sustaining environmental systems are harmed (Kapp, 1970; Swaney and Evers, 1989; Stabile, 1993; Hu, 2013).

The representatives of the institutional environmental economics concern that the society has a fixed stock of what can be called social capital (environment, level of knowledge, etc.) in a given time horizon. In this concept, the more social capital is used, the faster it will depreciate. This depreciation of the social capital leads to social costs. The criterium for social value in relation to the analysis of social costs should be based on the continuity of human life and the normal recovery of the society through the instrumental use of knowledge. Because a society who does not preserve its social capital, risks not to survive. And the market always makes a trade-off between effectiveness and fairness, which calls into question its function as an arbiter (Tool, 1979; Stabile, 1993).

The examples of Kapp (1970) for „social costs“are:

Firstly, the disruption of the environment, which is related to „disruption over certain threshold levels of the sum of all external conditions and influences, affecting the life and development of people and human behavior and respectively the society”.

Second, the labor conditions: “such phenomena as occupational injuries and accidents, rhythms of work, harmful to human health, inadequate living conditions, harmful noise levels, forced and uncompensated adaptations to the structural changes, workers ‘compensation systems, which become inadequate because of inflation and last but not least the monopolistic determination of the real estate values and the rents in congested urban areas.

Thirdly, he concludes that this statement could be examined in a broad sense – disruption of the “human natural and social environment”. As part of this broad sense, the author accepts the reduction of “amenities” provided by “natural resources” as social cost along with the reduction of resources, provided by the “natural assets”. In such type of growth “the rise in the consumption or in investments is possible not because of the net production, but in the expense of decrease in our natural assets in the form of resources and amenities”.

In the mentioned cases of costs redirection, there is a risk the negative effect on the society to be multiplied. When the cost redirection is happening in a highly competitive environment, then the other companies will be forced to follow the example of leaving the sector. Then the rational decisions of the market subjects will not lead to the socially desired outcome, even in the presence of perfect competition conditions. Therefore, the assumption for autonomy of the economic system of neo-classical economists ignores this effect and does not find the right way to solve the problem (Swaney and Evers, 1989).

According to the institutional environmental economics, the reasons for emergence of social costs are complex interconnected phenomena, that are not sufficiently well understood. Kapp (1970) thinks, that „it is not enough to be indicated the obvious relationship between the increase of population and the concentration of population in the urban agglomerations, which arise under the impact of high productiveness rate of workers, which is a result of the development of science and technologies…“. In contrast to neoclassical economists, according to Kapp, the "negative externalities" are (internal) systematic permanent problems but not random and external (exogenous) problems for the economic system. It is a mistake to be considered that the economic system is a closed and autonomous system, and it is not a part of the natural and social environment of the human society. That is why, he uses the term „cost-shifting", but not "externality". In contrast to the understanding in the neo-classical economic theory, the institutional environmental theory perceives social costs as a problem that affects to a large degree the entire society, but not just its individual representatives.

The understanding of social costs as internal costs for the economic system enables them to be defined as predictable ones and to be systematically regulated and managed (Swaney and Evers, 1989; Stabile, 1993).

Kapp, presented by Moreau et al., (2017) concerns that “externalities” are mostly an institutional problem, paying attention to three important aspects:

Firstly, their valuation leads to limited understanding of their heterogeneous character. According to him, it is a mistake to work with static process output pollution coefficients and linear correlations of the national and regional production concerning the environment disruption. The big problem, according to Kapp is that part of the damaged resources and values do not have market price and they are in a great extent public goods. In order to prevent their damage, the society needs high level of information for their utility value. However, according to him, for this purpose it is not enough to determine the monetary equivalent of human health and life.

Secondly, the legislation determines the formal border between costs, done by the private agents and those, which could be passed to the society.

Thirdly, the competitiveness forces private agents to shift costs to the others, but not vice versa – the competition to solve the problem.

Kapp (1970) declares that internalization by the economic “subsystem” of costs is related to the “damaging non-market effects”. Thus, according to him, there should be sought an achievement of social efficiency and optimality of the entire macrosystem. Because the drive to rational maximization of the utility at microlevel leads to lack of social optimality at macrolevel. When social optimality is sacrificed, then from the viewpoint of the macrosystem one will sacrifice „*with impunity these values and goals, which from the viewpoint of the macrosystem could be highly important and actually are the foundations of the individual well-being and survival* “. The disruption of the environment and the perception of the economic, social and natural systems as opened ones, gives the opportunity to determine the socially desired macroeconomic purposes, which are oriented to the sustainable equilibrium between them by concerning the interests of the entire society. To prevent the current status of the environment, when the social costs are present, it could be late and even to have hard social consequences (Kapp 1976; Swaney and Evers, 1989).

The institutional environmental economics have perceived the following solutions as possible ones: preventive measures at the earliest investment stage, but not ex post facto removal of the pollution or punishing the guilty party; regulations for command and control; stimulating the investments in polluting abatement technologies; ecological standards and institutional framework (Berger, 2008; Hu, 2013).

According to Swaney and Evers, (1989) and Stabile, (1993) the changes in the economic structure are appropriate for analysis of social costs, because there is a built-in tendency the market system to create new institutions (and new technologies), which generate externalities. Given this, the institutional economics will be an appropriate tool for conducting such type of analysis and for finding the suitable solutions.

The theory of social economics also tries to make an analysis of the social costs through the relationship between social costs and social values. According to Schweitzer, (1981) the right of free choice does not mean that the consumers have the right to make mistakes, which are harmful to the others. There are areas of freedom, but also such areas, which are controlled by the public authorities. The boundary between these two areas is determined by anything which creates a risk to the society and should be forbidden or put under public control. This analysis could be hardly accepted as an individual one and rather represents a supplement to the institutional environmental economics.

The main difference between institutional environmental economics and neo-classical economic theory is the understanding of social costs as internal ones for the economic system. In this way, the perception of these costs as systematic and constantly generated gives opportunities they to be determined as predictable and to be systematically regulated and managed. On next place, the definition of institutional environmental economics for social costs is much broader than this of neo-classical theory. This enables their comprehensive analysis, although it complicates it.

The presentation of the general vision of neo-classical economic theory and of the institutional environmental economics for social costs aims to build a theoretical foundation for analysis of social costs of circular economy.

**The following conclusions about social costs from the analysis of theoretical problems could be made:**

Pigou-type taxes and subsidies are not accepted as a possible efficient decision for internalization of social costs in the current research.

The Pigou's analysis is criticized as overly simplistic and that contains implicit bias towards constant state intervention. In this approach a comparison of the total social product, obtained under alternative social arrangements, is missing. In many cases state intervention to remove externalities itself creates other externalities (Cheung, 1978; Demsetz, 1996; Frischmann and Marciano, 2015).

Some of the critics, concerning the viewpoint of Pigou (Pigou, 2013) for internalization of the social costs, could be summarized as follows:

**Firstly,** social costs are reciprocal in nature. The use of taxes (subsidies) for correction of the externalities, concerning their reciprocal nature will require bilateral taxation, which is administratively very complex and expensive.

According to Frischmann and Marciano, (2015) the damages, related to negative externalities are co-generated and they are caused by reciprocal, interconnected actions, and participants, which fact is not considered by the neo-classical analysis. The reciprocal nature of social costs requires not only the causer of externalities to consider the costs imposed to the affected third party. It requires also the "affected third party" to consider the costs, imposed by the producer (i.e. lost profits) of any reduction of his actions as a result of the imposed adjustment – tax.

The use of taxes (subsidies) for correction of the externalities in considering their reciprocal nature will require bilateral taxation, which is based on universal access to perfect information. Neither the government, nor the companies have access to free and complete information, and the costs for imposing adjustments could not be determined as negligibly low (Cheung, 1978; Schlag, 2013; McClure and Watts, 2016). This will make such type of taxation too expensive and complex.

**Secondly,** the neo-classical model with which Pigou works, accepts transaction costs as equal to zero and this model does not consider them (Coase, 1960; Medema, 2011; Ménard, 2013).

While the „existence of a problem with social costs itself shows the presence of transaction costs„ (Cheung, 1978). In this way, transaction costs are partly responsible for the market failures, and they affect the markets functioning. The perfect competition requires public and complete information about prices and technologies. Transaction costs could be accepted as a barrier for access to such type of information, which fact will limit competition. This cannot be solved by Pigou's taxes.

**Thirdly,** Pigou does not specify the property rights (for example in the case of roads), which casts doubts on his theory. And the property rights are a necessary constraint on any economic decision (Cheung, 1978).

The lack of determination or misunderstanding of the property rights in externalities is of key significance (Coase, 1960; Cheung, 1978; Lichfield, 1996; Demsetz, 1996; Frischmann and Marciano, 2015; Mohrman, 2015). According to this understanding, what is traded on the market is a set of rights (enabling the performance of certain activities), but not physical goods (including factors of production). Consequently, a key issue in the analysis of social costs internalization is the allocation of rights and the mechanisms of their transfer (Ménard, 2013).

For persistent externalities, there are other possible solutions, which exclude state intervention. According to „Coase theorem “(Coase, 1960), if property rights are well-defined, the number of people involved is small and the transaction costs could be insignificant, the private bargaining will lead to efficient outcome, regardless of the initial allocation of property rights. In these circumstances the government should limit its role to encourage the bargaining amongst stakeholders (Cheung, 1978).

**Fourthly,** according to the theory of public choice, the idealization of the state intervention is wrong. This is the concept of the so called „government failure“ (Cheung, 1978; Lichfield, 1996). There exist the so called „coordination costs“, which are very often significant and could make government intervention inefficient alternative of the market (Medema, 2011; Allen, 2015; McClure and Watts, 2016). In most cases governments are not guided by higher social goals, but by asserting their power and influence. In this process lobbying intervenes and influence on the government decisions gives the party with the biggest possibilities to put pressure (companies, causing the externalities or affected parties-voters). After all, there are no guarantees that the government intervention will solve efficiently the problem with social costs and will not lead to additional problems in the allocation of resources and incomes (Coase, 1960; Mueller, 1976; Cheung, 1978; Lichfield, 1996; Frischmann and Marciano, 2015). Also, the efforts for receiving information about taxation over externalities could not be financially evaluated without the clear idea for future improvement of the net social product (Mohrman, 2015).

**Fifth,** the existing waste taxes create wrong price signals, which influence the private costs. They are always determined as the second-best alternative, because of their inherent inaccuracy and their perception as an element of competitiveness in attracting investments (Domenech and Bahn-Walkowiak, 2019).

The problem with private costs, according to Porter (2010) is that many of the participants in the process of waste management face prices which are below the marginal private costs, let alone marginal social costs. In most cases, the price of waste collection is covered by the municipality funding, based on local taxes. The amount of the tax is uniform and flat, and it is independent from the quantity of waste generated. Thus, the marginal private cost of disposing an additional unit of waste is zero. It is only ensured that the total revenues need to equal the total costs, which is a stimulus to generate too much waste or it to be disposed in inappropriate way. According to Porter (2010) a disadvantage is that the government must calculate accurately the size of the marginal eco-tax, which should equal the sum of the marginal externalities. And this is very complicated. Otherwise, either an indue burden for the business will be created or the pollution of the environment will be stimulated.

Given Pigou's rejection of taxes as a possible decision for internalization of the social costs, the present research will be searching for a solution, based on the institutional economic theory.

## 1.2. Viewpoints for the social costs of circular economy

The present research for social costs of circular economy is focused on the EU model, but not on the concept for circular economy as a whole and in global perspective. The choice of the research focus can be presented by the difference between two models for circular economy, that of EU and that of People's Republic of China:

The concept of EU for circular economy and that of China differs significantly, as do their economic systems. The main differences must be sought in the market structure, regulations, technologies, and provision of critical raw materials. The only similarity between them is the desire for more sustainable usage of the resources McDowall et al., 2017 ; Luo et al., 2021).

The term "EU circular economy concept" is introduced in the scientific literature by Winans et al. (2017) and its purpose is to distinguish the EU circular economy model from other existing models in the world. In the scientific literature the concept for circular economy is called also a "paradigm shift", which includes the transition from linear economy to closed sustainable system of production and consumption. In this system, the society uses nature as inspiration to respond to the social and economic challenges. The concept of the paradigm is studied also as a way for reaching sustainable development (Luo et al., 2021; Prieto-Sandoval et al. 2018).

The first distinction is related to the reasons and emergence of the circular economy concept in EU and China:

EU. The circular economy concept has been introduced in EU since the end of 70 years of XX century, when the normative hierarchy of waste management was discussed as a legislative act in the Netherlands. Germany is the next country, which could be given as an example of early applying the circular economy concept (McDowall et al., 2017)

Although the earliest examples of introducing circular economy concept are from individual EU member-states, its introduction to the Community level is very late. In 2011, the European Commission has proposed a Roadmap to a Resource Efficient Europe (European Commission, 2011). Later, this roadmap was replaced by the so-called Circular economy package, which is named: Closing the loop - An EU action plan for the Circular Economy“ (European Commission, 2015). In 2020 this plan is updated (European Commission, 2020).

The main priorities of the plan are sustainable production, eco-design, sustainable consumption, sustainable pricing.

They foresee: changes in the public procurement rules; new rules for landfill, including new mandatory targets; they are proposed significant changes in the principle of extended producer responsibility; changes in the definition of what is “waste”; new standards to promote resource markets (of recycled and other goods); defining the priority areas of specific waste flows; funding of scientific research and innovations, related to the circular economy in the “Horizon 2020” program and etc McDowall et al. (2017).

The circular economy is forming the foundations for the so-called „European Green Deal“(European Commission, 2019).

China. The circular economy concept was first introduced in China in 90 years of XX century, as it was following the ideas for industrial ecology, which were applied in Japan and the USA, as well as in individual EU member-states. The US President’s Council on Sustainable Development established three eco-industrial parks after the UN Conference on environment and development in June 1992 (Winans et al. 2017).

At the beginning, this concept was developed through the idea of eco-industrial parks and subsequently moved to the vision for “harmonious society” (Luo et al., 2021).

In the Cleaner Production Promotion Act from 2003, for the first time it was normatively defined the term circular economy in China. The circular economy concept was officially introduced by the Chinese government in 2004 as a new strategic development framework and it was listed in the $11^{th}$ 5-years plan (2006-2010) for national economic and social development of the country (Winans et al. 2017 ; Luo et al., 2021).

In 2009 the Circular Economy Promotion Act was implemented and after that this Act was included in various Action Plans of the State Council of China, the earliest known of which was from 2013. (McDowall et al. 2017; Winans et al. 2017; Luo et al., 2021). As of 2017, normative standards for cleaner production, concerning more than 30 industries was adopted in China (Winans et al. 2017).

As a conclusion from the analysis of the reasons and emergence of the circular economy concept in EU and China, it could be summarized that the Chinese model is based on the idea for eco-industrial parks and this fact makes the model significantly different from the European one.

The second distinction is related to the vision of both models.

The Chinese’s vision for circular economy is broader than that of EU and it includes other ecologic problems, related to the industrial pollution of all environmental aspects alongside waste and resources management. It is constructed as a counteraction of the environmental challenges of the fast economic industrial growth and in lesser extent to the insurance of critical resources, that China has (McDowall et al. , 2017; Winans et al. , 2017; Luo et al., 2021).

The Chinese concept for circular economy is based on the vision for continuously increasing economic growth rate, which creates ecologic and social deficits, which need to be reduced and managed. In this regard, the Chinese concept accepts economic growth as permanent and guaranteed phenomenon, and the circular economy is not perceived as its catalyst. Rather, it should be related to an economic growth with "harmonious development" – a term, which according to McDowall et al. (2017) is very often used in the Chinese government documents. That is why, a conclusion is drawn about the wider context of the process.

The Chinese model for circular economy is mainly focused on the production phases and to a lesser extent to the consumption phase. It underestimates the necessity of efforts for reducing the consumption and the creation of circular culture of consumption and this fact strongly reduces the positive environmental effect. This model concentrates on the wastes of production processes, reduction of the industrial pollution, efficient usage of resources in the production process.

These specifics of the Chinese model could be explained by the export orientation of the Chinese economy. Unlike the EU economy, which works for a highly solvent domestic market, the Chinese economy is geared towards global exports. The scale of production is different, and it influences the degree of industrial pollution. The consumption models are also different, and the Chinese export could not force consumers from other countries to the ways how to consume its production McDowall et al. (2017).

The Chinese model, unlike the EU model, relies on regional pilot and demonstrating projects for circular economy and this also influences the process of funding, taxation policy and other instruments. This creates risk of unevenness of the process and lagging of separate regions, formulation of different fiscal policies and etc. In contrast, the EU model is universal, and it uses binding targets for EU member-states, coupled with penalties for non-compliance.

In North America, Japan and South Korea, the approach for the development of the circular economy concept is bottom-up and it comes from companies and their purposes to optimize resource efficiency, to test the production of new goods, to establish new relationships with the consumers. While in China and EU, the approach is top-down with the active participation of public authorities, which normatively set regulations and obligations to the companies (Winans et al. 2017 ; McDowall et al., 2017).

The emerging economies in Africa and Asia develop high extensive economic growth and they have a progressive increasing population. This leads to increase in the generated production and household waste. The transition from property over products and waste to provision of services is still at an early stage in the emerging economies. In these countries, the repair and reuse of classical products is common, and this does not stimulate the development of circular design of products. And the recycling sector is a part of the shadow economy, and it is developing without any rules, which hinders the creation of high-tech and environmentally secured sector for recycling. That is why the models for circular economy, that are implemented by the developed countries could not be directly used in the developing countries (Patwa et al., (2021). This distinguishes the EU model for circular economy from those of emerging countries like Africa and Asia.

Unlike Chinese, the EU concept for circular economy is significantly more narrowly focused on waste and resources. Its purpose is the opposite, it should stimulate economic growth and competitiveness through completing the environmental targets and eco-innovations development. That is, EU perceives the circular economy concept as economic possibility for reaching competitiveness. Also, as a means for access to critical raw materials which member-states do not have in the context of geo-political competition for them, and they are important for the development of high-techs (McDowall et al., 2017; Luo et al., 2021; Prieto-Sandoval et al. 2018).

The European documents are much more focused on innovative consumption models and sustainable design of products in comparison with the Chinese documents. Thus, EU develops the circular economy concept through a balanced supply and demand model, while the Chinese model relies much more on the supply management McDowall et al. (2017). The legislation framework of EU is much more developed, systematic, and ambitious, compared to that of China, while the Chinese model relies much more on various experiments and is much more flexible (Luo et al., 2021).

The analyzed reasons are enough to limit the focus of the present research to social costs of EU model for circular economy and not on the circular economy concept as a whole and in global aspect. The model of EU circular economy is specific, and it distinguishes from the other main models, especially the Chinese one. At the same time, it is assumed that the export-oriented countries will gradually adopt the EU model in order to have full access to the large and solvent EU market. Moreover, the EU model is mandatory for Bulgaria and the other EU member-states.

On the next place, the current barriers to the circular economy put pressure also on its institutional structure. That is why, they must be identified. According to Corvellec, Stowell and Johansson, (2021) although the circular economy concept is something revolutionary, it is not a new idea. Rather it represents a unifying concept of many existing old economic concepts from the 60 and 70 years of XX century, which are related to the dematerializing of economic growth and sustainability. This viewpoint is confirmed by other authors, who concern that the circular economy concept is just a rebranding of the idea for sustainable development but not an independent or a new paradigm (Friant, Vermeulen and Salomone, 2021; Grafström and Aasma, 2021; Hobson, 2021).

In the general term for circular economy, theoretically are unified the long known in science: “environmental economics”, “industrial ecology” “green capitalism” and many others. This is also the reason, according to the authors, that there are over a hundred definitions of circular economy that include different content according to the theoretical concept to which they belong. The only thing generally accepted is that the definitions include the circular movement of resources, which should lead to economic and environmental effectiveness. This strongly hinders the theoretical development of the concept, as well as its practical adoption by the business and governments. Corvellec et al., (2021) believe that its emphasis is largely economic and technological with a focus on growth and competitiveness, and the social and environmental focus remains in the background, being largely unclear. The biodiversity is completely absent from the circular economy concept, there are no targets and indicators, related to it. Then the EU circular economy package covers partly the UN sustainable development goals and could hardly be perceived as a completely new paradigm (Friant, Vermeulen and Salomone, 2021).

The construction of circular material flows is seen as isolated end, but not as a concept that have universal environmental character. On the other hand, the technological progress like industrial 3-D printing, work with big data and industrial technologies, based on artificial intelligence contribute to the development of concepts like that of the circular economy. However, this once again will highlight its technological character (Patwa et al., 2021).

The main barrier to the development of the circular economy is that of „path dependency“. According to Korhonen et al., (2018), the existing recycling technologies can be highly sustainable, even if they are more inefficient in comparison to the modern technologies of the circular economy. In this way, the circular and recycling economy will compete each other both for consumers and waste flows. This will delay the investments payback period of the circular economy and the technologies of the recycling economy will dominate the market as more efficient at scale. Thus, the circular economy technologies will be considered as financially risky, especially in terms of lower prices of the initial non-renewable nature resources (Ritzén and Sandström, 2017). Only the desire for better business reputation with customers and good image of the government and intermediaries in the supply chain could counteract to this barrier (Corvellec et al., 2021).

The effect of „path dependency“ is also confirmed by **Giannakitsidou, Giannikos and Chondrou, (2020).** According to them**,** Denmark**,** the Netherlands, and Finland incinerate a great percentage of their municipal solid waste (over 35%), which means that in order to reach the target of 65% of recycling till 2030 they must turn the waste incineration into recycling. And the waste incineration is cheaper than recycling. However, this means cancellation of the long-term commercial contacts with the incineration operators and changes in the policy for energy production, a component of which is waste-to-energy. At the same time, the waste incineration hinders the climate change mitigation (Friant, Vermeulen and Salomone, 2021).

The economic barriers are also a challenge. They are related to more complicated definition of the relationships in the supply chain of the circular economy. The modifications of the life cycle of the products in the context of sharing property rights amongst producers, intermediaries and clients could lead to improper allocation of their responsibilities. The locations of production, purchase, consumption, and recycling of a product are scattered all over the planet, making its circular management very difficult. The limitations of supply as well as the changing prices of recycled materials, make their usage uncertain for the producers and make primary resources preferred. At the same time, the costs for separate waste collection remain high so do the investment costs for main activity. The markets for recycled materials remain undeveloped, sufficient government initiatives for circular requirement about public procurements are missing (Corvellec et al., 2021; Salmenperä et al., 2021; Grafström and Aasma, 2021). The circular economy concept encourages marketing of recycled products, which lead to increases in the consumption and trigger a rebound effect (Friant, Vermeulen and Salomone, 2021).

The institutional barriers are studied by many authors. Corvellec et al., (2021) marked them as regulatory barriers, linked to the lack of favorable justice system or an insufficient institutional framework without entering into a detailed analysis. Salmenperä et al., (2021) indicate the fast changes of market regulations, related to import and export of waste, the definition and classification of waste, as well as the access and share of information. This effectively discourages investments in these activities. Here we could also indicate the existing differences in regulations at the EU member-states themselves. Grafström and Aasma, (2021) define institutional barriers as inconsistent political messages and bad infrastructure. "Path dependence" is marked as an example of poor institutional infrastructure. In addition, they indicate also the lack of standardization of the recycled goods, complicated regulations and weak enforcement of legislation, problems with the determination of the property right on waste.

The social barriers are also a problem to the development of the circular economy. Socially embedded understanding for modernity leads to a desire for constant growth of consumption and lack of will for long-term usage and repair of the same product. Circular economy systems are usually concentrated in the big cities, which lead to marginalization of villages and countryside and could lead to some types of inequalities (Corvellec et al., 2021). Companies, which have developed their conservative image through development of trademarks can be put to a test in the usage and development of recycled products (Salmenperä et al., 2021). The social element of the circular economy is largely implied or complimented in other EU policies, for example this for reaching sustainable development and inclusive growth. Only one indicator when concerning the circular economy in the EU is aimed at measuring the social component - that of employment (Friant, Vermeulen and Salomone, 2021). Other social barriers are conservative company culture, lack of consumer awareness (Grafström and Aasma, 2021)

In the scientific databases there are still few publications, which analyze the relationship between social costs and circular economy. Only a few are identified and they will be analyzed:

According to Moreau, Sahakian, Van Griethuysen, & Vuille, (2017)., the economic theory, mainly considering profitability from the viewpoint of the economic competitiveness, cannot deal with the institutional and social prerequisites, necessary for the transition to circular economy. They believe that in this transition should also be taken into account public issues, related to working conditions, allocation of wealth and management of systems. They also consider that in global aspect the structural increase of the service sector in comparison to the industrial sector has not led to a reduction in the use of raw materials in economic processes.

Moreau et al., (2017) has analyzed Kapp (1950), who has studied externalities mainly as an institutional problem. They confirm some of his viewpoints as a starting point in their analysis. First, that the monetary analog of external factors leads to inappropriate and underestimated reporting of their heterogenous character. Second, that the legislation defines the formal boundary between costs, borne by the private agents and those, which could be passed to the society. Third, that the competitiveness forces the shifting of costs to the others. As such, the resulting market asymmetries of the market power are reflected in the institutional conditions, resulting from the institutional strategies of the agents with high market power.

According to the authors, the boundaries between private and social costs are determined both by the development of property rights (in the value chain) and the institutional conditions, which fact influences the allocation of profits. From the viewpoint of the circular economy, three main institutional aspects are important: social embeddedness of the economy, legislative aspect of the costs allocation and political (strategic) process. This influences the profitability and competitiveness of relative actions and participants, as it determines their motives, values, and behavior. For example, the alternative models for property rights allocation in which the goods (containing potential waste elements) are leased rather than sold to the consumers, is a manifestation of institutional element of the circular economy.

In this context, the influence of the environmental regulations and management on the national competitiveness in the Balkan countries is studied by Marikina (2018). And the competitiveness from the viewpoint of institutional change, related to Bulgarian accession to the EU is studied by Ruscheva, (2005).

However Moreau et al., (2017) examines social costs from the perspective of "social and solidarity economy" (SSE), according to which the economy is implemented into the social sphere. They include in the term SSE the following: putting people in the center of the economic life through addressing social inequality and inclusive economy, as well as fairness to the cost of labor. They consider that in the conditions of SSE, there could be taken public decisions, related to the circular economy. For example:

- what materials need to be recycled or reused as a public priority. Regardless of economic profitability.
- public co-funding of green working places.
- preservation of materials value and the achievement of lower energy intensity should not be done in the conditions of higher labor intensity.
- the transfer of many labor taxes to the consumption of resources and energy.

Although the publication of Moreau et al., (2017) is related to the issue, concerning social costs, they are examined in other aspect. The authors have studied the possibility the circular economy concept to be examined through the SSE approach. This focus is interesting from the viewpoint of the possibility for higher interdisciplinary institutional view of the circular economy. However, in this way, the opportunities for internalization of external costs are analyzed, rather than social costs are measured.

Their view on the social and institutional issues of circular economy is too general and schematic, as it is limited to the labor conditions and wealth allocation. The study is based entirely on the qualitative analysis. An econometric or statistical analysis is missing. It could be assumed that this publication is a call to address attention to the institutional and social issues in the transition to circular economy, but not only to those, related to financial profitability. Also, as a call for interdisciplinarity and implementation of social instruments to the analysis of this problem.

Barrett and Kathleen, (2019) examine the circular economy model as mediated by the transition to renewable energy resources and as opportunity for building economic, environmental, and social capital. Social costs are studied through the relationship between externalities and mispricing. According to them, the marginal social costs should be internalized through taxes and fees, which should be charged to direct utility consumers (households, companies, or "things" – for example real estate). Also, subsidies for activities that do not meet the principles of the circular economy should be stopped. The amount of green taxes needs to have signal function, concerning the real scarcity of non-renewable resources and the real price of services, that are included in the circular economy system. There must also be sufficient alternatives so that consumers of taxed services are not disproportionately affected.

The authors make difference between the existing green taxes and proposed new taxes to the circular economy. According to them, despite the apparent similarity, the existing green taxes seek to change the behavior and to correct external factors, but they still leave the linear structure of the economy unaffected. While, the new taxes of the circular economy aim to radically restructure the economy, seeing it as embedded in the environment, rather than something separate from it. This should lead to a change in the economic paradigm. With the current green taxes, governments charge what they can, but not what is needed. According to them, a long-discussed topic in this regard, is shifting the emphasis from taxing labor to taxing the consumption of non-renewable resources. This, in addition, positive environmental effects should also lead to increased productiveness of the economy and to emergence of new "green" jobs.

And the so-called Trade Emissions Schemes should be reformed in order to be more efficient and not to be used as a means to avoid green taxation.

The publication of Barrett and Kathleen, (2019) only marks the potential relationship between social costs and circular economy. It does not either analyze or define this relationship. The term “social costs” is used at only one place. The publication states the standard views for the negative externalities and their possible internalization through a taxation approach. The accent of the publication should rather be sought in the comparison between existing green taxes and new green taxes, which are foreseen to be part of the circular economy.

The publication of Rathore & Sarmah, (2020) is also of interest for the purposes of the current analysis. The authors base their research on the fast growth of the population and urbanization, which increases the generation of solid waste and the demand for natural resources. According to them, the level of urbanization in 1950 was 30%, in 2014 it has increased to 54% and till 2050 it is expected to reach 66%. And the extraction of resources is three times higher, compared to the previous forty years and could cause serious environmental damage.

In the study, it is proposed a model for calculation of the total costs, related to solid municipal waste. They are calculated by summing of the following costs: functional costs, transportation costs, rental costs, environmental costs, social costs, and penalty costs.

It is of interest the distinction that authors made between environmental and social costs. According to them, environmental costs are related to the operating facilities and transporting of materials, which pollute air quality and release carbon into the atmosphere. The compensation of these effects is done through taxes on carbon emissions.

These taxes include emission costs, related to the transportation of waste, emission costs, related to incineration of household waste at landfills and emission costs due to the incineration of waste for obtaining energy.

And the social costs are defined as: „ *cost of negative impact on society due to various types of pollution* “. At the same time, the authors exemplify this definition for social costs with the mentioned environmental costs of air pollution from carbon emissions of waste transportation, landfills, and energy from waste.  As well as noise pollution. The negative effects of them are global warming, health problems, environmental disruption, traffic jams, road accidents. The calculations in both approaches (this of environmental costs and that of social costs) are analogous, using a conversion coefficient. Further in the calculation model as an example for another related determinant, could be included the unfavorable living environment in the areas with installed waste recovery or disposal facilities.

Evaluation of the model. The publication of Rathore and Sarmah (2020) consists of a model, which aims to analyze in-depth the social costs in waste management processes. However, in the proposed model, there are two unclear issues. The first one is related to the possibility of double reporting of environmental costs in calculation of social costs. Despite the obvious similarity of defining these two types of costs (environmental and social), the examples of social costs are slightly broader than those of environmental costs, but at the same time they cover them. In the model, social costs except for environmental costs, include also costs for road accidents, costs for traffic jams and costs for unfavorable living conditions in the areas of installed facilities for waste management. Regardless of this, a question for possible double counting of environmental costs arises, because these costs constitute a very large component of them.

The second issue is related to the unclear definition of the waste management system in the circular economy model. Obviously, according to the authors, this model differs from the EU circular economy package. In their model, municipalities and municipal solid waste management occupy a central place. A definition of a separate system, related to the extended responsibility of producers of waste, generated by the goods, produced by them, is missing. Also, there is a lack of collective systems of producers, who act in parallel with the municipality systems for waste management. This makes the model useless for analysis, concerning the circular economy concept. And given the unclear definition of the social costs, it calls into question how efficient the model is for making analysis.

Another publication, which is close to the topic for social costs of circular economy is that of Medina-Mijangos, R., et al., (2020).

The authors make an analysis, through which they conclude that in the municipal solid waste management process there could be generated various impacts (social and environmental), which are not considered in the economic analysis of their systems. In addition, in the developed countries the systems for municipal solid waste management are more complex than those in the developing countries, but in these countries an informal activity in the sector is observed.

The model used, is based on the well-known social cost-benefit analysis, which has been established in the economic research (Hoogmartens et al.,2014). Based on the analysis of scientific literature on the topic, the authors bring out the environmental and social impacts on the society, which are included in similar models.  According to them, these impacts are as follows: opportunity costs for land, external environmental costs due to air pollution and costs due to road accidents, caused by the activity in question.

The conclusion of Medina-Mijangos et al., (2020) is also of interest, as it says that in the scientific literature there are no publications that collect and group in methodological way identification and description of the most important impacts, which should be considered in the implementation of a project for solid waste management. They cite models in which the economic evaluation of solid waste management systems is based on the principles of LCC. They study only the following externalities: environmental emissions (pollution of the atmosphere, impacts on soil and groundwater, impacts on the quality of life) and the society’s willingness to pay for avoiding emissions. The compensatory effects are also studied, as external positive effects (revenues): usage or displacement of electricity and reducing fertilizer use from compost.

Interesting part of the model of Medina-Mijangos et al., (2020) are the following elements that have much in common with the social costs:

First, determinants, related to affecting human health. Any stage of waste handling, treatment and disposal may be involved. It can be caused either directly (through exposure to hazardous substances in waste or to emissions from incinerators and landfills, vermin, odors and noise) or indirectly (eg through ingestion of contaminated water and food). It can also be direct – in the conditions of serious accident, causing short-term exposure to high levels of potentially dangerous substances. And it could be also a chronic, with long-term exposure to low concentration levels of these substances (through inhalation, ingestion, or dermal contact). Damage to public health could be evaluated from the viewpoint of the workers (formal or informal sector) and the population, which lives close to the operating facilities for solid waste. For the population, which lives near the operating facilities for solid waste, most affected are kids, who are exposed to various infections and poisonings in comparison to the people who live far from the facilities.

The main impacts on the public health according to the authors are grouped in three aspects:

(a). Physical risks, related to the exposure to noise, ionizing radiation, and temperature. In landfills, the risks include surfaces and underground fires and a risk of explosion, associated with the processes of biodegradation.

(b). Chemical risks, related to exposure to gases, vapors and chemicals. The operating facilities for solid waste could emit some chemical polluters, like dioxins, volatile organic components (VOC) and heavy metals among others. The long-term exposure to their impact causes several toxic effects, including immunotoxicity, developmental and neurodevelopmental effects, and effects on thyroid and steroid hormones and reproductive function.

(c). Biological risks include the exposure to viruses, bacteria, blood, and blood products. Bioaerosols (organic dust) may act as infectious, allergenic, toxic, or carcinogenic agents to workers involved in the waste industry.

Second, determinants, related to environmental impact. In this group the following impacts are included (a) emissions in the atmosphere, (b) emissions to the soil and (c) emissions to groundwater and surface water:

(a). Emissions to the soil. The uncontrolled disposal of waste (fly ash) or leachate from landfills, which leads to soil pollution. It affects soil pollution, groundwaters and surface waters.

(б). Emissions to the atmosphere. These are the emissions of greenhouse gases, emissions of combustion gases with polluting compounds such as particles, heavy metals, organic compounds, and dioxins, among others, which cause environmental damage.

(c). Emissions in groundwaters and surface waters. The discharge of wastewater from incineration plants with wet flue gas cleaning systems (contain many pollutants including suspended solids, dioxins, and heavy metals). Also, leachate leakage from landfills (contaminates surface and ground waters). An additional hypothesis are the damages, caused by packaging waste in water bottles (plastic and others), especially in the ocean.

Third, determinants related to quality of life. Some municipal solid waste facilities (such as landfills and incinerators) are generally associated with nuisance and disturbances (warm, dust, eye contamination, odors, noise, traffic, vermin, flies), which arise due to the existence of such types of facilities. The effectiveness of each system for management of municipal solid waste depends on its acceptance by the local community. The nuisance and disturbances also affect the prices of local real estate and could lead to a NIMBY syndrome (not in my backyard) due to the impacts, generated on the quality of life (wellbeing) of the local community. The impacts depend on topography, distance, climate, etc..

Four, determinants, related to the opportunity costs concept (opportunity price). According to the authors, the opportunity costs concept, used in the system for municipal solid waste, could be explained by two ways. First, when there are several opportunities for waste usage, the opportunity costs will be borne by the usage, that provides the best economic results, if these yields are higher than those of the financial instrument. Second, when there are no alternative uses, the opportunity costs are borne by the effectiveness, which is some kind of financial instrument, when the investment and operating costs will be invested in it. Also of interest is the statement that traditionally opportunity costs have been aligned only with the maximization of profits. However, according to the sustainable development concept and its three pillars, the best opportunity will be the one that ensures not only the best economic performance, but also the best environmental and social performance.

Evaluation of the model. The publication of Medina-Mijangos et al., (2020) consist of a model, which tries to analyze in-depth the social costs in waste management. However, in the proposed model, there is one unclear issue. The model used, is based on the well-known social cost-benefit analysis, which is confirmed by the economic studies (Hoogmartens et al.,2014). The correlation of the presented model with the circular economy concept is not entirely clear. The presented model consists of the components of the municipal system for mixed household waste management. There is no definition of a separate system related to the extended responsibility of producers for the waste related to their products supplied to the market. Collective systems of producers, which operate in parallel with the municipal waste management systems are also missing. Waste recycling is not enough reason to be accepted that the model is related to the circular economy concept. With these critical notes, the model could hardly determine the right correlations between affected parties and will lead to efficient forecasting of their behavior.

The fifth publication, which has much in common with the topic of social costs of circular economy, is that of García-Barragán, Eyckmans, & Rousseau, (2019). In this publication, the authors share problems, related to the complexity of finding reliable and applicable indicators for circular economy. On the first place, they highlight that it does not exist a universally accepted definition for the circular economy, but there are more than 100 such. And since there is no consensus, which is the most correct, this creates uncertainty how the circular economy is measured. On second place, what the different definitions have in common is that they accept maximization of material's value as an explicit measure for effectiveness. And according to the authors, recycling itself is only one of the many indicators, characterizing the industrial activity. This discrepancy shows that the direct implementation of indicators, characterizing recycling activities as indicators for circular economic activities will not be methodologically correct.

Their econometric model is macroeconomic and includes a representative consumer with preferences over consumption of various functionalities (mobility, energy, communication services) in certain time limits model with a stationary utility function. As the representative consumer focuses on the functionality of the consumed goods, he does not focus on the materials from which these goods are produced (recycled or primary materials).

Social costs are presented in a limited way through the "social recycling costs" and their relation to the category "social welfare". Specifically, according to the authors: "the materials flow increases social welfare when present and future marginal benefits, considering durability, equal social marginal costs for recycling, including the impacts on material resource scarcity and landfills. The marginal benefit of each additional unit of material consists of the marginal value, attached by the user to the functionality, dimension of the marginal productiveness of good for ensuring functionality (product intensity of the functionality), the dimension of marginal productiveness of materials for goods production (material intensity of products).“ The marginal benefit of the recycled material is compared to the marginal benefit of the primary material. In conclusion, the authors summarize that recycling can be used as an indicator for measuring circular economy, only if the social welfare is considered.

Evaluation of the model. The publication of García-Barragán, Eyckmans, & Rousseau, (2019) is an interesting one from the viewpoint of the methodological problems, related to the complexity of finding a reliable and applicable indicator for circular economy measurement. The econometric analysis done in the publication is not explained in sufficient details. It is not clear how in a microeconomic model the isolated understanding for “social recycling costs” could embody all the existing social costs in the circular economy. The category “social welfare” used in the model also remains too general and not well specified to measure. The idea for comparison between costs and benefits of the primary and recycled resources is not a new one. It is analyzed by Porter (2010) and many other authors, even in the context of waste management. That García-Barragán, Eyckmans, & Rousseau, (2019) replace the term “waste management” with the term “circular economy” is not an enough reason to obtain sufficient clarity, regarding social costs. Separately, the accent on recycling in this publication contradicts to the circular economy concept, according to which recycling must be replaced by the reuse or long-term use of assets (goods). Still their conclusion that “the direct use of the indicators for recycling as indicators for circular economic activity is not correct in methodological viewpoint” and could be used as basis for future scientific research.

And other recent publications in a similar way examines the correlation between social costs and circular economy without adding anything different. In this way, they will be commented only briefly.

For example, Balasubramani et al., (2020) consider that the improper disposal of solid waste leads to serious impacts on the environment, like pollution of air, soil and waters, greenhouse gas emissions, infections, and others.  According to them all landfills leak toxic leakages, as even the most “state-of-the-art” landfills will eventually leak and will pollute the near groundwaters. The treatment and disposal of waste at conventional level includes costs paid to the waste collector, social costs of waste (disposed to landfills), the quantity of the toxic gases, like methane, released into the air and pollution of groundwaters. The other environmental (non-material) costs include unwanted odor and noise by heavy-duty vehicles. In conventional methods, the value of land is an additional price. Preventing odors or fly-breeding could be an additional economic cost.

The publication of Balasubramani et al., (2020) does not add anything new in the correlation social costs and circular economy, again leaving open the question of the distinction between social and environmental costs.

Wuyts et al., (2019) base their study entirely on qualitative analysis. An econometric or statistical analysis is missing. Insofar the publication deals with social aspects of circular economy, it examines the value attitudes and motivational factors of the Japanese society regarding the long-term use of residential buildings. These are non-technical factors which determine the consumers ‘attitude. The authors also comment social norms and cultural impact.

The issue of negative externalities is not explicitly addressed in their publication. Also, it does not apply to the EU and the European concept of circular economy including the principle of "extended producer responsibility".

Santos, Mendes and Teixeira, (2019) use models of social life cycle assessment -SLCA, which is widely known. It is applied to make strategic decisions based on an indicator matrix. The analysis is focused on only one element, which is oriented towards the circular economy concept – the illegal waste disposal. However, this issue concerns mainly municipality authorities and local communities, while leaving aside the principle of extended producer responsibility and the other participants and phases in the circular economy concept.

The purpose of the publication of Huppertz et al., (2019) is to evaluate the real value of a unit of resource for present and future generations. They use the well-known methods for life cycle assessment (LCA) and cost-benefit analysis (CBA). It is not clear in what way the authors propose to be developed both methods, other than by calling for an improvement in the quality of the input information. The question of determining the social discount rate, apart from the standard assumptions about its range, has not been explicitly resolved.

Boachie, (2012) have constructed their publication entirely on scientific literature from 80 and 90 years of the XX century. It could be assumed that the publication deals with the early theories for Life Cycle

Assessment - LCA. It applies entirely to developing economies, in which the environmental legislation it weak or it is missing, in comparison to the environmental legislation of EU, which is one of the leading in the world. Although the circular economy concept is commented on in several places in this publication, it is difficult to see the connection with its contemporary dimensions.

The publication of Boardman, Geng and Lam, (2020) also falls into this group. The publication does not refer to the European model for circular economy. However, it examines social costs as specific waste flow and that is why it will be presented in the current research. The purpose of the publication is to evaluate social costs from shadow e-waste processing in developing countries, taking China as an example. E-waste contains dangerous (toxic) substances, and their informal processing releases them uncontrollably to the environment. This process increases mortality and decreases human being's physical and mental functions. They use the well-known method for evaluation of the "social opportunity costs" (Hitchens, Thampapillai and Sinden, 1978). This, according to the authors is the values of resources of which the society must give up in order to adopt a certain program. The model is constructed in a standard way, through a cost-benefit analysis. Main determinants of the model are mortality (caused by exposure to toxic materials) per number of citizens in the studied area or in the entire country; value of the human life and lost labor productiveness (missed incomes), volume of the waste proceeded. The effects are measured both on those who proceed in uncontrollable way the waste and on the entire local community.

The following conclusions can be made as a result of the analysis of publications, concerning the correlation between social costs and circular economy:

- First, one of the approaches for studying the correlation social costs and circular economy is through the application of the theory for social and solidarity (Moreau et al., 2017). However, this understanding is limited only to the labor conditions and allocation of wealth. Its original source could be found in the publications of Kapp (1970). It represents a call to pay attention to the institutional and social issues in the process of transition to circular economy, but not only to those issues, related to the financial profitability. This focus is interesting from the viewpoint of the opportunity for greater institutional view to the circular economy. But in this way, the possibilities for internalization of external costs are analyzed rather than for measurement of the social costs. Studies of this type could also use as an accent the term “social capital” (Wuyts et al., 2019).

- Second, in some of the publications there exist a risk of double counting of environmental costs in the measurement of social costs – where environmental costs are defined as individual determinants in the model and at the same time they are defined as part of the social costs which are another separate determinant of the model (Rathore and Sarmah, 2020; Balasubramani et al., 2020). This risk does not change, even though social costs are defined more broadly than environmental costs.

- Third, in some publications there exists a definition for the system for waste management, which differs from the system, described in the EU package for circular economy (Rathore and Sarmah, 2020; Medina-Mijangos et al., 2020). This could lead to errors in the calculations. In these models, municipalities and municipal management processes of solid waste occupy a central place. There is no definition of a separate system related to the responsibility of producers for the waste related to their products supplied to the market. Collective systems of producers to act in parallel with the municipal waste management systems are missing. If this aspect is included in the model, it will become more complicated, but it will also become more accurate.

- Fourth, a complexity in the analysis of social costs is the determination of the level of impact of various negative externalities. There is a wide range in scientific estimates of the value of human life, human health, the value of various environmental services, the effects of global warming, the public concern about waste treatment facility location (Miranda & Hale, 1997; Porter, 2010).

- Fifth, many of the models used, are based on well-known principles for analysis. For example, the model for social cost-benefit analysis or the model for social life cycle assessment (SLCA), which are confirmed in the economic research (Medina-Mijangos et al., 2020; Santos, Mendes and Teixeira, 2019; Huppertz et al., 2019; Boachie, 2012; Jamasb and Nepal, 2010;

Ferrão et al., 2014). The standard models of this type include opportunity costs for land, external environmental costs due to air pollution and costs associated with accidents due to the main activity studied. However, it could also be included the determinants, associated with affected public health; with affected environment; determinants, associated with the quality of life (warm, dust, odor, noise, and other from waste facilities); determinants, associated with the opportunity costs concept, applied to the systems for solid waste management. Some of the models rely on a method for evaluation of the „social opportunity costs ", examined as the value of resources from which the society must give up in order to adopt a certain program (Boardman, Geng and Lam, 2020). There are also models for cost benefit analysis of alternative policies and scenarios for assessment of social benefits.

- Sixth, some of the publications use microeconomic models for explanation of the correlation between social costs and circular economy (García-Barragán, Eyckmans, & Rousseau, 2019). These models use some form of generalized summaries with a particular-to-general approach. For example, it is not completely clear how in a microeconomic model the isolated understanding for "recycling social costs" could represent all social costs in the circular economy. The category "social wellbeing" which is used in such type of models remains too general and not well specified for measurement. The idea for comparison between costs and benefits of the primary and recycled resources is not a new one. It was analyzed by Porter (2010) and many other authors in the context of waste management. And placing it in the context of the circular economy does not bring new clarification of the social costs.

- Seventh, in some publications the correlation between social costs and circular economy is examined through the standard concept for negative externalities and their possible internalization based on a taxation approach (Barrett and Kathleen, 2019). A novelty could be sought in relation to the comparison between existing green taxes in the context of linear economy and the new green taxes, which are foreseen to be part of the circular economy. According to these publications, despite their apparent similarity, the existing taxes aim to change the behavior and to correct the externalities, but they leave the linear structure of the economy unaffected. While new taxes of the circular economy are trying to radically restructure the economy as they assume it as embedded in the environment, rather than something separate from it.

# SECOND CHAPTER

# CRITICAL ANALYSIS OF THE SCIENTIFIC LITERATURE, DEVOTED TO THE CORRELATION SOCIAL COSTS AND WASTE MANAGEMENT

The critical analysis of literature, devoted to the correlation between social costs and waste management could expand the theoretical basis of the study, conducted in the previous chapter. Given the few numbers of scientific publications in the field of correlation of social costs of circular economy, this would appear to be a logical transition from induction to deduction, which aims to deepen the theoretical focus of the research.

## 2.1. The correlation social costs and waste management as a component of the circular economy concept

Hamilton (1993) examines social costs in the context of hazardous waste management. In his publication, there is not a stand-alone definition of the social costs. It is assumed that the author examines them in the meaning, given by the „Coase's theorem"(Coase, 1960). According to that definition, when the property rights are well-defined and transaction costs equal zero, the company that causes externalities will find where, other things being equal, it can cause the least harm.

To the analysis are added also the political costs as discussed by Becker,(1983). His concept is an alternative to the Coase's theorem, but it is based on analogous reasoning. The accent is put on the affected party by the externalities, but not on the cause. The analysis covers possible (political) competition amongst affected communities, which power of counteraction to the externalities leads to its positioning where it will cause the least negative effect to the society. This theory depends on the activity of the affected persons, their opportunity to realize the problems and to unite and solve them.

In both theories (that of Coase, 1960 and of Becker, 1983) it is a matter of correcting the market failure, related to the inability of the market to minimize externalities and to determine their most socially tolerable location. Hamilton (1993) unites both theories and examines them in the context of hazardous waste management. The publication analyzes the dependence between externalities internalization level of companies, managing hazardous waste and the ability of the endangered population to participate in collective counteractions.

The main determinants of Hamilton's model (1993) are:

(1). Volume of the generated waste in the region and free capacity for its management. This covers the level of industry development in the region, its added value and export potential. The higher it is, the greater is the necessity of new landfills and installations for their recovery. On next place, the higher industry level for development in the region means potential lobbying (political) support for construction of new waste management facilities.

(2). Production factors prices. The Earth is the main factor of production in the process of construction of landfills and installation facilities for waste recovery. Its price is important for determination of capital costs, as well as for the interest of choosing a suitable place for landfills location.

(3). The number of potential compensations for the damages caused by the external factors. The company, which causes external costs, according to the Coase's theorem will consider several things. First, the demographic determinants (incomes, real estate prices, education, population density). The higher the value of these parameters are, the higher will be the risk of legal action and political resistance. Second, the potential costs of legal actions, due to the cause of future environmental and health damages (value of property at risk, number of affected people, regions with environmental damages).

(4). The potential of local community for collective actions. The magnitude of this potential determines the risk of legal and political actions against the investment intentions. The main determinant here is the percentage of adult population, which has voted at the last national elections, compared to the national voter turnover. An additional determinant here is the history of the region, concerning the previous oppositions to investment intentions.

(5). Reactions of the public officers to election positions in the region. In the standard case, the election public officers will defend the positions of citizens, who have voted for them in order the citizens to re-elect them in the next elections. However, in regions with low voting activity, where the corporate voting decides the elections, they may side with investors.

Evaluation of the model. Given the aforementioned, it could be assumed that Hamilton (1993) completes the „Coase's theorem" (Coase, 1960) for social costs by adding to it the theory of Becker (1983) for political costs. And after that, he uses both theories to create a model that analyzes the hazardous waste management, based on the political activity of the affected local communities. Based on the results of the model, he derives the level of internalization of the external factor, which is assumed to predetermine the sum of the social costs. However, at the same time, the hazardous waste management is special case (element) of the circular economy concept of EU. This is only one of approximately ten specific waste flows in the circular economy, which management aims to minimize the damages for environment and human health rather than preserving the primary natural resources. Thus, the model of Hamilton (1993) apart from predating the implementation of the circular economy concept in the EU, it cannot be taken as representative because the hazardous waste flow cannot be taken as representative of other waste flows.

Another publication which is relevant to the current research is that of Jamasb and Nepal, (2010). The authors have not given an independent definition of the social costs. The authors adhere to the well-established model in the economics of welfare for social cost-benefit analysis of the alternative policies and scenarios (Harberger, 1978). The analysis is oriented to the comparison of various scenarios for waste management in Great Britain and the production of energy from coal, carbon intensity of the separate scenarios and carbon emissions pricing. The reduction of risks, related to climate change, the security of energy supplies and the expansion of renewable energy resources are considered as benefits of the exploitation of waste for energy production. The costs are calculated based on comparable amount of energy, produced from coal and from waste. The private costs are the standard ones, used in the neo-classical economic theory – fixed and variable costs. They are also supplemented with some opportunity costs for the assets used, as well as with the vision for possible economy of scale of the activity.

Sulfur dioxide, lead, and dioxins, released from the installations for waste energy production are indicated as externalities. As well as traffic jams, increased number of road accidents due to the transportation of waste to the installations. Also, possible disdainful attitude of the others to the citizens, who live in the neighborhood, because the existence of such waste facilities is associated with unpleasant smells. The recycling is the lowest-emission option in comparison to landfills and waste-to-energy production.

Regarding, the publication of Jamasb and Nepal, (2010) it cannot be attributed to the circular economy concept. Obtaining energy from waste occupies a second place in the circular economy concept, because it leads to low preserving of primary natural resources. This method relies on the reuse or long-term use of assets (goods). Obtaining energy from waste requires primary energy supply to the installation, which amount of energy could equal the energy obtained from waste, because all hazardous and harmful chemical and biological substances, which are dangerous for the population, need to be neutralized. And if this is not done, then obtaining energy from waste will create its own externalities and accordingly its own social costs. That is why, this model could not be used as a representative one for the circular economy concept of EU.

The publication of Ferrão et al., (2014) is an interesting one from the viewpoint of that it examines the existing system for waste management of an EU member-state (Portugal). This system covers the present model of EU legislation for waste management, which is a basis for the European concept for circular economy. The expectations by reading the title of the publication are the social costs to be studied in this context, but nowhere in the publication "social costs" are explicitly mentioned. The term "social impacts" is mentioned in several places in the publication. In this context, it is also used the standard model for evaluation of the life cycle. It has been calculated how much the emissions from CO2 are reduced by the utilization of waste and energy. They are analyzed the economic and social benefits of the high economic added value of the activities in waste management and the creation of additional employment. In conclusion, according to this publication, social impacts of the waste management system of EU as of 2014, are associated with the creation of green jobs and increase in the employment rate. According to the authors, a contribution of the publication is the analysis of concrete profiles of these green jobs.

Given the lack of a focused studying of the correlation between social costs and waste management process, as well as the lack of a definition and targeted analysis of social costs, this publication and its presented model cannot be accepted as representative ones.

Haraguchi, Siddiqi and Narayanamurti, (2019) do not present an independent definition of the social costs. The authors adhere to the model of cost-benefit analysis of alternative policies and scenarios, also including an evaluation of the social benefits. The contribution of their publication is the application of a simulation model like Monte Carlo. The model includes stochasticity and uncertainty of key parameters, such as volume of the waste generated (growth of population, GDP and others) technical effectiveness of waste recovery (future technologies), fixed and variable costs, policies of different nature that influence the activity. The social benefits are considered through reducing the social costs of atmospheric air emissions (greenhouse gas emissions and nitrogen oxides (NOx), sulfur oxides (SOx), particulate matters (PM)). This stems from the limited scope of the study, which is associated with the systems for waste-to-energy production, but not to the entire waste management. The avoided external costs from greenhouse gas emissions are monetized and examined as revenues. The reduction comparison is against the GHG emissions trading mechanism baseline. This has the appearance of partial accounting for social costs.

As a result of the above, this model could not also be a representative one for the correlation social costs and circular economy. On one hand, because it studies obtaining energy from waste, which according to the circular economy concept should be avoided. On the other hand, this model is oriented to the social benefits and their monetization, but not to measurement of social costs, associated with waste management.

The publication of Miranda & Hale, (1997) analyze private and social costs of waste production for energy production. The waste-to-energy production plants are studied in their complex functionality: they simultaneously produce energy and recover household waste. According to the authors, if these plants are examined only from the viewpoint of energy production plants, then their comparison with the conventional plants would define them as inefficient. In the view of their second function (their participation in the waste management process), they should be compared to the conventional energy plants from the viewpoint of the social costs. Their impact on the environment, including and through their effects on the use of conventional fuels. The authors reach the conclusion, that from the viewpoint of the private costs for energy production, the waste-to-energy production plants are from two to five times ineffective than conventional fossil fuel plants.

Of interest is the conclusion of Miranda & Hale, (1997), associated with the possibilities for externalities evaluation. They consider that a lot of authors are trying to define private and social costs for fossil fuel energy production. However, most of the studies rather than assessing externalities, they more often analyze the possibilities for actions to mitigate externalities. This once again highlights the relevance of the problem and necessity of conducting research with appropriate models for externalities evaluation.

From the viewpoint of fossil fuel energy production, the authors examine particulate matters emissions as an external costs' driver (particulate matters, sulfur dioxide, nitrogen oxides, carbon dioxide and dioxins). According to them some of the externalities (such as acid rains) are analyzed in-depth, while others (such as global warming) are still being discussed and they are disputable, regarding their level of impact. The valuation (monetization) of this impact remains particularly controversial. There is a wide range of scientific estimates of the value of human life, human health, the value of various ecological services.

The social benefits of the waste-to-energy technology are its possibilities to protect human health from the risks of unutilized waste (by killing pathogens and other bacteria, which cause infections in the waste).

The social costs of waste to-energy technology are related to the risks for the human health and environment, as well as to aesthetic impact. These plants, like energy power plants, operating by fossil fuels, release: particles of nitrogen oxides, heavy metals, including lead and mercury and other toxins to the air. After considering the ever-increasing strictness of environmental government regulations and the development of technology over time.

The authors determine also “aesthetic external costs”, related to the deployment of such facilities as an external cost. They suggest that people are uncomfortable with living near such facilities. Especially if that facility receives waste from other communities to which residents feel no loyalty. Locals may even assign higher values to these costs over others.

Marginal damages functions, according to the authors, include mortality effects, diseases, material effects, crop destruction, visibility effects, contribution to the global warming from organic and non-organic air pollutants. Although the quantifiable risk of toxic air emissions from WTE plants is low, public concern tends to increase. This concern is related to the great uncertainty about the risks arising from these pollutants. Therefore, when measuring external costs from these installations, one must work with a large variability in estimates.

According to Miranda & Hale, (1997) when determining the social costs of waste disposal, there must be considered: emissions, natural resources prevention, use/production of energy and every increase in the dangerous materials because of recovery processes. And the external costs, associated with functioning of landfills are high where water masses are high or when the energy available from methane emissions is not utilized. As waste disposal facilities must be evaluated in the context of the local political (strategic), economic and environmental context.

Impressive in their publication is the considering of "aesthetic external costs" from the waste recovery facilities, which slightly increase the narrow focus of their vision for social costs. Also, their analysis of many publications indicates that most of the studies rather than evaluating external costs, they analyze the possibilities for performing activities for their mitigation. This once again highlights the relevance and necessity of scientific research in this field, which consist of models for external costs evaluation.

Evaluation of the model. The conclusion, concerning the publication of Miranda & Hale, (1997) is that they examine social costs in an isolated context, which makes their analysis limited. Social costs are put in the context of comparison between waste-to-energy plants and conventional fossil fuels plants. In this limited context they examine them as environmental costs after they consider their impact on the human health. This is also highlighted by their vision for evaluation of the waste disposal facilities. According to them, they must be evaluated in the context of local political (strategic), economic and environmental aspect. The economic aspects in their study are presented by the classical understanding for private costs and they do not add anything new in this regard.

The publication of Rathore and Sarmah (2020) contains a model, which aims to analyze in-depth the social costs in the waste management. However, in the presented model, there is a risk of double counting of the environmental costs in measuring the social costs. There is an obvious similarity in the definition of both types of costs (social and environmental), however the examples for social costs are somewhat broader than those for environmental costs. In the model, social costs cover except for environmental costs, but also the costs from road accidents, traffic jams. Also, the unpleasant living environment in the regions of waste management plants. Nevertheless, a question, concerning the possible double counting of environmental costs in the measurement of the social costs, arises, as they constitute a huge component of them. Separately, one can comment on the issue that according to the definition used in the model, the social costs should be fully covered by the environmental ones, and only the examples given in the exposition of the model imply that the authors may have other elements in mind (mainly caused by transportation problems or aesthetic costs by waste management installations).

Specifically, the model of Rathore and Sarmah (2020) measures total costs for solid waste management. They are calculated through the sum of the following types of costs: functional costs, transportation costs, rental costs, environmental costs, social costs, and penalty costs. Of interest is the distinction which authors make between environmental and social costs. According to them environmental costs are associated with the operation of facilities and transfer of material, which pollute air quality and release carbon in the atmosphere. Compensation of these effects is done through taxes on carbon emissions. These taxes cover: costs for emissions, associated with the activities for waste transportation, costs for emissions, associated with the incineration of household waste at landfills and emission costs due to the incineration of waste for the purposes of obtaining energy. And the social costs are defined as: „price of negative impact on society, because of various types of pollution “. At the same time, the authors exemplify this definition of social costs with the mentioned environmental costs of air pollution from carbon emissions from waste transportation, landfills, and energy from waste. Also, noise pollution. Their negative effects are global warming, health problems, environmental damage, traffic jams, road accidents.

The calculations of these two groups (environmental and social) are analogous, as a conversion coefficient is used. Further in the calculation model it is added also another example, concerning the unpleasant living environment around the waste recovery or disposal plant. But this does not change the risk of double counting, based on matching air quality pollution and atmospheric carbon emissions, which is included in both the environmental and social costs of the model.

In the publication of Balasubramani et al., (2020) the issue, concerning the distinction between social and environmental costs remains open. According to the authors, waste treatment and disposal on conventional level includes costs paid to the garbage collector, social costs of waste (dumped in landfills) and cover the number of toxic gases, such as methane, released in the air and pollution of the groundwaters.

The publication of Medina-Mijangos et al., (2020) contains a model which aims to analyze social costs in depth in the waste management process. However, in the presented model, there is an uncleared issue. The model used, is based on the well-known social cost benefit analysis, which is confirmed by the economists. The relationship of the presented model with the circular economy concept is not entirely clear. The model is more like a municipal system for waste management. A definition of a separate system, associated with the producers' responsibility for waste, generated by the products, supplied to the market, is missing. Collective systems of producers, which operate in parallel with municipal waste management systems, are also missing. The inclusion of waste recycling in the model is not enough to be assumed that it embodies the circular economy concept. In this form, the model will have difficulties in determining the correct correlations amongst the affected stakeholders and in predicting their behavior.

Based on the critical analysis, we could identify various types of social costs in the waste management process:

**Types of social costs in the waste management process, defined in the scientific literature:**

- Social costs of waste disposal.
- Social costs of incorrect or illegal disposal of waste.
- Social costs of presence of shadow sector in the waste treatment activities.
- Social opportunity costs.
- Social recycling costs.

These social costs will be studied in-depth, as they are of interest from the viewpoint of establishment of theoretical basis for construction of a model, which is one of the tasks of the present monograph.

The first type of social costs in the waste management, derived by the analysis of the scientific literature: „**social costs in waste disposal“:**

These costs could be associated to several components: risks for health and environment, consequences of accidents, caused by waste treatment activities, opportunity costs for land, used for activities for waste management, as well as the aesthetical changes (Miranda & Hale, 1997; Balasubramani et al., 2020; Medina-Mijangos et al., 2020; Hu, 2013; Porter, 2010; Rathore and Sarmah, 2020). Here what matters are the released emissions, risk for the natural resources, use/recovery of energy, creation of dangerous materials in the process of waste disposal, unwanted odors, and noise from heavy duty vehicles (Miranda & Hale, 1997; Balasubramani et al., 2020; Hu, 2013).

For example, social costs for waste-to-energy technology could be associated with the risks for the human health and environment, as well as the aesthetic changes. These plants, like energy power plants work with fossil fuels and they release nitrogen oxides particles and dioxins, heavy metals, including lead and mercury and other toxins for the atmosphere. In this regard, the stringency of environmental regulations and technologies is constantly increasing over time, which affects all components of the costs associated with this activity (Miranda & Hale, 1997).

The second type of social costs in the waste management, derived from the analysis of the scientific literature are: **„social costs, associated with the improper disposal of solid waste“** (Balasubramani et al., 2020)**:**

The authors think that this leads to seriously impacts on the environment, such as air, soil and water pollution, greenhouse gas emissions, infections and etc. According to them, all waste landfills leak toxic filters, and even the most "state-of-the-art" landfills will eventually leak and contaminate nearby groundwater. The other environmental (non-material) costs include unwanted odor and noise of heavy-duty vehicles. In the conventional methods, the value of land, occupied by improperly disposed solid waste is an additional (opportunity) cost. Preventing odors or flies breeding could be additional economic cost.

The third type of social costs in the waste management, derived from the analysis of the scientific literature: publications, that analyze waste management process in developing countries derive that „**social costs of shadow sector presence in the waste treatment activities“ (**Boardman, Geng and Lam, 2020; Hu, 2013)**:**

Such a shadow sector for waste treatment activities exists in all countries, even in those which are high developed, albeit in a minimum percentage. This means that in the models for social costs measurement, this sector must be considered. The influence of this sector is mostly on the health of workers, as well as on the local population in the areas, where such type of waste treatment systems is built. Mainly their impact is associated with increased mortality, morbidity, and loss of income. But also: environmental damage, transport accidents and traffic jams, loss of tax revenues, stimulation of a black market for useful products extracted from waste.

The fourth type of social costs in the waste management, derived from the analysis of the scientific literature: Boardman, Geng and Lam, (2020) raise the question for **„social opportunity costs“** in the analysis of social costs, associated with the waste management:

This, according to the authors is the value of resources, from which the society must give up in order to adopt a certain program. For commodities, that are bought and sold on well-functioning markets, the market price is a reasonable measure for the possible costs. However, there are no market prices for many impacts of the government programs, for example for lives protected or for various types of pollutions. For such social impacts, the authors use "shadow prices", which is the price that would have an impact in a well-functioning market, if such exists.

The fifth type of social costs in the waste management, derived from the analysis of the scientific literature is **„marginal social costs for recycling" [MSCr], (**Porter (2010). These costs are composite, and they consist of several elements:

„*Marginal social costs for primary natural resources*" is denoted by **MSCv** - which stands for primary natural material saved.

In addition, recycling saves also „*the costs for disposal of the primary natural material after its use*" - **MSCd**. Then, the total marginal social price for production and later for the disposal of the primary natural material is the sum of: **MSCv + MSCd**.

And the „*marginal social costs for recycling*" **(MSCr)** will increase with an increase in the volume of the recycling itself, because in order this increase to be achieved, it will be necessary more and more new primary materials to be included. The more types of materials we recycle, the more the ineffectiveness of scope will increase. Recycling must increase to the limit: **MSCr = MSCv + MSCd**. Failure to observe this equality will result in inefficiency. Either we will waste primary natural resources instead of recycle them optimally, or we will recycle too much losing financial resources to do so. Given this, the optimal size of the marginal social cost of recycling according to Porter (2010) will be:

$$\mathbf{MSCr = MSCv + MSCd} \qquad \mathbf{(3)}$$

The various types of social costs in waste management, which are derived from the analysis of the scientific literature, could serve as basis for determining the borders, concerning what has been done in this thematic scientific field. On the next place, they present a variety of approaches, used in studying this issue. And at the same time, they hardly justify enough arguments to study the social costs of circular economy, which brings out the need of searching for a new approach.

## 2.2. Critical analysis of Porter's (2010) baseline model for measurement of the social costs in waste management processes

Despite the lack of fundamental economic model, which reflects the specifics of the circular economy concept and social costs, one exists for the waste management. This is the key monographic study of Porter (2010). Its focus is mainly on the economic analysis of the USA system for waste management, but also consists of certain comparisons with the waste management systems of EU member-states. The more important is that in this analysis of the economic aspects of waste, social costs are included, however partially. And its volume exceeds 300 pages and could be accepted as basis for any subsequent research in this field. Here, this analysis is summarized in few pages and it is presented abstractly with the purpose to help the contribution of each publication, related to the topic, to be deducted.

According to Porter (2010) external costs must be reduced to an optimal level. In this regard, he follows the well-known model of the neo-classical marginal analysis (Opocher, 2017). His contribution is the adaptation of this model in the field of waste management, so as to be carried out an evaluation of the marginal benefits and marginal costs of all activities and effects, related to waste. Schematically, this model assumes that the marginal utility is negatively sloped and decreasing, as the environment is getting cleaner and the pollution decreases. Then, each subsequent cleaning of the environment brings less and less marginal utility. At the same time, marginal costs are getting higher, because they are positively sloped and increasing. The efficiency is sought in: **MR=MC.** The complexity of the analysis comes from the inclusion of external costs in this well-known neo-classical equation, which are related also to the social costs.

The definition for „social costs", which he has adopted as basis for his studies in relation to the benefit analysis and cost efficiency analysis in the waste management was established by the neo-classical economic theory. This definition examines social costs as a sum of the private costs and the external costs (Gruber 2012). Porter (2010) determines „private costs" (for example for wages, for fuel, for machines, for land), as easily measurable, based on statistical, accounting and tax data for monetary expenditures incurred by relevant economic agents.. He accepts also the definition for „external costs", according to which they are borne by someone else, who has neither given a permission for them to be imposed on him, nor is he compensated for the damages occurred (Buchanan and Stubblebine, 1962). But he indicates that in most of the cases, it is very difficult to understand who bears the external costs and also to implement adequate value on the burden incurred.

As examples for externalities, Porter (2010) determines most of the problems, associated with waste: noise during operations with waste, air and groundwaters pollution in the disposal of waste, toxic and in some cases radioactive danger in the improper recovery of waste. Such costs, he has also identified in the process of transportation of waste to landfills or to the waste recovery plant, such as reduced safety in transporting the waste on highways, additional noise on the streets and increased traffic jams. The analysis of the product lifecycle could also be used for identifying the external costs. The complexity comes from accurately valuing of the identified externalities.

Starting from the neo-classical definition for social costs, Porter (2010) defines private and social costs in the field of waste management, in order they to be calculated. According to him, private disposal costs are those that are paid by the generator of waste. And the social costs are those that the society pay for disposal of waste. From this viewpoint, the marginal private price is the costs of the producer to produce something. His marginal social price is what costs of the society this thing to be produced. The difference between them both is the marginal external costs – costs of the society, which are not costs of the producer. That is, marginal social costs exceed marginal private costs **(MSC> MPC).**

According to Porter (2010) many of the participants (in the process of waste management) face prices, which are below marginal private costs, not to mention marginal social costs. In most cases, the price for waste collection is covered by the municipal funding, based on the total fees and local taxes. The amount of the fee (tax) is uniform and flat, and it does not depend on the volume of the generated waste. Thus, **MPC** for disposing additional unit of waste equals zero. It is only important the total revenue to equal the total cost **(TR=TC)**. Which is the important incentive: more waste to be generated or the generated waste to be disposed improperly.

This inefficiency is a signal for government to intervene and to start regulating this market. The question here is to what extent, this could happen. To solve this problem economically, according to Porter (2010) this means change of prices. Because waste is collected, disposed, recovered, recycled and in all these operations it goes through at least **MR=MC**. However, this is not enough, because there are externalities. According to him, the correct price of everything is the marginal social cost **(MSC)** for its production. People will buy something only if their willingness to pay **(WTP)** is higher than the price (P). Consequently, the function **P = MSC** guarantees that **WTP ≥ MSC**. In this way, consumers must be ready to pay for the desired product a sum that equals social costs for its production.

Here must be accounted also the impact of the market failures, which arise from the improper evaluation of waste. There are two types:

The first market failure is the “hidden subsidy”, which means underestimation, that means the price does not cover the marginal private costs **(P <MPC)**. Then **WTP ≥ P**, but **P < MPC < MSC**. This represents an incentive and subsidy for households to generate more waste. It happens when the waste handling costs are calculated due to the operation costs of the landfill after it has been built. In this way, the costs for its future decommissioning and monitoring afterwards, as well as the costs for construction of a new landfill are not accounted. Calculating only the operation costs for landfill construction is a hidden subsidy.

The second market failure is external costs due to waste disposal. External costs mean that the marginal social costs are higher than marginal private costs: **MSC> MPC**. If, either **P <MSC**, or **MPC <MSC**, there is no guarantee that **WTP ≥ MSC.** That is why, when the price covers external costs in which **MSC> MPC**, we offload the waste generators of costs and direct these costs to third parties through common taxes and fees. Thus, we encourage increase in the waste disposal. Only if **WTP ≥ P = MPC = MSC**, then **WTP ≥ MSC.**

Porter (2010) derives another social costs in the waste management – the "marginal social costs for recycling". The derivation of this social costs is provoked by two issues. The first one is that market prices of recycling materials do not reflect the social benefit of recycling, what is the reduced production of energy or the reduced extraction of primary natural resources. These prices account only the reduction of private costs of buying cheaper production resources. The second issue is related to how much quantity must be recycled. The recycling of many types of materials would lead to higher marginal costs because of inefficiency of scale, which inefficiency starts from the separate waste disposal of households and their collection by the recycling company (many types of waste containers, complex trucks and etc.). Also, the marginal costs of the recycling process itself vary in the various types of materials (such as plastics, glass, aluminum). It can be accepted that all types of materials can be obtained either through extraction of primary natural resources or through recycling. Thus, both alternatives are perfect substitutes, and the society must make a choice between them. Of course, this is a theoretical model because no material could be completely recycled.

„Marginal social costs for primary natural resources“ are denoted by **MSCv**. This means that recycling saves **MSCv**. In addition, recycling saves also „costs for disposal of primary natural materials after their use“ - **MSCd**. Then, total marginal price for production and later for disposal of primary natural material sill be the sum of: **MSCv + MSCd**. And the “marginal social costs for recycling” **(MSCr)** will grow with the increase in the volume of recycling itself, because in order to be achieved this increase, it will be included more and more new types of materials. The more types of materials are recycled, the more the inefficiency of scale will increase, as it was stated in the paragraphs above.

Recycling must be increased to the limit: **MSCr = MSCv + MSCd**. Failure to observe this equality will lead to inefficiency. Either we will waste primary natural resources, instead of recycling optimally, or we will recycle too many materials and loose financial resources for that.

Separately, we must take in mind that no pricing scheme is not completely effective, because it could not ignore completely the drive to illegally dispose of waste. Each tax for collecting and recycling of waste, even it is minimum, it drives to that. Especially in the case when it equals marginal social costs, or it is higher than them. The effective market must send correct price signals to households, which should tell them that recycling leads to lower social costs than landfilling or illegal dumping. However, the real market does not send such price signals. And households decide how to act on the basis of their moral beliefs and habits. When relying on non-economic factors in the decision-making process, it always leads to high social costs. Options with the creation of various deposit systems are expensive and complex for execution opportunity to prevent illegal dumping, which is why they are hardly implemented in practice.

From the viewpoint of demand, the price also plays important role through the signals it sends. That is why, public authorities are trying to create high price of the recycling materials to encourage their supply. They set minimum thresholds for the content of recycled materials in new products or conduct green public procurement through which they seek recycled products for the public administration. For example, recycled paper and others. In this way, they motivate companies to look for ways to encourage households to hand over waste for recycling. However, the government cannot buy all the recycled products and in this way, it intervenes and distort the market.

According to Porter (2010), there exists an illusion that recycling creates jobs. The necessity of many jobs for it only proves that this is a labor-intensive and expensive activity. On the next place, jobs for recycling are not new and additional ones, but they just replace jobs in other budget policies. If they are additional jobs, then they are borne by the increase in the state or municipal taxes, which is not a good idea according to the neo-classical economic theory. It should be borne in mind that the recycling operations itself also create external costs, even if they are lower than those in the processes of incineration and landfilling.

Porter (2010) also derives some general relationships:

Elasticity. Waste generation has positive elasticity with income and waste increases with income. However, this elasticity is lower than 1, because a significant part of the increased income is directed to the service sector.

Evaluation of the external costs. The evaluation of external costs with the financial burden could be done through comparable values. For example, when we analyze the strictness of the safety measures of landfills in the viewpoint of the risk of serious diseases of people living in the vicinity of them, it could be used the difference between prices of used and brand-new cars. This difference must account the preference and weight, which people give to their own safety and health. After that, the sum must be compared to the price of safety measures of landfills, measured in one human life saved from averted (cancer) disease. Another possibility is to be accounted the lowest prices of real estates as an indicator for the value of external costs imposed by the landfill (noise, polluted air and etc.). For example, a difference of 25 000 US dollars in the prices of real estates in accordance with the distance from the landfill, given the annual real interest rate. This is calculated as the readiness of households to pay 1000 US dollars per year, to move 6 miles further from the landfill. Then, this amount of money will show the willingness of the inhabitants of the municipality to pay for the expected life saved from a potential cancer disease. This would help with various calculations, including the fees for waste.

Internalization of external costs. There exist several variants for internalization of the external costs, which have their pros and cons:

(1) Through negotiations of the type, proposed by Ronald Coase, bringing together all those affected parties and they negotiate with the causer of the external costs. Thus, there will be accounted the interests of all affected parties in the negotiations. A disadvantage of this approach is that the affected parties must be few in order to organize each other and to reach a unified position to defend. But for the external factors, associated with waste, this could not be reached, because the number of the affected parties is large, and this will make more expensive such type of negotiation.

(2) To file a lawsuit, if the affected parties could invoke to a violated legislation act or another governmental or municipal regulation. The polluter will be fined a monetary penalty and thus his private costs will become very high and will demotivate him to continue polluting. A disadvantage of this approach is that the lawsuit can be prolonged in time and its result is always uncertain. Especially when there are many polluters in the same living area, and it is difficult to calculate what is the contribution of each of them in the pollution.

(3) Public authorities could implement regulations which aim to prohibit or reduce external costs generation, providing the relative penalties for their violations. But as is well known, the model for administrative reduction and control is inefficient because of its bureaucratic and political disadvantages. This leads either to very high or very low fines. And there probably won't be enough control resources to monitor its compliance.

(4) Pigou taxation of each unit of pollution. A disadvantage of this approach is that the government must accurately calculate the amount of the marginal tax, which must equal the marginal external costs. However, this is very difficult. Otherwise, either there will be a very high financial burden for the business, or this process will encourage pollution.

(5) Trading with pollution permits. The advantage of this approach is a reduction in the pollution ceiling. The disadvantages are less than in the other models. The inefficient primary permits allocation is corrected by the trading process of them. However, it remains in a great extent an administrative market.

(6) Nationalization of the polluting industry. Thus, it will be managed in a beneficial for the society way. In most cases, companies for waste management are municipally owned, which classified them in this group. A disadvantage of the approach is that they are managed in an inefficient way, and this leads to higher community costs.

The marginal private costs of the landfill. When an additional unit of waste is introduced to the landfill, two types of costs arise. First, variable costs for landfilling – digging, pushing, lining, covering and others. Second, when this waste block is already landfilled, then it moves forward the date when the landfill is due to be closed. This includes also monitoring costs after its closure, as well as costs for construction of a new landfill. The depreciation costs must be also accounted, as well as the opportunity costs of the land. Any additional costs are also possible: when the potential neighbors fight against the construction or the expansion of the landfill or the plant for waste incineration. Usually, they are motivated by concern for future external costs or possible decrease in the value of their property. Then, they add many real costs to the project: for negotiations, meetings (opportunity costs of the time lost) and legal fees, lawsuits.

Comparison of external costs and private costs of waste incineration plants with those of landfills:

Comparison by private costs:

a) the primary capital costs of waste incineration plants are much higher than those of landfills.

b) waste incineration plants show economy of scale to a much extent rather than landfills do.

c) costs for land are insignificant part of the waste incineration costs.

d) incineration once operational, is very expensive if it does not receive a steady stream of refuse derived fuel (RDF).

e) incinerators extract more energy from waste than landfills do, which reduces their operating costs on this indicator to a greater extent than landfills.

The difficulties in comparing private costs come from the fact, that they have a time horizon form 30 and more years, which requires forecasting and calculations with a discount models. And for such time horizon, it is very difficult everything to be predicted, including the long-term prices of energy. The comparison of external costs is also not easy: if the incineration plant is plasma and expensive (A) and pollutes few units of polluter (a). At the same time, the landfill (B) costs less, but emits more units of pollutant (b) – then the equation will be as follows: **(A - B) / (b - a)**. According to the results, it will be judged if the marginal utility due to the reduction of this pollutant is sufficient. However, when the pollutants are many and they have a complex effect, the task is much more difficult. There must be accounted that in the long run, both private and external costs will decrease because of the advances in the technical progress. This also leads to a fierce competition from new investors and technologies, as an investment made now could turn to be unprofitable in the future.

Effects of recycling on costs for collection and disposal of general household waste. On the one hand, separately collected waste is not necessary to be collected by the system for household waste. This reduces the volume of waste which must be collected and transported to the landfill. On the other hand, the system for recycling and separate waste collection operationally and technically duplicates the system for general household waste. It uses its own containers next to each other container for general household waste and uses trucks for collecting the waste from its containers. Unlike municipal waste trucks, the separate waste collection trucks cannot compact the separately collected waste in order not to damage it, and this means more courses and a larger amount of trucks. Thus, the variable costs for collection of recycling materials are much higher than the variable costs for collection of traditional solid waste. And the marginal cost savings here are certainly much less than the average cost. Of course, this only applies to the collection stage of both types of waste. The reduced costs of delayed landfill capacity must then be considered. And also, the revenues of the sale of recycling waste, which could be accepted as social compensation (social benefit), when the costs for the recycling process are deducted.

Market structure. The collection of solid waste has a market structure – a natural monopoly. Presumably, this is because of the large scope of the activity and the possibility of economies of scale.

Porter's (2010) analyzed publication is over 300 pages long and has been summarized very briefly for the following reasons. First, because it can be accepted as basis for any subsequent research in the field of waste management. Although it is not specifically referred to the circular economy concept, it also analyzes social costs from the viewpoint of waste management. In the absence of such analysis, specifically oriented to the circular economy, this is a starting point. On next place, Porter (2010) examines this problem very strictly only from the viewpoint of the neo-classical economic theory. This gives an opportunity to be used another theoretical approach for analyzing the problem.

## 2.3. Critical analysis, based on the elements in the neo-classical modelling of social costs in the waste management process.

Most of the scientific publications, analyzing the correlation social costs and waste management are based on the neo-classical economic theory.

Porter (2010) uses this approach in the benefits analysis and in the cost efficiency analysis in the waste management. There he analyzes social costs based on **Equation (1).** In the context of waste management, Porter (2010) accepts that: social costs are those, which are the real price for the society to dispose waste as opposed to the usually lower price that the waste generator pays for doing so. In order social costs in the waste management process to be in balance with the private costs, according to him, the following equation must be achieved:

$$\mathbf{MSC = MPC = P} \qquad \textbf{(4)}$$

This means that marginal social costs must equal marginal private costs and, they must equal the price of the operations, related to waste management. However, it should be considered that this equation is efficient only in the conditions of perfect competition are complied, which in the waste management sector is questionable.

This gives opportunity to be made a critical analysis, based on the elements of the neo-classical modelling of social costs in waste management processes.

**The first element are the „external costs".**

An integral part of the neo-classical analysis of social costs in waste management is the definition of „social costs". In his analysis of waste management, Porter (2010) accepts the neo-classical definition for „external costs", according to which these costs are borne by someone else, who has neither given a permission for them to be imposed on him, nor is he compensated for the damages he suffers as a result of them (Buchanan and Stubblebine, 1962). But he indicates that in most cases, it is difficult to understand who exactly bears the external costs and to assign an adequate value to them in relation to the burden incurred. Porter (2010) uses the neo-classical definition of Gruber (2012), of which he derives external costs as the difference between social and private costs.

As an example, for "external costs in the waste management", Porter (2010) indicates: noise in the operations with them, air and groundwaters pollution in their disposal, toxic and in some cases radioactive danger in their improper recovery. Such costs are identified in their transportation to the landfills or to the waste recovery plants, safety at highways, additional noise at streets and increased traffic jams. According to him, for the identification of the external costs could be used the analysis of the product lifecycle if there is an opportunity to value them.

According to Medina-Mijangos et al., (2020) in the scientific literature, there are no publications, which collect and group in a methodological way identification and description of the most important impacts that must be considered in the implementation of projects, concerning the management of solid household waste. They cite models for economic evaluation of solid waste management systems, which models are based on the LCC principles, including only the following external costs: environmental emissions like air pollution, influence on the soil and the groundwaters and impact on the quality of life. In these analysis, compensatory effects (revenues) are also examined: use of substitute (green) electricity and reduction of fertilizers use by compost. As the publication of Medina-Mijangos et al., (2020) is relevant, their conclusion about the limited nature of the existing research on external costs sets the stage for a scientific contribution in this area.

In the review of the scientific literature, there could be identified the following categories of "external costs" in the waste management process, which are included in various measuring models (Miranda & Hale, 1997; Balasubramani et al., 2020; Medina-Mijangos et al., 2020; Hu, 2013; Porter, 2010; Rathore and Sarmah, 2020):

First, determinants, linked to the public health affecting. Any stage of waste handling, treatment and disposal could be related to this fact. It could be done either directly (short-term intensive direct impact) or indirectly (long-term indirect low-intensive impact, which could lead to chronic diseases). It includes mainly: physical risks (exposure to noise, ionizing radiation, temperature); chemical risks (exposure to gases, vapors, fumes, and chemicals); biological risks (exposure to viruses, bacteria, blood, and blood products). Damage to public health could be evaluated from the viewpoint of workers (formal and informal sector) and the population, which lives close to the solid waste management plants.

Second, determinants, related to environment affecting. This group of impact includes emissions to atmosphere, emissions to soil, groundwaters and surface waters. There are also costs for affecting tourism and agriculture.

Third, determinants, related to the quality of life. Nuisance and disturbances (warm, dust, visual intrusion, smell, noise, traffic, traffic accidents, vermin) caused by the infrastructure and facilities for waste management (landfills, waste incinerators). They generate resistance from the side of the local community, because of their very existence, even no pollution is proven (NIMBY syndrome). They affect the prices of local real estates and the image of the residents in comparison to the other regions. Miranda & Hale, (1997) determine them as „aesthetical external costs“. According to them, local citizens could even assign higher values of these costs, compared to the others.

Fourth, determinants related to the opportunity costs concept (opportunity price). There exist few opportunities in the scientific literature. First, calculations, based on various opportunities for waste use. Second, if opportunities for waste use are missing, then the calculations are done according to the standard neo-classical model for opportunity costs compared to the yield of the financial instrument if the resources for waste management are invested in it. Third, calculation of the opportunity costs not only according to the principle for profit maximization (the best economic alternative), but also according to the principles for sustainable development (alternative which best ensures their achievement). Fourth, opportunity costs for land, used for waste management purposes.

What are the opportunities external costs to be measured according to the review of the scientific literature. According to Miranda & Hale, (1997), most of the studies rather than assessing the external costs, they analyze the opportunities to mitigate them. They concern that in the external costs' measurement of waste treatment plants one must work with a large variability in the estimates. The external costs of these waste disposal facilities must be evaluated in regard with the local political (strategic), economical and environmental context. Medina-Mijangos et al., (2020) believe that very often it is necessary to translate environmental and social aspects into monetary values in order to work in uniform units that allow adding of total costs and revenues in municipal solid waste systems. Consequently, it is necessary to be determine the units, which these environmental and social aspects have for each of the studied impacts. After that, these units will be the foundations for the economic evaluation. All these units of quantification must be referenced to a set time in the impact frequency. According to the authors, there are several methods, which share the common feature of using market information and behavior to infer the economic value of the externality. These procedures are known as evaluation techniques. They represent a classification of the methods and techniques, where they are considered amongst other methods, as an averting behavior method, benefit transfer method, compliant assessment method, control cost method (abatement method) and cost of illness. They are used methods that study the value of statistical life (VSL) or the years of life lost (YOLL) for external costs evaluation, related to the human health.

In summary, the analysis of external costs in waste management in the context of the neo-classical economic theory, have some disadvantages:

- In most cases, it is very difficult to understand who bears the external costs and also to assign an adequate value to them in relation to the burden incurred. Porter (2010);
- According to the current review, done by Medina-Mijangos et al., (2020) in the scientific literature there are no publications, which collect and group in a methodological way the identification and description of the most important impacts that must be taken into account when implementing a management project of solid household waste.
- According to Miranda & Hale, (1997), most of the studies rather than evaluate the external costs, they analyze the opportunities to mitigate them.

In the scientific literature there exist various views of the possible **internalization of external costs** in the waste management process, which are criticized because of their disadvantages:

1) Through negotiations of the type, proposed by Coase (1960), as all affected parties unify and negotiate with the causer of external costs. However, according to Porter (2010) a disadvantage of this approach is that the affected people must be few in order to organize themselves and to achieve a unified position to defend. To the externalities, related to waste, this could not happen, because the number of affected individuals is great, which could make this type of negotiations expensive (high transaction costs).

2) Hamilton (1993) complements Coase's theorem (1960) for social costs as he adds the theory of Becker (1983) for political costs. And after that he jointly uses them to create a model, analyzing the hazardous waste management, based on the political activity of the affected local communities. And based on the results achieved through the model, he derives the level of internalization of externalities, which is assumed to predetermine the amount of social costs.

In this regard the main determinants in the Hamilton's model (1993) are:

(a). Amount of the potential compensation for the damage caused by the externalities. A company, that causes externalities according to Coase's theorem will consider several things. First, the demographic variables (incomes, real estate's prices, education, density of population). The higher these parameters are, the higher the risk of legal action and political resistance is. Second, the potential costs for legal actions due to the cause of future environmental and health damages (value of property at risk, number of affected individuals, an area with environmental damages).

(b). Potential for collective actions of the local community. The magnitude of this potential determines the risk of legal and political actions against the investment intention. The main determinant here is the percentage of adult population, which has voted on the last national elections compared to the national voter turnover. An additional determinant is the history of the region, related to previous oppositions to the investment intentions.

(c). The reaction of the public officers, who are employed on elective vocations in the region. In the standard case, elected public officers will defend the positions of citizens, who have voted for them in order to reelect them. However, in regions with low voting activity, where the corporate vote decides the elections, they could take the side of the investors.

(d). On next place, the high level of industrial development in the region means potential lobbyist (political) support for construction of new waste management facilities.

3) To file a lawsuit if the affected individuals can invoke a violated law or other governmental or municipal regulation. The polluter will be fined, so his private costs will become very high, disincentivizing him to continue polluting. According to Porter (2010), a disadvantage of this approach is that the court case can prolong for a long time and its outcome is always uncertain. Especially when there are many polluters in the same area and it is difficult to determine what is their contribution to the pollution.
4) The public authorities may adopt regulations that prohibit or limit the generation of external costs, providing appropriate penalties for their violation. But as is well known, the model of administrative restraint and control is ineffective because of its bureaucratic and political shortcomings. This leads to either higher or lower fines. And there probably won't be enough controlling public resource to monitor their compliance (Porter, 2010).

5) Pigou taxation for each unit of pollution is studied by Barrett and Kathleen, (2019). They try to discuss the possibilities of applying green taxes in the circular economy model. According to them taxes and fines, which internalize external costs must be charged to the direct recipients of utility (households, companies or „things“– for example real estate). The amount of green taxes should have a signaling function about the real scarcity of renewable resources and the real price of services which are part of the circular economy system. There must also be sufficient alternatives so that consumers of taxed services are not disproportionately affected.

   According to Porter (2010), a disadvantage is that the government must calculate exactly the amount of the marginal eco-tax, which should equal the marginal external costs, and this is very complicated. Otherwise, there will either be an undue burden on business or pollution will be encouraged.

   Barrett and Kathleen, (2019) distinguish existing green taxes from proposed circular economy taxes. According to them, despite the apparent similarity, the existing taxes aim to change the behavior and to correct externalities but leave the linear structure of the economy not affected. While the new circular economy taxes aim to radically restructure the economy, seeing it as embedded in the environment rather than something separate from it. This will lead to a change of the economic paradigm. With the current green taxes, governments charge everything they can, not what they really have to. A long-discussed topic in this regard is shifting the emphasis from taxing labor to taxing the consumption of non-renewable resources. This, in addition to positive environmental effects, it will also lead to an increase in productivity in the economy and to the opening of new "green" jobs. And the so-called emissions trading schemes must be reformed to be more efficient and not serving to avoid green taxation. Also, subsidies for activities that do not meet the principles of the circular economy should be stopped.

6) Trading with pollution permits. According to Porter (2010), a positive side is the reduction of pollution ceiling. Inefficient primary allocation of permits is corrected during the trading process. However, there are some negative sides because in a great extent it remains a type of administrative market.
7) Nationalization of the polluting industry. Thus, it will be managed in a way beneficial for the society. According to Porter (2010), in most cases waste management companies are municipally owned, which categorizes them into this group. A disadvantage is that they are always managed in an unprofitable manner and result in high costs to the community. The standard criticisms of public choice theory mentioned in the previous part of this chapter also apply here.

In parallel with the reporting of external costs in the waste management process, there must be considered also the **external benefits** of this activity.

According to Medina-Mijangos et al., (2020) this is primarily the possibility for resources usage, obtained because of the waste management activities. This is the extraction of energy from waste, raw materials from recycled products or compost for agricultural needs. Second, the additional effects that rise are better energy security, reduction of emissions from greenhouse gases, creating jobs and extending the life of the landfill. The energy security is linked to the guarantee of supply of resources and energy from waste in the conditions of their scarcity. For countries, where this scarcity is very high, then the guarantee itself is an external benefit. This also includes the appropriate quality according to the purpose of the recycled or recovered resources, as well as energy recovery. This required quality may also require specification and standardization. In all these cases, however, the resource extraction is also associated with the consumption of resources, because recycling does not lead to a complete recovery of the primary resource, but only to a partially one. Separately, the recycling process itself has its own external costs and separate energy consumption. Given this, the external benefits of waste recycling or energy extraction from them do not fit into the circular economy concept.

**The second element are the „private costs“.**

Both approaches for social costs determination (both this of neo-classical economic theory and this of the institutional environmental economy) include measurement of **private costs** in the waste management process. That is why, it is necessary to be analyzed the nature and elements of the private costs in the models for social costs determination in the waste management process. From the scientific literature review, it could be summarized the following:

Defining the private costs. According to Porter (2010), private costs of disposal is what the generator pays for its waste, as opposed to the social costs which show how much it really costs to the society to dispose this waste. Private costs in the waste management process are for example: wages, fuel, machines, land, and others. They could be easily measured, based on statistical, accountancy and taxation data for the monetary expenditures done by the relevant economic agents. According to Medina-Mijangos et al., (2020), private costs could vary depending on the type of waste management system chosen.

Price signals affecting the private costs. The problem with private costs according to Porter (2010) is that many of the participants (in the waste management process) face prices which are below the marginal private costs, let alone marginal social costs. In most cases the price of waste collection is covered by the municipal funding, based on general fees or local taxes. The amount of the fee (tax) is uniform and flat, and it is not dependent on the amount of waste that is thrown away. Thus, **MPC** for disposing an additional unit of waste equals zero. It is only ensured that the total revenue is equal to the total cost **(TR=TC)**. This is an incentive to be generated too much waste or dispose of it in inappropriate way.  Porter (2010) defines also the elasticity in the waste management process. The waste generation has positive elasticity to incomes and the amount of waste increases with its increase. This elasticity, however, is below 1, because a significant part of the increased income is directed to services.

Structure of private costs. According to Jamasb and Nepal (2010), the construction and operation of waste treatment facilities includes the following private costs:

- Direct (variable) costs include costs for operation and maintenance, and they vary in accordance with the production quantity, such as raw materials, labor costs, facilities maintenance, equipment, and training programs. Also, taxes for disposal of unwanted residual materials in waste incinerators or hazardous materials for recycling. Medina-Mijangos et al., (2020) examine the same costs by supplementing them with the costs for sanitation and cleaning of roads and piping in the municipal areas.
- Indirect or fixed costs do not vary with the quantity of output. These are primarily "costs for land", for installation construction and a place to store waste (which will be recycled or incinerated). "Capital costs" in the various stages of installation construction are high and they are part of the financial costs.
- Opportunity costs (interest costs or lost revenues), associated with delays in the building permit and operating licensing processes.
- Losses or economies of scale. Technical progress. The effective scale of installations affects costs. For example, doubling the size of waste incinerations could increase capital costs by only 70% and achieve even greater savings in labor costs.

  In the model of Hamilton (1993) the efficiency of scale of the installation is set as a function of the volume of waste generated in the region and of the available free capacity for their management. This covers the level of industry development in the region, its added value and export potential. The higher the level is, the greater the necessity for construction of new landfills and waste recovery plants is.

Jamasb and Nepal, (2010) found that the technological progress decreases the total private costs of waste recovery facilities by 1,5% per year.

- Structure of the industry. According to Porter (2010), municipal solid collection waste has a market structure – a natural monopoly. Presumable this is because of the large scope of the activity and the possibilities for economy of scale.

Rathore and Sarmah (2020), construct a function for the total costs in the waste management process, which function is calculated by summing functional costs, transportation costs, rental costs, environmental costs, social costs and penalty costs. All these costs are covered by the local authority, which is responsible for the solid waste management, without generating profits and revenues.

According to them, private costs are:

- Functional costs: they are obtained by the sum for daily current expenditures of the waste treatment facilities. They include mostly the operation costs, maintenance, and resources costs, such as charges for electricity, water, wages, maintenance and facilities charges, fees and taxes and others.
- Transportation costs: these are additional costs for transportation of the municipal solid waste between places of origin and landfills. Medina-Mijangos et al., (2020) also pay special attention to the transportation infrastructure and the associated costs, distinguishing them from other categories of costs.

- Rental costs: these costs are applied in most of the cities, where municipalities and local authorities rent vehicles or a company to collect and transport the municipal solid waste.
- Penalty costs: in many cities, the municipality imposes a penalty for delays of companies operating municipal activities. The company must pay the penalty if it exceeds the given time for collection and disposal of municipal solid waste according to the deadline of the contract signed with the municipal authorities.

Several things in the model of Rathore and Sarmah (2020) for determining the private costs are of interest for the present research. With the term „functional costs“ they denote the mixed category, which is a sum of both variable and fixed costs. On the other hand, they subtract from the set of fixed and variable costs and single out several types of costs that, according to economic theory, are part of them. These are: transportation costs, rental costs, and penalty costs. It could only be supposed that they put a special accent on these types of costs, because they consider them as very important for the waste management. On the other hand, this makes disputable the efficiency of such model from the viewpoint of the neo-classical theory.

Medina-Mijangos et al., (2020) emphasize mainly on the chosen waste collection system when determining the private costs. Their model examines the costs by elements of the chosen system infrastructure; costs, associated with the waste collection infrastructure; costs, associated with the transport infrastructure; costs, associated with the infrastructure for preliminary and subsequent waste treatment; costs, associated with the infrastructure for final waste disposal.

Porter (2010) has defined „marginal private costs of landfilling". According to him, when one additional unit of waste is introduced into the landfill, two types of costs are incurred. First, variable costs for landfilling – digging, pushing, lining, covering and other. Second, landfill decommissioning charges: each new amount of waste that is landfilled moves forward the date when the landfill must be closed. This also includes costs for monitoring after its closure, as well as costs for construction of a new landfill. Depreciation costs should also be taken into account, as well as the opportunity costs of the land. Additional costs could also incur when potential neighbors are fighting the construction or extension of the landfill or the waste incineration plant. Medina-Mijangos et al., (2020) use an analogous approach to determine the costs, associated with the infrastructure for final waste disposal.

According to Medina-Mijangos et al., (2020) and Hu (2013) when determining the private costs, there must be considered also the private benefits. The private benefits are obtained when private costs are deducted from the private revenues. These private revenues are a result of the selling price of recovered products per annual volume of waste treated or energy generated. When the obtained private benefits are greater than zero this will guarantee that the project for solid waste works economically and financially from the private agents' viewpoint.

**"Market failures" are the third element in the waste management process.**

Social costs in the waste management process are associated with a market failure. According to Porter (2010), market failures in the waste management are two types:

**The first market failure is the „hidden subsidy“**, which means underestimation. These are the cases when price does not cover the marginal private costs (**$P < MPC$**). Then willingness to pay will be **$WTP \geq P$**, but **$P < MPC < MSC$**. This represents an incentive and subsidy for households to generate more waste It happens when waste handling costs are calculated only by the operation costs of landfill after it has been built. Thus, the costs for future decommissioning and monitoring afterwards, as well as the costs for building a new landfill are not considered. Charging only the operation costs for landfilling is a type of hidden subsidy.

$$[WTP \geq P] \text{ when } [P < MPC < MSC] \quad (5)$$

**A second market failure is the external costs for waste disposal.** External costs according to Porter (2010), mean that the marginal social costs are greater than the marginal private costs: **$MSC > MPC$**. If either **$P < MSC$**, or **$MPC < MSC$**, there is no guarantee that **$WTP \geq MSC$**. Therefore, when the price does not cover the external costs representing **$MSC > MPC$**, we offload waste generators from costs and direct these costs to third parties through general taxes and fees. This is how we stimulate the increase in waste disposal. Only when **$WTP \geq P = MPC = MSC$**, then **$WTP \geq MSC$**.

Effectiveness in the waste management from the viewpoint of the social costs will be present:

$$\text{When } [WTP \geq P = MPC = MSC] \text{, then } [WTP \geq M \quad (6)$$

According to Porter (2010) external costs must be reduced to an optimal level. In this regard, he is following the famous model of the neo-classical marginal analysis (Opocher, 2017). Schematically this model suggests that the marginal benefit is negatively sloped and it is decreasing as the environment becomes cleaner and pollution decreases. Thus, each subsequent cleaning action (waste treatment) brings smaller and smaller marginal benefit. At the same time the marginal costs rise because they are positively sloped and increasing. The effectiveness is sought in the equation: **MR=MC.** The complexity of the analysis is determined by the introduction of external costs in this popular neo-classical equation. The evaluation of the external costs in relation to the financial burden incurred could be done with comparable values. For example, according to Porter (2010) when we analyze the strictness of security measures of landfills in view of the risk of serious diseases to people, living around them, could be used the differences between prices of new and used cars. Another option is to be considered the lower prices of real estates as an indicator for the value of external costs imposed by the landfill (noise, polluted air, and others).

**The fourth element – modern neo-classical definition of social costs in the waste management.**

The social costs in the field of environmental policy are defined by the Ministry of environment of USA (EPA-USA, 2000a). According to **Nozharov (2018),** EPA defines "total social costs" as the sum of opportunity costs incurred by the society because of a new regulation policy. They are measured by the value of the lost by the society products and services because of the resources usage to comply with the regulation, as well as because of the reduction of the final output. They include five main components (Real-resource compliance costs, Government regulatory costs, Social welfare losses, Transitional costs, Indirect costs).

In the analysis, it is considered the demand elasticity and supply of affected products and services. Transaction costs in this model are included as part of the transition costs, together with unemployment, companies ‘closure and others. They are presented as a very small part of one of the five main components of social costs. According to EPA transaction costs arise from the implementation of new incentive-based policies, such as the tradable permits program.

All subsequent scientific publications, studying the social costs in waste management use the model of EPA-USA (2000a): Kinnaman, Shinkuma and Yamamoto, (2014); Jamasb and Nepal, (2010); Dijkgraaf and Vollebergh, (2004) and others. In this way, according to the purpose of the current research, social costs in the waste management are completely or partially accounted, because of the inaccurate accounting of transaction costs.

To a large extent the neo-classical theory is an instrumental one. It has many disadvantages from the viewpoint of the focus of the present study. Such type of instrumental measurement of social costs requires accurate evaluation of the negative externalities. Very rarely their measurement is exhausted only with the observed momentary effects and their monetization. Often, in a long-run period additional effects arise, which are accumulated with the previous ones, but they can be hardly predicted and measured. To what extent the measurement of their presence and internalization is possible in the presence of the relevant economic and political institutions (through the methods of the new institutional economy), it will be studied in the next chapter of the research.

One of the main problems, related to the social costs' internalization is associated with their predicting and valuation. According to Saltelli et al., (2015) in the conditions of „uncertainty ", which exists about predicting the negative consequences associated with social costs: causal chains or networks are open. There is neither enough information about what the likely result will be, nor what the mechanisms of its reproduction will be, nor what the probability of its occurrence will be. A further complication is that the forecasts must simultaneously incorporate aggregate modeling of the physical, socio-economic and political effects of climate change (Farmer, Hepburn, Mealy and Teytelboym, 2015).

A disagreement continues to exist in scientific assessments of various elements affecting social costs, such as the value of human life and health, the value of various ecosystem benefits, the effects of global warming (Miranda & Hale, 1997; Porter, 2010). According to Moreau et al., (2017). There is a lack of credible information about the socially beneficial value of the damaged global public goods, as for that purpose it is not sufficient the monetary equivalent of human health and life to be determined.

This calls into question the credibility of the models for evaluation of the real value of a unit of resource for the present and future generations such as the Life Cycle Assessment (LCA) and Cost-Benefit Assessment (CBA) methods (Huppertz et al., 2019). Other similar methods are the method of avoiding undesirable behavior, the transfer of benefits, the method of evaluation of complaints, the method of cost control (reduction costs) and costs related to affected illness. Indicators that consider value of statistical life (VSL) or years of life lost (YOLL) are used to estimate external costs related to human health (Medina-Mijangos et al., 2020).

Also, when consumers ‘decision for protection of certain rights includes a real monetary payment, measured by their willingness to pay (WTP), in many cases they refuse to pay.  Although as members of the society with constructed values according to the cultural, educational and normative environment, they have previously supported a given social position (Söderholm and Sundqvist, 2000).

On the other hand, to take actions in relation to the social costs incurred, accounting the present conditions of the environment, it could be late, and it could have negative social consequences (Kapp 1976; Swaney and Evers, 1989). The following solutions in the institutional environmental economy are perceived as possible ones; preventive measures at the earliest investment stage but not ex post facto removal of the pollution or punishment of the guilty party; regulations for control; encouraging investments in technologies for pollution reduction; environmental standards and institutional framework (Berger, 2008; Hu, 2013).

According to Berta & Bertrand (2014) with clearly defined property rights, transaction costs equal zero and the institutional arrangements are optimal, which fact will lead to no externalities Such definition of property rights is possible in the emission markets trading permits.

Only investments in new knowledge are reliable instrument for uncertainty reduction in the field of forecasting the internalization of social cost (Libecap, 2014).

In conclusion, it could be said that most of the authors, analyzing the correlation between social costs and waste management, base their studies on the neo-classical economic theory.  All of the components of **Equation (1)** were presented in the present analysis from the viewpoint of the waste management. This means, that from the viewpoint of the neo-classical approach, this issue is sufficiently clarified. Defining social costs in the context of waste management from the viewpoint of institutional economics is still a challenge. There is a lack of sufficient publications, containing statistical and econometric models. The purpose of the present second chapter of the research is to serve as theoretical basis for the next (third) chapter in which there will be presented the vision for a new model of studying the social costs in the context of circular economy.

# THIRD CHAPTER

# MODELLING THE RELATIONSHIP BETWEEN SOCIAL COSTS AND TRANSACTION COSTS IN THE CONTEXT OF CIRCULAR ECONOMY

## 3.1. The institutional ineffectiveness as possible basis for construction of an analytical model of the relationship between social costs and transaction costs

The main task of the research is to propose a model for deriving social costs only by transaction costs. Till that moment, such a hypothesis has not been studied in the scientific literature, as it is supposed that it is impossible to be proved (Berger, 2017). The only open access publication, which points to such possibility is that of Nozharov (2018b). The author steps on the understanding of the institutional economic theory, that overcoming market failures could happen through replacement of free markets with hierarchies (companies and the state). However, he finds that the state as a public hierarchy could not replace entirely the company as a private hierarchy, because there exist problems with the effectiveness of public institutions. The author also considers the low intensity of incentives, that underlie the hierarchies in contrast to the market incentives.

The further development of the cited analysis could be basis for clarification of the relationship between social and transaction costs through applying the theory of institutional inefficiency. There are number of issues which should be addressed here.

The dilemma, posed by Dahlman, (1979) is whether markets or the state are more efficient in solving market failures. That is, if markets do not function well in neutralizing of negative externalities, does this mean that the government will do better. Wouldn't active intervention of the state cause greater problems rather than the ineffectiveness of markets cause. To answer this question through the institutional economics methodology, there must be defined the terms institutional environment and inefficient institutions. This issue is also important for the analysis of the effectiveness of the circular economy concept of EU. According to Grafström and Aasma, (2021), without efficient institutional configuration the relevant market will not emit efficient price signals, which is relevant to the market of recycled materials. This will push investors back and they will not see opportunities for profits and this market will suffer. The lack of effective price signals will lead to inefficient allocation of resources and goods, which are traded on this market. That is why, the institutional effectiveness turns out to be of key importance.

Swaminathan and Wade, (2016) define the institutional environment as set of normative and regulatory pressure, exerted on organizations by the state, society, and professional communities. This pressure can be coercive and direct (courts and regulations) and indirect (through the implementation of standards that must be met by the organizations). Companies' compliance with institutional regulations could give them legitimacy and insure them better access to resources.

Acemoglu, (2006) defines ineffective institutions as those that enrich a narrow social group and at the same time cause stagnation or low grow of the social development and society as a whole. Ineffective institutions are also those, that do not maximize the growth potential of a society. According to him, Pareto's efficiency is not a good definition for ineffectiveness. In this context, a set of institutions will be ineffective according to Pareto, if another opportunity set of institutions would make everyone better off. In addition, Platteau, (2008) puts effective institutions in dependence of the criterion for effectiveness (according to Pareto) under the following assumptions: „the possibility for transfer payments, zero costs for negotiations, ability to complete and implement decisions, achieved in the bargaining process and absence of wealth effects „. This is practically in the absence of transaction costs.

Main issue in the study of Acemoglu, (2006) is why the ineffective institutions could be sustainable and what are the reasons for their emergence. The analysis covers the policies for balance and allocation of the various types of institutions and the preferences of the different groups and individuals in these institutional frameworks. The author concludes that the ineffective institutions are persistent when the groups, which prefer the generated policies by them are sufficiently powerful (de jure and de facto) and when other institutional regulations, which would generate more effective allocations could be found.

In the analysis of ineffective institutions, their structure should be studied. Acemoglu (2006) presents institutions in two dimensions. The political institutions govern the distribution of rights in the society (through the decision-making process), and the economic institutions regulate the economic interactions (with a focus on the fiscal policies – mainly taxes, as well as the regulations of technologies and investments). Bodmer, Borner and Kobler, (2004) also study the institutions along these two dimensions. The political institutions, which regulate the process of creating formal normative rules and the legal system, as well as the economic institutions, which regulate the coordination of economic activity, related to the property and contractual relationships. According to them, the political institutions must establish such a public order in which neither the majority, nor the minority to be able to unjustifiably encroach on the property rights of other individuals (economic agents). Effective institutions will be those in which property rights are clearly defined and their protection could happen under every circumstances.

A similar statement is made by Bodmer, Borner and Kobler (2004), according to whom there exist various forms of transaction costs according to type of the institutions that generate them, considering the distinction between economic and political institutions. According to them in the economic institutions – the problems are rising mainly due to the asymmetric information between sellers and buyers, as well as due to the (possibility of enforcing) the execution of agreed transactions. These are "weak" property rights, which generate transaction costs. On the other hand, they think that the state is structured by the political institutions, which coordinate the processes of creation and implementation of the legal environment. That is why, in the political institutions, transaction costs arise from the definition and enforcement of property rights and bargaining rights. In order to have effective institutions, the state must have the necessary instruments to enforce property rights. However, it must have legitimacy, which is done through efficient public control. According to Bodmer, Borner and Kobler (2004) this is called „power" and „commitment" of the state. The state will be committed with effective institutions, if the mechanisms of public control are at such stage that they can make non-market (undemocratic) behavior on its part too costly (loss of elections for the ruling party or other consequences). The state itself is examined as an organization, which is a set of individuals, political groups, and political institutions, which interact one to another. The greater is the power of the state, the lower are its opportunity costs for “production” of effective institutions.

Which are the possible approaches for methodological studying of the ineffective institutions.

In searching for the causes for institutional ineffectiveness, Platteau (2008) examines four possible approaches: approach of transaction costs; the principal–agent approach; the game equilibrium approach and the evolutionary approach.

The first approach for analysis of the ineffective institutions is that of the transaction costs. Bodmer, Borner and Kobler (2004) consider that the ineffective institutions cause high transaction costs and in this way they limit the economic effectiveness by creating unstable environment for accumulation and growth of the capital. Given this, the transaction costs approach will be accepted as a main approach in the present analysis.

The term “transaction costs” includes not only the costs for signing and implementing the contracts, but also the costs for outlining and controlling the “exclusive rights” – including those of institutional arrangements such as legislative acts. This definition is very important as it is difficult to separate expenditures for market volumes from the expenditures for defining the rights over the resources – which are the subject of the market transaction (Cheung, 1978). The essential emphasis here is related to the costs, associated with the legislative acts directing to determination of the rights on the resource, because they are related to the participation of the state in the regulation of economic processes. There are some exceptions: Allen, (2015) examines a particular hypothesis. According to him, it is possible to have a case, in which there are certain fixed transaction costs, but the marginal transaction costs are zero. Then the property rights are perfect, although the transaction costs are positive.

Zerbe Jr & McCurdy, (1999) consider that the net value of externalities represents the lower boundary for the associated transaction costs. That is, transaction costs will never be less than the net monetary impact of the externalities representing the net benefits, which must be obtained when the problem is eliminated. Transaction costs, associated with the state intervention, concerning the public goods, consist of two components. The first one is the high exclusion costs (related to the problem of the free rider - non-excludability and non-competitiveness). And the second one – costs for determining a separate level of damages and those necessary for pricing or taxation. In addition, here must be considered also the realistic assumptions for the existing cultural norms defining the acceptable behavior, as well as the costs for monitoring of the implementation of the institutional norms.

From the viewpoint of the present analysis it is interesting the understanding for transaction costs as costs for system management because this focus covers the systematic institutional structure and effectiveness. Analyzing the approach of transaction costs, Platteau, (2008) defines them as: „costs for system management, representing the costs for coordination and motivation“. In this context the author determines the effective institutions as a memorandum that aims to reduce transaction costs in order to allow economic agents to take advantage of the economic opportunities. He places effective institutions in dependence on the specifics of the institutional environment in which they must function. In this way, the institutional choice comes down to a trade-off between technological and transaction costs. The systematic approach to transaction costs gives opportunities to be also made conclusions about the circular economy. As it is more complex than the linear one, it is more static and hardly reacts to changes (Dossa et al., 2020).

Marinescu, (2012) following Arrow, (1969) also determines transaction costs as costs for management of the economic system from the viewpoint of the institutional economics. According to him transaction costs from objective perspective are associated with the "measurement of the valuable characteristics of what is being changed". As the effectiveness is inversely proportional to the sum of the transaction costs, however the problems are related to their measurement and determination.

Of interest is the view of Marinescu, (2012) according to which transaction costs are rather imposed by state institutional constraints (on entrepreneurship). In this context, the author makes a distinction between two types of transaction costs. The first type are the familiar market transaction costs. These are the ones, which participate in establishing and maintaining existing property rights and limiting opportunistic behavior that threatens the efficiency of market exchange. The second one are the transaction costs, imposed by the state, which have exogenous character for the economic activities:

> *„According to this viewpoint, the imposed transaction costs would be the sum of expenditures and efforts, made by the entrepreneurs and market agents as a whole in order to conform to the official institutional framework, which the governing political system imposes to the society."*

The author gives an example for such expenditures, like complicated regulatory framework, burdensome taxation, legislative instability, and corruption.

In both contexts transaction costs could be defined also as costs for coordination and in this context the alternative institutional-political arrangements that minimize them will be efficient ones.

The second approach for analysis of the inefficient institutions is that of the “principal-agent approach”. Platteau, (2008) examines the “principal-agent” approach as analogous of the transaction costs in the determination of the institutional ineffectiveness. According to him, the losses of effectiveness incurred by the institutions or by the contractions in the “principal-agent” approach are equivalent to the transaction costs. However, transaction costs approach analyzes the costs incurred explicitly related to market transactions and based on motivational and coordination problems. And the “principal-agent” approach analyzes the created mechanisms for self-fulfillment by the principal, which should oblige the agent to fulfill his duties when he is outside of the direct supervision.

The understanding of Platteau, (2008) is also of interest, as it is associated with the game equilibrium approach, according to which a formal rule even if established by law, could not be a true institution if agents did not mutually believe in it. This approach in great extent is based on the Nash’s equilibrium and emphasizes the behavior-dependent nature of institutions rather than their environmental, technical, or other specifics. On the other hand, this approach explains the existence of inefficient institutions with risk of uncertainty about potential future institutions, which could replace them and whether they will not be even worse than the current ones. According to Swaminathan and Wade, (2016) in the past, the institutional environment was presented as exogenous force that shapes and constrains organizational actions and policies. However, the modern views suggest that organizations could take steps to form the institutional environment, in which they are embedded and to influence it. That is, they can act strategically and can coordinate their actions.

From the viewpoint of the evolutionary approach, Platteau, (2008) believes that it is impossible to measure the ineffectiveness of a certain institution if there is no dynamic basis for comparison. This approach is examined as a constant movement towards an ideal organizational system in which uncertainty is reduced to a minimum, but this movement creates transformation costs (Kuzmin & Barbakov, 2015).

The measurement of ineffective institutions according to Bodmer, Borner and Kobler (2004) and Henisz and Delios (2000) is possible through:

- capturing the distortions of the foreign exchange market arising as a result of actions or inactions of the state (the premium for the exchange rate on the black market compared to that on the official market);
- share of the currency in the monetary aggregate M2;
- interviews with the business about the property rights status;
- various measures against corruption;
- share of deposits of population in the bank system;
- level of inequality;
- the legal traditions (according to the empirical studies the Anglo-Saxon has a positive sign, and the continental one has a negative one);
- level of education; lack of social discipline because of due to serious gaps in legislation, its compliance and implementation;
- the ability of the government to discipline its administration (bureaucracy) and the private groups for pressure (lobbyism);

- risk of non-service of the government debt;
- level of foreign direct investments;
- the openness and competitiveness of the procedures of the recruitment of state servants.
- whether politicians do what citizens expect from them (the implementation of the pre-election programs and others);
- level of the general fiscal burden, allocation activities and size of the public administration;
- cultural traditions and rules (culture embedded in society) that shape the behavior of economic agents and expectations from the state;
- high openness of the society leads to better institutional environment;
- freedom to vote at political elections, free competition amongst parties during elections, freedom of speech.
- number of murders per 1000 persons of the population, political crisis per year.
- complexity of the institutional system (the more complicated is the system, the higher is the risk of increasing of the transaction costs, based on the complicated system coordination).

When transaction costs are positive, legal regulations can affect the prices, level of production and effectiveness with which the resources are allocated (Medema, 2011). According to Regan, (1972) changes in the legal regulations will not lead to re-allocation of the resources, but to a re-allocation of the economic approach. In the general occasion, there may be no party which earns any income at all from the regulated activity. All parties could earn only their opportunity costs for working in this field. That is why, an important issue for the current analysis is in which cases the effective resources allocation could be done through normatively implemented rules, but not at the free market.

For example, there are cases in which the reform of the institutions rests neither on fairness nor on effectiveness, but on protecting the public interest, as in the case of the COVID-19 pandemic, when economic activity was severely curtailed (Hoffman & Hwang, 2021).

When bargaining is expensive, the efficient resources allocation may require that the law create rules that give parties incentives to act efficiently—rules that direct parties to outcomes that imitate those that the market would produce if transaction costs were low (Schlag, 2013). The law must reduce those transaction costs, which companies or the market could not reduce by themselves or when the companies could not adapt to the transaction costs. It is also important to define in the given market: what goods are wanted and who are the interested parties. Hypotheses about the functions and properties of markets (for example price setting, resource allocation, efficiency) (Regan,1972; Schlag, 2013).

Approaches for change through the law according to Schlag, (2013) and Medema, (2011) can be:

- intervention in the market – sanctioning of certain activities and subsidizing of others (production costs), which is a type of direct state intervention;

- facilitating markets (if they are not competitive – reducing the transaction costs or circumventing their effect with compensating actions) as long as they comply with Kaldor-Hicks rule;
- mechanisms for more efficient dissemination of information;
- or legal rules, which reduce transaction costs;

The institutional theory for the so called "missing markets" occupies an important place in the present analysis. A change in rights, brought by the law, could be identified not only by the observed changes in the law itself, but also by the efforts for its implementation (Cheung, 1978). In this context of interest is the view of Dahlman, (1979) according to whom the alternative of the government actions (inefficient public measures and inefficient public organizations) is the eestablishment of appropriate markets which interact to one another. They must force economic agents to account externalities they generate. The issue, which is not clarified by Dahlman, (1979) is who must establish these "appropriate markets" when private and social costs diverge.

Berta & Bertrand, (2014) citing Arrow, (1969) define externalities as missing market. They justify this definition with the understanding for externalities as interactions, which avoid the parametric pricing system induced by the assumption of perfect competition. The lack of price signals generates ineffectiveness according to Pareto in a competitive environment, which requires correction and determination of the externalities as a market failure. The acceptance of transaction costs in such environment is based on the impossibility for excluding participants and next on the presence of information costs. One option to overcome the market failure, related to the externalities is to be created the missing competitive market and this requires the construction of special institutional organization. It supposes the creation and property rights allocation, which however, does not in itself guarantee competitive exchange. Such there will exist in setting of parametric prices of the externalities, which could be traded and accepted by the participants at such market.

That is, when markets could not ensure effectiveness, then there must be established rights to stimulate or imitate the market, which generate the results, which could be achieved if the missing market existed. The stimulation of a market could lead to results that will be efficient according to Pareto, insofar as the optimal reduction of the negative externality could be determined (Coleman, 1980). However, here must be excluded the so called Posner's rule – the imitating of market should be done by court order, but we should have in mind imitation of a real market through which the information asymmetry and the level of willingness to pay can be overcome at least partially. And the state intervention in the resources allocation should happen only in addition to efficiency or when some other result is sought, for example justice.

At the same time, if we create an imitating market on which we can trade the negative externality, then this will not lead to its elimination. The pollution will not become zero, but it will eventually be optimal according to Pareto (Berta & Bertrand, 2014; Hurwicz, 1995; Dahlman, 1979).

Another option for creating a missing market is related to the vision of Coase. The Coase's theorem determines externalities as a manifestation of missing or not well-defined property rights, but not as an effect, which is associated with the lack of competitive exchange. Berta & Bertrand, (2014) citing Farrell (1987), examine the „Coase's theorem", which according to them abandons the binding requirement for perfect competition of Arrow, (1969). However, it replaces it with the firm assumption that a mutually beneficial agreement will always be reached in zero transaction costs. In practice and in the case of the so called "dual monopoly" this will require a lot of coordination and negotiations. In this regard, the determination of property rights through the market may require such large costs, that the level of optimal production will never be reached. Then it will be more efficient this determination to be done through the judicial system if this alternative of market form of organization will require less costs. The choice of such type of policy must be done based on comparison amongst the results of various institutional arrangements. According to Berta & Bertrand (2014), these various arrangements include the market, but also the company, public regulations, and the status quo. In the Coase's theorem the concept for transaction costs replaces the concept for externalities. This is because the presence of externalities does not prevent achieving a Pareto optimal allocation, but positive transaction costs do (Dahlman,1979). That is the presence of externalities indicates presence of transaction costs. In this context transaction costs lead to social costs.

Berta & Bertrand (2014) indicate that Coase (1960) is considered as the founder of markets for emission trading permits, as he is one of the first authors who indicates the necessity of trading property rights. When property rights are clearly defined, the transaction costs are zero and the institutional arrangements are optimal, which would result in no externalities. According to the authors, "the effectiveness of permit markets is usually justified by the assumption that, in equilibrium, firms will equalize their marginal costs of reducing carbon emissions to the price of emission permits." In this way they will adopt price-taker behavior in a competitive environment". However, these markets (in EU and in the USA) are built as centralized competitive Arrow-type markets, not as decentralized Coase bargaining markets. That is why, transaction costs examined in the context of solving problems, related to market failures imply a comparison of the market with alternative ways of organization (companies, public intervention, and other hierarchies). This also implies a comparison between hypothesis of a market with transaction costs and one with zero transaction costs, as well as the determination of an equilibrium point according to the subjective view of a competitive environment, if such a competitive environment exists (Dahlman, (1979). But this according to the author would achieve optimal (approximate) rather than competitive equilibrium. When there are transaction costs, it is important to whom duties and rights are assigned, which influences the processes of resources re-allocation, benefits, and revenues. In conclusion, Dahlman, (1979) states that it is wrong to focus on elimination of the externalities. He considers that if transaction costs are eliminated, the problem with the externalities will be solved on its own, as the market will internalize them.

Contrary to the generally accepted understanding that the Coase's theorem excludes state intervention in the problem solving with negative externalities, there are some authors who believe in the opposite. According to Medema, (1994) such type of understanding is incorrect, "because the government participates in the creation and transfer of rights in the same way as it is in the policies of taxation" and this fact is confirmed by Coase (1970b, 99). Starting that "….. there is no reason, according to which sometimes … the government administrative regulation should not lead to improvement of efficiency" (Coase 1960, 18). He gives an example for externalities, associated with pollution. According to him, the approach to compare regulatory results with the optimal conditions is wrong according to the economic theory for achieving a socially optimal result. It is more accurately to compare the expenditures spent for achieving the intended result and the effect of these expenditures in comparison to the existing real situation. Given that, according to him the different institutional structures may lead to a lot of different results. In the context of the present research, this requires to be considered the specifics of transaction costs, associated with the waste management in order to be analyzed the effectiveness of the related institutional framework.

Regan, (1972) also believes that Coase does not exclude government intervention a priori, but only in the cases when transaction costs are zero, which fact is impossible to happen in the real life.

But on the other hand, according to the Coase's concept, presented by Schlag (2013), if transaction costs for rearranging of rights through private bargaining are so high as to preclude, then the bad-chosen regime of law will remain final. The total value of production (total for both sides) will be lower than the optimal one. If we consider the total product of both activities (activity causing the externality and affected activity) when comparing alternative social arrangements, the total social product obtained from these different regimes should be compared. Then the comparison between the private and social output will not be correct. The government intervention (implementing reliability) will influence the level of production and costs not only of the affected activity but also of other activities associated with it. Therefore, emphasizing the comparison between private and social output will produce wrong results from an economic point of view, including in the case of adjustment through the law (Coase, 1960; Swaminathan and Wade, 2016; Schlag, 2013).

According to Mohrman (2015), Coase questions the abstract idea of laissez faire, as he indicates that there is no “state of nature” without institutional arrangements, which the economists could study and there exist at least a minimum state intervention in the economy.

The imitating market must lead to the correct social outcomes – outcomes that could occur if we did not have two different contracting parties and the two activities affecting the subject of the transaction (affecting activity and affected activity) were managed by the same person (Medema, 2011).

According to Zerbe Jr & McCurdy, (1999) the existence of market failure is necessary but not sufficient condition for state intervention in economic processes. When the benefits of public intervention exceed the risks, then it will be justified. One possible explanation is that, based on the theory of transaction costs, analyzing the persistent existence of externalities, when the transaction costs for their elimination are too high. In this context, according to the authors whenever we talk about externalities, we are talking about transaction costs. Transaction costs call into question not only the market optimality, but also the ability of markets to direct the resources towards more valuable goals (Coase, 1960; Medema, 2011).

Specifying transaction costs in the context of the current research. The issue of the nature of transaction costs related to the management of global public goods is very specific and complex. There are over 100 international conventions, related to the protection of these goods.

According to Libecap, (2014) four factors increase transaction costs for transferring of property rights:

(a) scientific uncertainty concerning the costs and benefits for mitigation;

(b) different perceptions and preferences in the different populations (countries);

(c) asymmetric information;

(d) the degree of conformity and new implementation (something that appears for the first time).

In this context Libecap, (2014) concerns that institutions, related to the property rights, affecting the global public goods management are expensive for implementation and they are always incomplete. The access to the global public goods is difficult to be limited at reasonable costs and the problem with the free rider rises. The key assets must be measured, and certain consumption, which affects them must be limited or prevented. The prices must reflect social costs, which will decrease the incomes of consumers. Also, it should be considered that despite the general welfare-enhancing effect generated by the established property rights institutions in terms of the management of global public goods, the resulting distribution is not equal. In some cases, it is compensated by the transfer payments. There must be considered also the possible strategic interactions between the individual parties, which are affected or participate in this process. The investments in new knowledge decrease the uncertainty in the level and allocation of benefits and costs. If the investments in new technologies and knowledge are unsuccessful, then in future the society may have less ability to deal with the particular negative externalities. An interesting aspect in this process is the possible participation of the civil society and NGOs. Therefore, according to Libecap, (2014) transaction costs related to the global public goods are the costs of establishing, maintaining, and exchanging property rights to the revenues and benefits derived from the mitigation of negative effects within and between participating and affected parties. Property rights are granted when restrictions are placed on access to global public goods that are normally available to all. Also, the property rights are established by assigning responsibilities to reduce certain economic activity. The benefit of these two opportunities for establishing property rights is for the entire society because of the reduction of externalities. Given this, property rights are implemented in the rules of the community, which determines when and how the global public goods are used.

The issue with determining the effectiveness of institutions is further complicated by their multi-level nature (national, regional, and local). There exist also regional institutional environments, such as EU and ASEAN. We have better connectivity of individual countries' markets and many multi-national companies. Informal institutions, such as culture are more durable than formal ones and influence them by creating institutional substitution (informal norms developed to compensate the ineffective formal institutions). This creates institutional complexity in the form of combined effects of many institutions and their diversity, including polycentricity (Hitt, 2016). This finding is interesting in the context of the present research, as it accounts for the multi-national institutional environment within the EU.

Amongst the Bulgarian authors, who have examined similar issues, could be listed the publications of Gechev, (2008) who identifies the institutional framework for sustainable development in Eastern Europe, as well as Tchipev, (2009) who analyzes the institutional effectiveness of corporate government in Bulgaria.

The most important conclusions in this section are:

**First,** in the economic theory there is no definite answer to the question, set by Dahlman, (1979) – if markets do not deal good enough with the neutralization of externalities, does this mean the government will do it better. There exist various findings and the present research could add contributions to the literature in this field.

**Second,** the companies are dependent on the institutional environment and their legitimacy largely depends on it, as well as their access to resources. This dependence can be direct (normative provisions and court decisions) or indirect (creation of expectations and standards for their activities).

**Third,** it is disputable in the scientific literature to what extent the Pareto efficiency criterion can serve as a framework for analyzing inefficient institutions. And the model of Kaldor-Hicks has largely been rejected by the scientific literature and its use is not recommended in this regard.

**Fourth,** the effective institutions can be set in dependence on the specifics of the institutional environment in which they must function. In this way the institutional choice is boiled down to a trade-off between technological and transaction costs (Platteau, 2008).

**Fifth,** another issue, which is not examined in detail in the scientific literature in this field is why ineffective institutions can be persistent, as well as what are the reasons for the emergence of persistent ineffective institutions.

**Sixth,** the two dimensions identified in the literature review can be used in the analysis of inefficient institutions. The political institutions, which govern the distribution of powers in society and the economic institutions, which regulate economic interactions. To the extent, these dimensions are distinguished, this could help make analysis of clarifying the concrete reasons for the emergence and existence of inefficient institutions.

**Seventh,** according to Platteau (2008) there exist four possible approaches in searching for the causes for institutional ineffectiveness: the transaction costs approach, the principal -agent approach, the game equilibrium approach, and the evolutionary approach. The present research accepts this classification, and the author will use it in the section, devoted to the empirical analysis.

**Eight,** the term "transaction costs" also includes the costs for outlining and controlling of "exclusive rights" – also those of institutional arrangements, such as legislation. In this way, the costs for market exchange are distinguished from the costs for resource property determination (Cheung, 1978). Thus, one can try to study transaction costs, which arise because of the institutional actions of the government.

**Ninth,** from the objective perspective, transaction costs are associated with "measurement of valuable specifics of what is being exchanged", examined in the context of the theory for opportunity costs. Efficiency is inversely proportional to the amount of transaction costs, but the problems arise from their measurement and determination (Marinescu, 2012).

**Tenth,** transaction costs, associated with the state intervention, concerning public goods, consist of two components: high exclusion costs (related to the problem with the free rider – non-excludability and non-competitiveness) and costs for determining separate levels of damages and necessary for pricing or taxation.

**Eleventh,** interest from the viewpoint of the correlation modelling between social costs and transaction costs is the statement of Zerbe Jr & McCurdy, (1999). They believe that the net value of externalities represents a lower bound on the associated transaction costs. That is, transaction costs will never be lower than the net monetary impact of the externalities, representing the net benefits, which will be obtained because of eliminating the problem.

**Twelfth,** the understanding for transaction costs as costs for system management, encompasses system institutional structure and efficiency. Platteau, (2008) determines the costs for system management as costs for coordination and motivation. Such view could be used in the present study in the analysis of the inefficiency of public institutional models, such as that related to the waste management. In addition, according to Marinescu, (2012), transaction costs are imposed by the government institutional constraints, representing the efforts made by entrepreneurs and market participants in general to conform with the formal institutional framework. For example, the complicated regulatory framework, burdensome taxation, legislative instability, and corruption.

**Thirteenth,** the theory of so called "missing markets" occupies an important place in the present study. Overcoming market failure, related to externalities by creating the missing competitive market, requires the construction of a special institutional organization. It presupposes creation and allocation of property rights, which, does not in itself guarantee competitive exchange. Such will exist in setting a parametric value of the externalities, which could be traded and accepted by the participants at such a market (Berta & Bertrand, 2014; Arrow, 1969). At the same time, however, if we create an imitating market on which to trade with the externalities, then this will not lead to their elimination. Pollution will not become zero, but it only be possibly Pareto optimal (Berta & Bertrand, 2014; Hurwicz, 1995; Dahlman, 1979).

**Fourteenth,** another option with the creation of missing market is related to Coase. The Coase's theorem determines the externalities as a manifestation of a missing market or not well-defined property rights, rather than as effect, associated with the absence of competitive exchange. In this regard, determining the property rights through the market may require such large costs that the optimal level of production is never achieved. Then it will be more effective this determination to be done through the judicial system if this alternative market form of organization will require less costs. The choice of such policy must happen on the basis of a comparison amongst the various institutional arrangements: the market, the companies, the public regulations and the status quo. Emission trading markets are based on this concept (Berta & Bertrand, 2014).

**Fifteenth,** emission permits trading markets have efficiency, based on the assumption that in the equilibrium the companies will balance their marginal costs for carbon emissions reduction to the price of the emission permits and in this way, they will be price-takers in a competitive environment (Berta & Bertrand, 2014).

**Sixteenth,** Libecap (2014) determines transaction costs, associated with the global public goods management as costs for establishing, maintaining, and exchanging property rights over the revenues and benefits, obtained because of the negative externalities' abatement within and between participating and affected parties. Property rights are granted when restrictions are placed on access to global public goods that are normally available to all and by assigning responsibilities to reduce certain economic activities. The benefit of establishing property rights is for the entire society, because of the reduction of negative externalities. That is why, the property rights are embedded in the community rules, which determine when and how the global public goods are used. This concept of Libecap, (2014) could be used in the analysis of the institutional model, related to the waste management.

## 3.2. Model for studying the relationship between social costs and transaction costs in the context of circular economy

### Area of model analysis, theory, and existing models

The direction of model analysis is determined by the purpose of the present study and its main hypothesis. The purpose of the study is to derive an alternative concept for the specifics of social costs as a structure, a relative share and efficiency in the adoption of the circular economy concept of EU. And its hypothesis is that it is possible to model social costs only through transaction costs. It is related to the assumption that the leading definition in the economic theory, according to which social costs equal private costs plus external costs, does not have a universal character (Pigou, 1954; Kapp, 1953; Berger, 2017). In this regard, it could not be applicable in relation to the circular economy concept of EU. In addition, one of the main tasks of the study is to consider the possibilities social costs to be derived through the transaction costs.

From the viewpoint of the existing methods, the main alternative general approach to the dominant neo-classical model for deriving social costs (presented in equation (1)) is that of the institutional environmental economy (presented in equation (2)). It examines social costs as difference between social opportunity costs and private costs. According to this model, social costs are a category that is much larger than the external costs (Kapp, 1970). These are all those harmful consequences and damages, which other individuals or the society suffer (direct or indirect) because of the production processes and for which the private entrepreneurs are not easily held reliable. They cover the broad category of impairment of “the human natural and social environment”, as the social environment also includes such categories as level of knowledge and others. That is, the reduction in the “amenities”, provided by “natural assets” constitutes social costs, along with a reduction in the resources, provided by the natural assets (Kapp, 1971; Swaney and Evers, 1989; Stabile, 1993).

As it was presented in the second chapter, Porter (2010) uses the dominating neo-classical approach (equation (1)) in the benefits analysis and in the analysis of cost efficiency in the waste management. In this model – social costs are those, which represent the real costs to society of discarding or disposing of waste as opposed to the usually lower price that the waste generator pays for doing so. In order social costs for waste management to be in balance with the private costs, according to him, the equality, presented in equation 4 [marginal social costs = marginal private

costs = price of activities, related to waste management] must be kept. However, according to the present study, this equation will only be efficient in the conditions of perfect competition, which in the field of waste management is questionable. In most of the cases in this field we are talking about local monopolies or local oligopolies.

Of interest from the viewpoint of the present model analysis are “the social opportunity costs” in the waste management, derived by Boardman, Geng and Lam, (2020). They are defined as the value of resources which the society must reject in order to adopt a certain program. Comparisons, involving impacts of government programs for which there are no market prices, for example for saved human lives or for various types of pollutions, they are used “shadow prices” (the price that would have an impact in a well-functioning market, if one existed). There are several options for measuring the opportunity costs in the waste management:

- Calculations based on different opportunities for waste usage;
- In the absence of alternative options for waste usage, calculations against the standard neoclassical model for opportunity costs versus the yield of a financial instrument if waste management resources are invested in it;
- Calculating opportunity costs not only against the principle of profit maximization (the best economic alternative), but also against the sustainable development principles (an option that best ensure their achievement);
- Opportunity costs for land, used for waste management activities.

Secondly, it is important to determine with what the current model for waste management in EU characterizes from institutional point of view. The concept for collective implementation of extended producer responsibility in EU circular economy regulations can be defined as an imitating (administrative) market.

According to Nozharov (2018b) in the field of waste management EU has introduced a special hierarchy so as to overcome the existing market inefficiency. It is based on the "collective funding schemes" for fulfillment of the obligations, arising from "extended producer responsibility", defined for the first time in Directive 2012/19/EC. In the legislative package for "circular economy" of EU (COM/2015/0595) these schemes for collective responsibility are defined as "organizations for fulfillment of the extended producer responsibility" on behalf of the companies, generating waste. According to the cited publication, the newly created special hierarchy is a combination between the capabilities of private and public hierarchies.

It is likely that the EU thereby wanted to overcome the defects in both types of hierarchies. From the viewpoint of the private hierarchy if a certain company aims to fulfill its "extended producer responsibility obligations on its own, it will not have sufficient financial resources or will face high transaction costs" (Dubois,2012). In cases, where the public hierarchy would implement obligations on behalf of private firms, there is a risk that this implementation will be compromised to an even greater extent (Mueller, 2003). But this creates a risk, EU instead of combining the positive sides of the private and public hierarchies, to combine their negative sides. Moreover, the circular economy has the potential to create even greater transaction costs than the linear economy does, because of the complex construction of its supply chains (Dossa et al., 2020).

The disadvantages of quasi-market sector are well known. In this regard, the publications of Le Grand, (1991), Beev, (2013) and others could be mentioned – concerning the system of education, social policy, and public procurements.

Thirdly, what does a model of imitating market should include in the field of waste management, and what is created by public intervention (as far as this model is analogous to the type trading of emission permits). We consider a model of imitating market because the present research classifies the concept for collective implementation of the extended producer responsibility in this way from the viewpoint of the EU policy for circular economy:

- Clear specification of property rights;
- Clear definition of the externalities: who causes them and what is the amount of their negative effect;
- To what extent this market will be publicly regulated and to what extent there is potential government participation in it. An important role here plays the concept for so called „government failure"(Cheung, 1978; Lichfield, 1996). "Coordination costs" must be considered as in most cases they are significant and could make the government intervention an inefficient market alternative (Medema, 2011; Allen, 2015; McClure and Watts, 2016). In most cases, governments are interested in asserting their power and influence, for example inflating their budgets. Lobbying by companies, causing the externalities, or companies, which are affected constituencies should also be considered. Not always the government has the incentive to possess more information (or knowledge) than the affected parties.

- It is important to be defined what types of goods are searching for at the relevant markets and who are the stakeholders (Regan,1972; Schlag, 2013);
- The hypothesis about the functions and properties of markets – for example, setting prices, resources allocation, effectiveness (Regan,1972; Schlag, 2013). The allocation of property rights does not in itself guarantee competitive exchange. Such exchange will exist when the parametric price of externalities is set, which could be traded and accepted by the participants at such a market (Berta & Bertrand, 2014);
- The possibility of excluding participants and subsequently reducing or eliminating information (Berta & Bertrand, 2014);
- The efficiency of such type of market implies that "in equilibrium the companies will balance their marginal costs for reducing carbon emissions with the price of the emissions permits and, in this way, they will become price-takers in competitive environment" (Berta & Bertrand, 2014);
- Doesn't the public intervention in one sector disturb the competitiveness of another sector. Does public intervention have an impact on the supply chain;
- To what extent the behavior of the economic agents participating in this imitating market is influenced by the

pursuit of the goal of increasing public welfare (to what extent do they interact with public funds, what are the opportunity costs of these actions);

- What exactly is the nature of the exchange in this imitating market (towards the satisfaction of whose need it is aimed at: society by reducing waste, private companies by a new source of raw materials, consumers by long-term use of goods or something else);
- In parallel to the development of such type of market, there must be considered that investments in new knowledge reduce uncertainty in the level and allocation of costs and benefits (Libecap, 2014).

**Model, calculations and results.**

According to Swaney and Evers, (1989) and Stabile, (1993) the changes in the institutional structure are appropriate for the social costs analysis because there is a built-in tendency the market system to create new institutions (and new techniques), which generate externalities. Given this, the institutional economics will be the most appropriate tool for completing such an analysis and for finding solutions.

**Descriptive method**

There will be done a descriptive analysis for the evaluation of the status quo of the waste management in Bulgaria in order to assess where it stands against comparable EU member states. The selected countries for comparison are:

- Germany as an EU member-state with highly developed market economy;
- EU-27 to be assessed the average level for EU;
- Croatia and Romania, because they are EU member-states from the Balkans (neighbor countries of Bulgaria) they were admitted to the EU in a relatively close period with Bulgaria and they had planned economies before 1990;
- Hungary, because it is an EU member-state, it is from Central Europe, and it was a planned economy before 1990.

The first comparison will be done through the analysis of the volume of generated waste. In this way, there will be assessed to what extent the components of the circular economy concept work. It requires consumers to be encouraged to use products that are not disposable, products that are provided with warranty and service support and such products that could be used for a long time. This is also related to the introduction of green culture for sustainable consumption. In order to make the comparison between countries with different territory and population effective, there will be used the measuring unit kg of generated waste per person of the population.

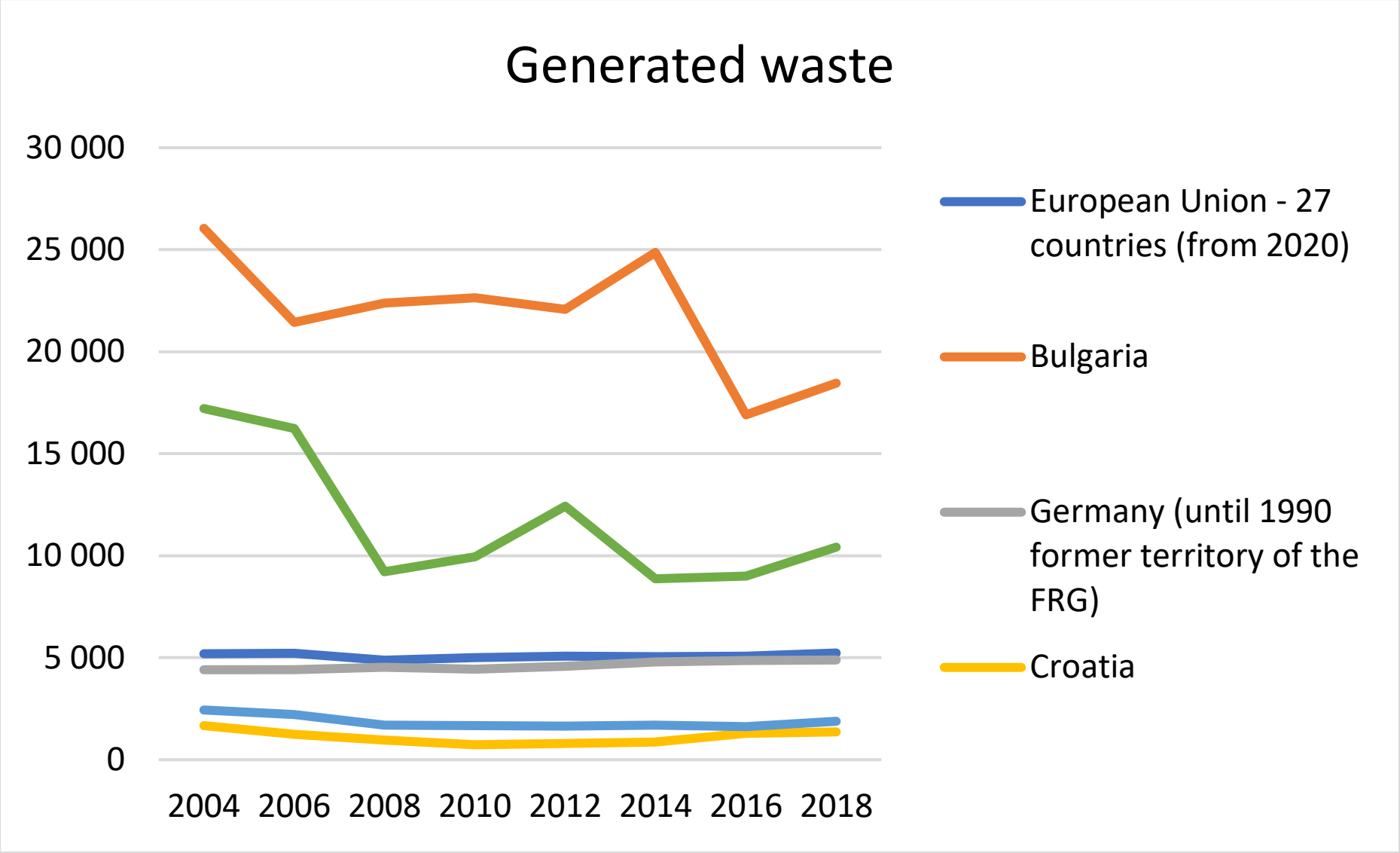

**Figure 1 Generated waste (kg per person of the population; Total-by all waste categories)**

**Source:** EUROSTAT (2022). Generation of waste by waste category, hazardousness and NACE Rev. 2 activity [env_wasgen]

From figure 1, one can see that in Bulgaria the generated waste is twice as much as its nearest comparable country, Romania, more than three times that of the EU-27 and Germany, and more than nine times that of Croatia and Hungary. This fact indicates that the elements of the circular economy, concerning the infrastructure do not work properly in Bulgaria. It also shows that the circular design of products does not function efficiently.

The second comparison will be done through the analysis of the volume of waste disposal. According to the circular economy concept disposal must be completely avoided. Waste must be turn into production resources or goods must be repaired and reused.

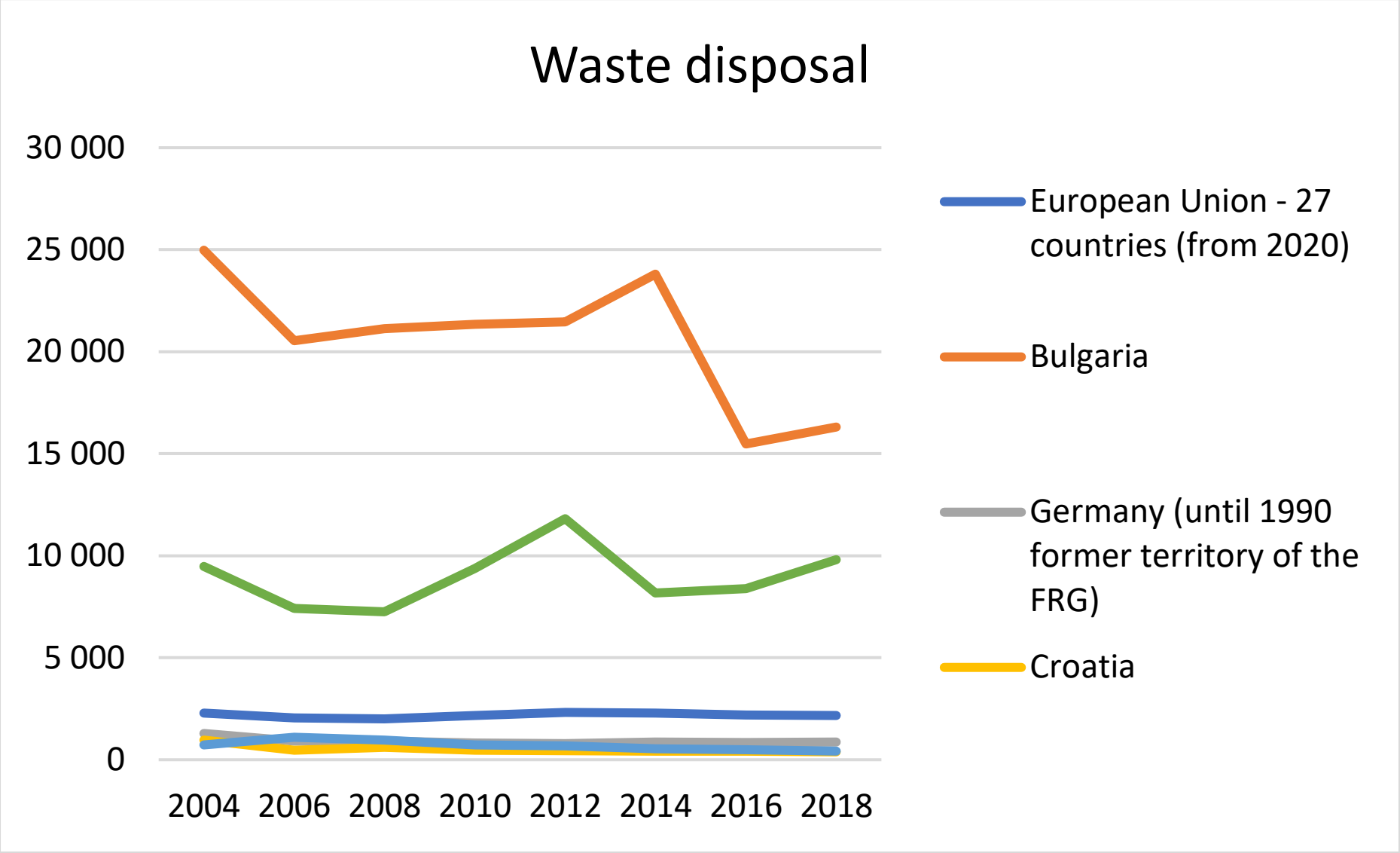

**Figure 2 Waste disposal (kg per person of population; Total-by all waste categories)**

**Source:** EUROSTAT (2022). Disposal - landfill and other (D1-D7, D12)

On figure 2, one can see that in Bulgaria eight times more waste is deposited than in EU-27, eighteen times more than in Germany, forty times more than in Croatia and Hungary and almost twice more than in Romania. The derivative effect of the large amount of generated waste is also considered here, but it is not the only reason for the observed inefficiency. The graph shows that the components of the circular economy in Bulgaria do not work as a whole. However, the emphasis falls on the lack of sustainable product design, as well as defects in the infrastructure of consumption.

The third comparison will be done through the analysis of level of energy recovery from waste. This is one of the elements of the circular economy, which enables beneficial economic usage of waste in a way that prevents its disposal. Especially in periods of increasing prices of energy resources in the EU region in which part of the member-states do not have access to own raw materials, this is an important indicator.

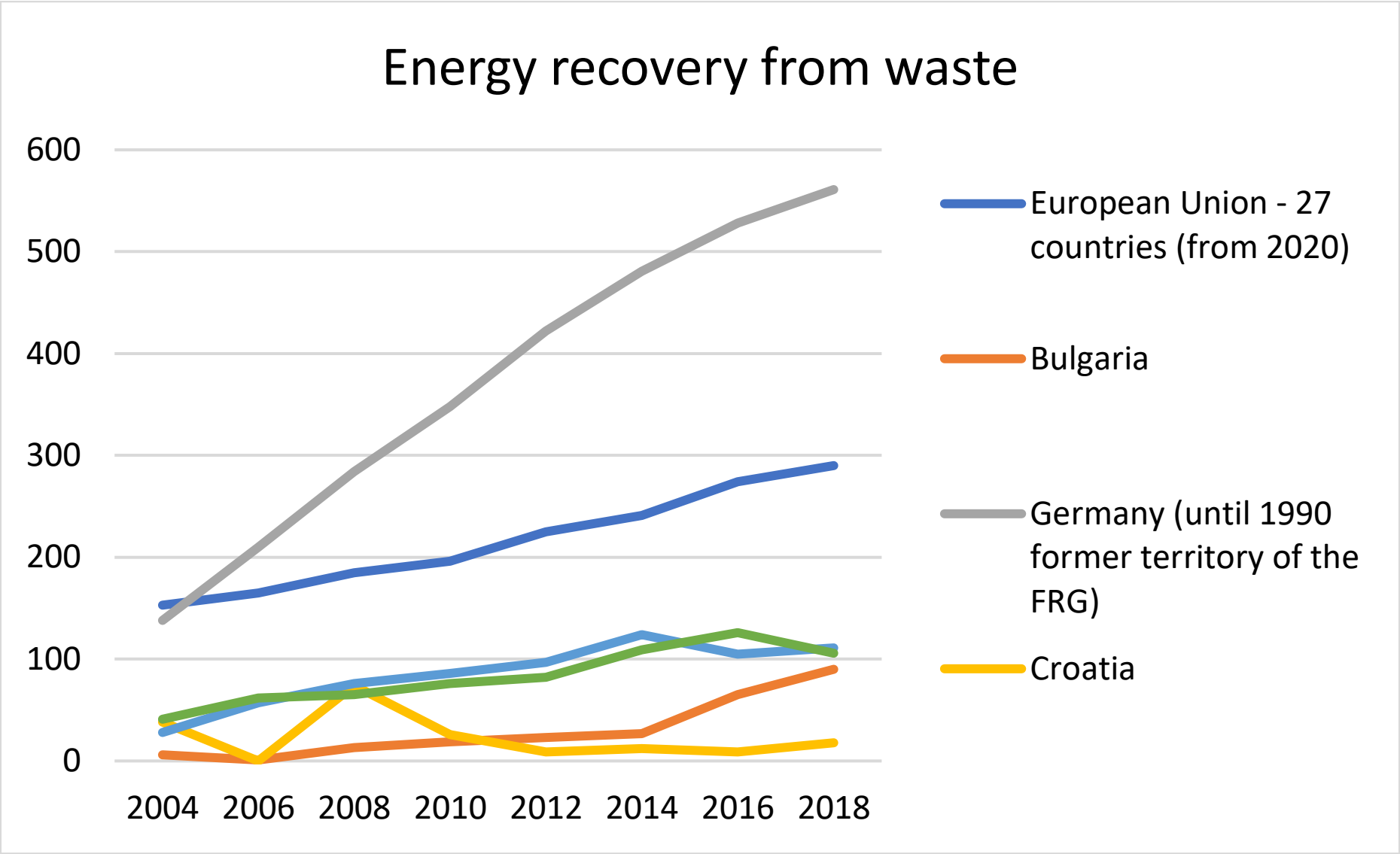

**Figure 3 Energy recovery from waste (kg per person of population; Total-for all waste categories)**

**Source:** EUROSTAT (2022). Recovery - energy recovery (R1)

On figure 3 one can see that Bulgaria recovers three times less energy from waste in comparison to EU-27, six times less than Germany, close to Hungary and Romania, and five times more than Croatia. It should be noted here that Croatia is a member-state of EU after Bulgaria does and probably the country has not still built the necessary infrastructure for this process. The general conclusion is that the former East-European countries lag behind both the average EU-27 level and the highly developed member-states, such as Germany.

The fourth comparison will be done through the analysis of the level of recycled materials and their overall recovery in a manner, suitable for their continuous use. This is part of the core components of the circular economy concept. It is related to the idea of raw materials independence, because a large part of the critical raw materials is located outside the EU borders. They are usually located in developing countries with high risk of political and military crisis, which makes the supply of these resources uncertain, both in price and physically.

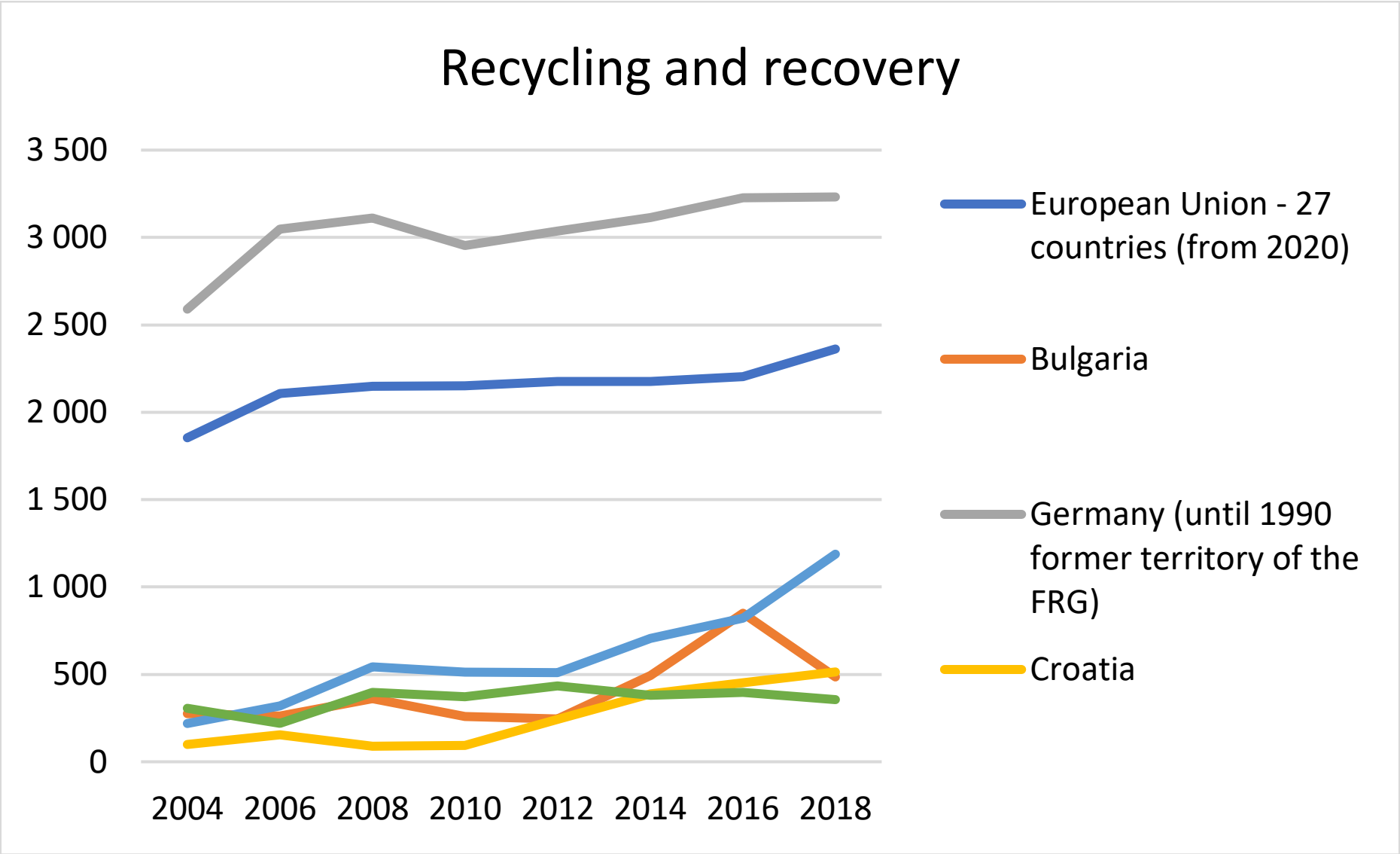

**Figure 4 Recycling and recovery of resources (kg per person of population; Total-for all waste categories)**

**Source:** EUROSTAT (2022). Recovery - recycling and backfilling (R2-R11)

From figure 4, one can see that Bulgaria recycles and recovers five times less amount of materials than EU-27, nearly seven times less than Germany, two and a half times less than Hungary and a comparable amount with Croatia and Romania. This indicates that the country is strongly lagging behind the EU average level and the highly developed economies in EU. As can be seen from the example, in 2004 Hungary started from the same point as Bulgaria, Romania and Croatia, but in 2018 it realized two times higher efficiency. Given the comparable territory and population, the example with Hungary obviously deserves to be analyzed.

The fifth comparison will be done through the analysis of the resource efficiency from the perspective of the productivity of the economy. The purpose is to be assessed to what extent the used industrial technologies are resource efficient. In order to make the comparison between countries with different territories and populations correct, the unit of measurement is euro per kilogram of input resource.

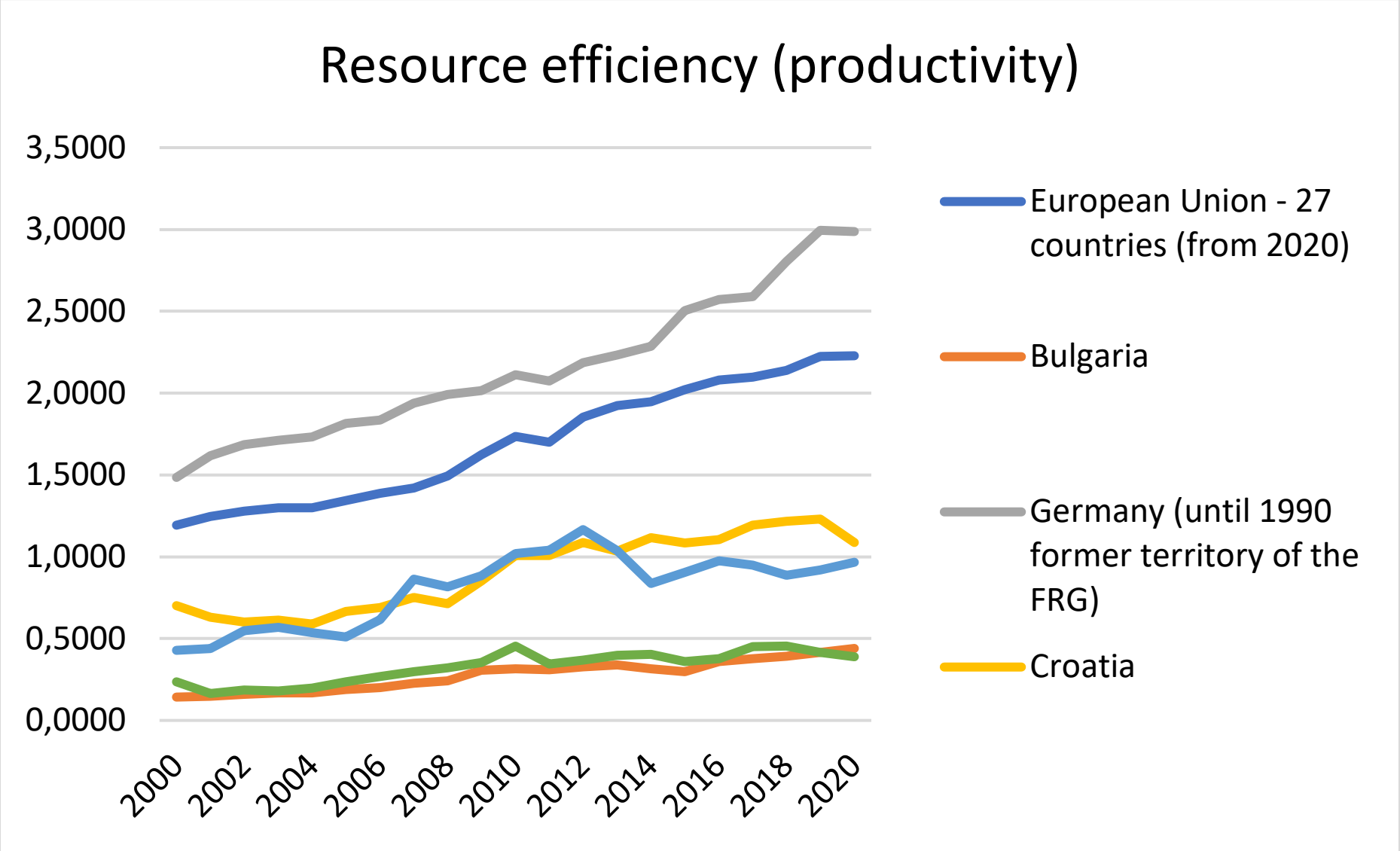

**Figure 5 Resource efficiency (productivity) (euro per kg of input resource)**

**Source:** EUROSTAT (2022). Resource productivity [env_ac_rp]

Figure 5 shows that the resource efficiency in Bulgaria by 2020 is over five times worse than in the EU-27, seven times worse than in Germany, two and a half times worse than in Croatia, over twice worse than in Hungary and comparable to that of Romania. This means that the industrial technologies in Bulgaria are outdated and inefficient, which is a signal for a necessary review of the national innovation policy.

A general conclusion of the analysis of the presented statistical data and graphs is that Bulgaria is significantly lagging behind both the EU average level (EU-27) and the highly developed EU member-states, such as Germany regarding the selected indicators for circular economy. By most of the indicators, Bulgaria also lags behind the Eastern European member states.  Even if there are some comparable indicators with Romania, the high level of waste generation and disposal in Bulgaria makes these isolated similarities unrepresentative. On the other hand, Bulgaria and Romania started from the same base, which requires a more detailed analysis of the reasons for this lag.

The results of the descriptive analysis are confirmed by the conclusions of other authors **Mazur-Wierzbicka, (2021)** who found that for the period 2010-2018, the progress of the EU member-states towards circular economy confirms the theory for the existence of “Europe at two speeds”. On the one hand, they are the highly economically developed "old" member states, such as Germany, Belgium, the Netherlands, and France. On the other hand they are the member-states from Eastern, Central and Southern Europe that are lagging behind, amongst which is also Bulgaria. According to the authors, the basis of this difference is in the socio-economic development, but above all the is the effectiveness and presence of the adopted strategies at the national level for the transition to a circular economy. In the countries from Eastern and Central Europe, they are formal and do not give a significant result. Other reasons, which the author has identified are the bad results of the undeveloped waste processing infrastructure and the lower public awareness of the circular economy concept.

According to **Giannakitsidou, Giannikos and Chondrou, (2020),** Germany has yet achieved the ambitious goal for 2035 for waste recycling, having exceeded 65%. And EU member-states, like Belgium, the Netherlands, Austria, and Slovenia have exceeded 50% of recycling, which has been the goal since 2020, while the member-states from Eastern and South Europe are still landfilling over 50% of their waste. The results of the **Giannakitsidou et al. (2020)** publication find that the worst performing EU member-states in the field of waste recycling, are Bulgaria, Romania, Croatia, Greece, Cyprus, and Slovakia. The difference between the more advanced member-states and those which are lagging behind, indicates that the last ones have a potential to make great progress.

**Marino and Pariso, (2020)** indicates that Bulgaria as an EU member-state performs poorly in achieving a circular economy. What is interesting in their analysis is that they point as a main problem in this field, the inefficient waste management processes of companies. This points to the principle of extended producer responsibility, although this is not developed in their analysis. They recommend the implementation of taxation incentives, which would force companies' investments and innovations in the field of circular economy. Croatia, Hungary, and Romania are also mentioned as poorly performing EU member-states in the introduction of circular economy with partial and uncoordinated measures between the private and public sectors at national and local level. This points to institutional inefficiency. Moreover, these countries receive high level of financing from EU funds in relation to the cohesion policy and a significant part of it is directed in the field of environment and infrastructure. Obviously, this funding is not used efficiently enough.

Germany, Finland, and the Netherlands are the countries which are performing best in this field. The authors conclude that the EU member-states with higher GDP per capita, are recycling more waste. In these member-states, they are used and traded more goods, produced by recycled materials. On the other hand, amongst the member-states, which generate higher amounts of waste per capita, there are such countries which have different socio-economic development.

Indications for institutional defects are identified in the analysis of Nozharov (2021), concerning the waste management on municipality level in Bulgaria. It turns out that in opposite to the world tendencies, in Bulgaria, with every increase in the scale of municipal waste management activities, the costs, instead of decreasing, they are increasing. For example, the Metropolitan Municipality collects with 14% less waste per capita in comparison to other large municipalities in 2016, while at the same time its expenditures in this field are with 40% higher. It has the lowest level of waste submitted for treatment per capita, while its separation and treatment costs are about 30% higher. That is why, the municipal waste management (mixed household waste) in the system of waste management in Bulgaria shows high ineffectiveness. This partially explains the high quantity of waste, which goes to landfill, however not only municipalities are responsible for that fault. This data gives signals that the system for collective implementation of the extended producer responsibility is also not efficient, and it does not reduce the amount of mixed household waste.

In order to be identified the reasons for lagging behind of Bulgaria in the field of circular economy, it is necessary to be made a statistical analysis by constructing an appropriate model. Having in mind the focus of the present research, these reasons must be sought in the field of institutional ineffectiveness.

**Quantitative method**

The analysis, based on quantitative methods has the purpose to analyze the level of social costs, that result from institutional ineffectiveness of collective waste recovery systems. In this way, they will be revealed the institutional failures, that underlie the EU circular economy concept. The presence of social costs is indicated by the qualitative analysis done in the publication of Nozharov (2018b).

**Model hypothesis:**

**Social public costs (SPC)** in the waste management through collective systems, based on the principle for extended producer responsibility, are assumed to be 0. Actually, **SPC** should be social benefits, but according to the qualitative analysis in Bulgaria (Nozharov, 2018b), it equals 0. We have even expenditures of companies, which are members of collective waste recovery organizations, which have certain opportunity costs from the viewpoint of their investment in other efficient public activities.

The term „social public costs"aims to distinguish them from the dominating neo-classical concept for social costs, which examines them as a sum of private costs and externalities (**Equation (1)).** It was discussed in detail in the second chapter of the study (Berger 2017, Pigou 1954).

**Transaction costs (TrC)** in the waste management through collective systems, based on the principle for extended producer responsibility are measurable costs, and they are always a positive value.

Then, if **(SPC)** is 0, which must be the benefit for the society by the system and at the same time the society and companies bear costs due to institutional ineffectiveness **(TrC)**, then **(SPC)** = **(TrC)**. That is, **social public costs** will equal **transaction costs**.

**Balancing the model (description):**

The highest level of transaction costs equals the lowest level of social public costs. Collective waste management systems cannot have classical externalities because their main purpose is to counteract to them, to eliminate and to reduce them. Then social costs of these systems cannot be measured by the negative externalities of some activity on the environment and human health. The only social costs, which these systems create, are related to the institutional ineffectiveness (direct and opportunity costs for the society). In the model, when the left side (social public costs) of the equation is zero because of the absence of effectiveness of the system (administrative imitating quasi-market), then SPC equal transaction costs (the costs for the society due to ineffectiveness and the costs for creating the system – an administrative imitating quasi-market).

**Structure of the model:**

In the creation of the model's structure, they are used the publications of **Nozharov (2018a и 2018b)**. The empirically established dependencies in these publications are the basis for construction of author's own model for analysis of the social costs in the context of circular economy, based on the understanding for institutional ineffectiveness. The model is rooted in the traditions of the institutional economics.

| | | |
|---|---|---|
| **Social Public Costs (SPC) =** | **Transaction Costs (TrC)** | **(7)** |

At which:

| | | |
|---|---|---|
| **Social Public Costs (SPC) =** | ***the quantity of regenerated waste oil with collective systems – the quantity of regenerated waste oil without collective systems*** | **(8)** |

as:

| | | |
|---|---|---|
| **SPC =** | ***the quantity of regenerated waste oil with collective systems – the capacity of technology in the private processing companies in the relevant country in accordance with Directive 2010/75/EU*** | **(8.1)** |

or

| | | |
|---|---|---|
| **SPC =** | ***the quantity of regenerated waste oil with collective systems – consumer quantity demanded of produced goods from regenerated waste oil according to the available statistical data*** | **(8.2)** |

Also:

| | | |
|---|---|---|
| **Transaction Costs (TrC)** = | *Fixed Tr Costs (FtrC) + Variable Tr Costs (VtrC)* | **(9)** |
| **FtrC** = | *Administrative fixed costs + Market fixed costs* | **(9.1)** |
| **Administrative fixed costs =** | *costs associated with the bank guarantee required for the permit + costs associated with the annual performance audits + costs for keeping the required control documentation and assistance during inspections from the public control bodies* | **(9.2)** |
| **Market fixed costs =** | *costs associated with the control over contractor's performance + costs associated with the communication among the members of the collective system* | **(9.3)** |
| ***VtrC* =** | *Performance costs + Alternative costs* | **(9.4)** |
| **Performance costs =** | *the agreed price for the services rendered by contractors – the price if the collective system has its own recovery facilities* | **(9.5)** |
| **Alternative costs =** | *cost of regenerated base oils sold – income from producers participating in the collective system* | **(9.6)** |

Then:

| | | | |
|---|---|---|---|
| *The quantity of regenerated waste oil with collective systems – the quantity of regenerated waste oil without collective systems* | **=** | *Fixed Tr Costs (FtrC) + Variable Tr Costs (VtrC)* | **(10)** |

**Derivation of a logarithmic function:**

From **equation (7)** it follows that:

$$\log SPC = \log TrC \qquad (11)$$

From where

$$\log \frac{SPC}{TrC} = 0 \qquad (12)$$

As $\log 1 = 0$ ,

then **(13)**

$$\log \frac{SPC}{TrC} = \log 1$$

We use that

$$\log \frac{\textit{The quantity of regenerated waste oil with collective systems} - \textit{the quantity of regenerated waste oil without collective systems}}{\textit{Fixed Tr Costs } (FtrC) + \textit{Variable Tr Costs } (VtrC)} = \log 1 , \quad \textbf{(14)}$$

From where:

$$\frac{\textit{The quantity of regenerated waste oil with collective systems} - \textit{the quantity of regenerated waste oil without collective systems}}{\textit{Fixed Tr Costs } (FtrC) + \textit{Variable Tr Costs } (VtrC)} = 1 , \quad \textbf{(15)}$$

then **Equation (10)** is correct**.**

**Empirical testing of the model :**

Till 2021 there is no sufficient public data, concerning the empirical testing of the model. At this stage, there could be done only a limited quantitative analysis. For such a short period of time about which the Ministry of Environment and Waters uploads public data (five-years period) there is impossible to be proved a correlation or similarity with any distribution (p-value in the correlation test is approximately 0.25). However, certain trends could be deducted, which would be confirmed in the future when much more public data is available.

The studied period is from 2016 to 2021, as the primary data is taken from the public registries of Ministry of Environment and Waters and they are explained in appendices 1-6, attached to the present research.

The average value (mean) of C, (social public costs SPC) is 68005,7.

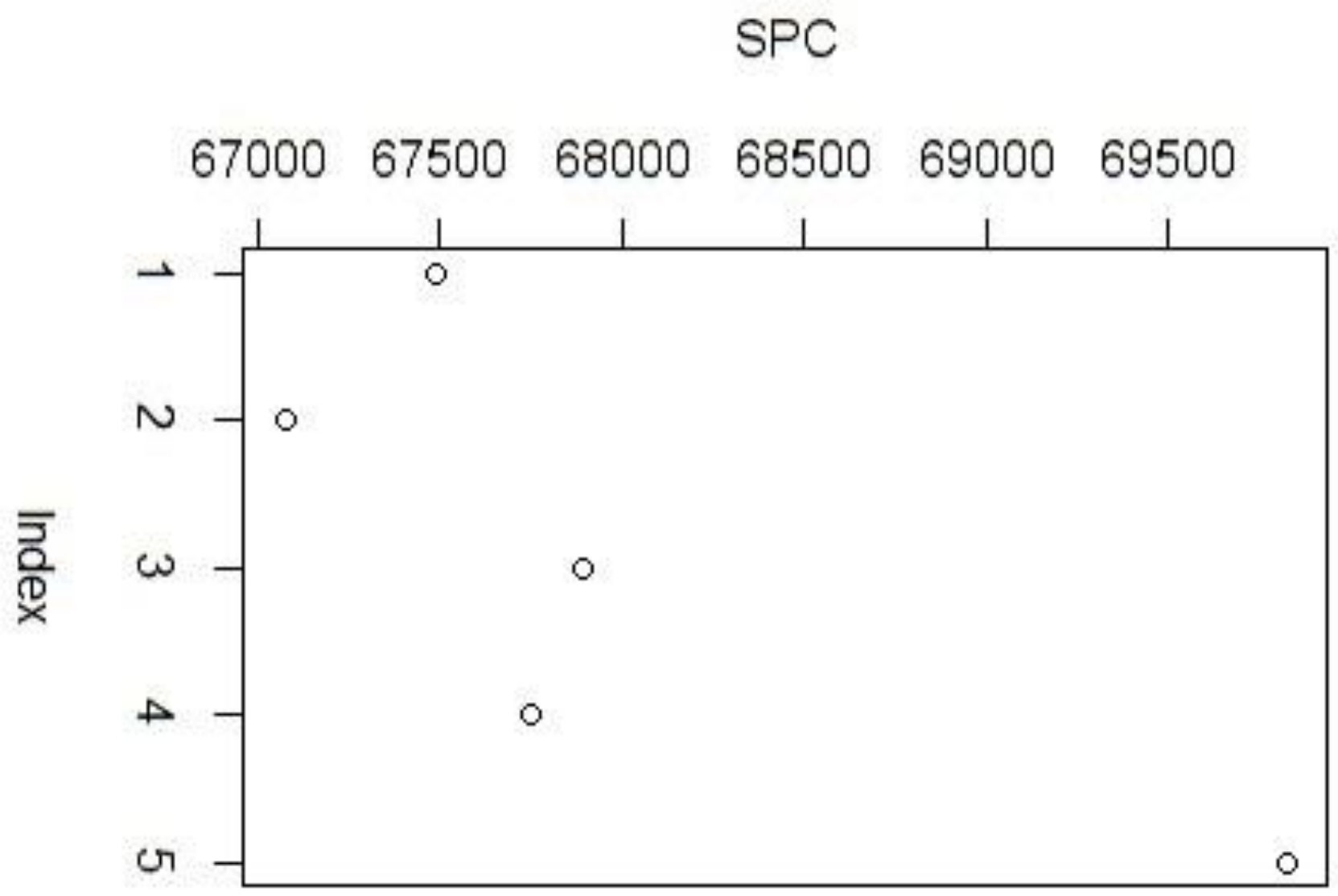

**Figure 6 Analysis of SPC 1**

Source: author's own calculations

The variance is 1134361 and the standard deviation is 1065,064. This is a relatively small standard deviation in comparison to the average value of SPC and it shows that SPC are almost constant, and they do not change year by year (1000 standard deviation on 68000 average value means that on average we expect the difference between the mean and the value in a given year to be 1000. That is, most of the years fall in: 67000-69000).

From SPC plot one can see that the average tendency of SPC is of increase. The correlation (Pearson) between the year and SPC is strong: 0.7963749. However, the correlation (Kendall) is only 0.6, indicating that there are often years in which the SPC is lower than the previous one.

Pearson correlation = 0.7963749

Kendall Correlation = 0.6

Sperman Correlation = 0.8.

The average value (mean) of transition costs: TrC is 13936791.

Standard deviation is 406255.3.

The correlation of the transaction costs for the five years is weak and negative -0.206247. This could indicate improvement in the efficiency of the administrative costs or reduction in the market costs.

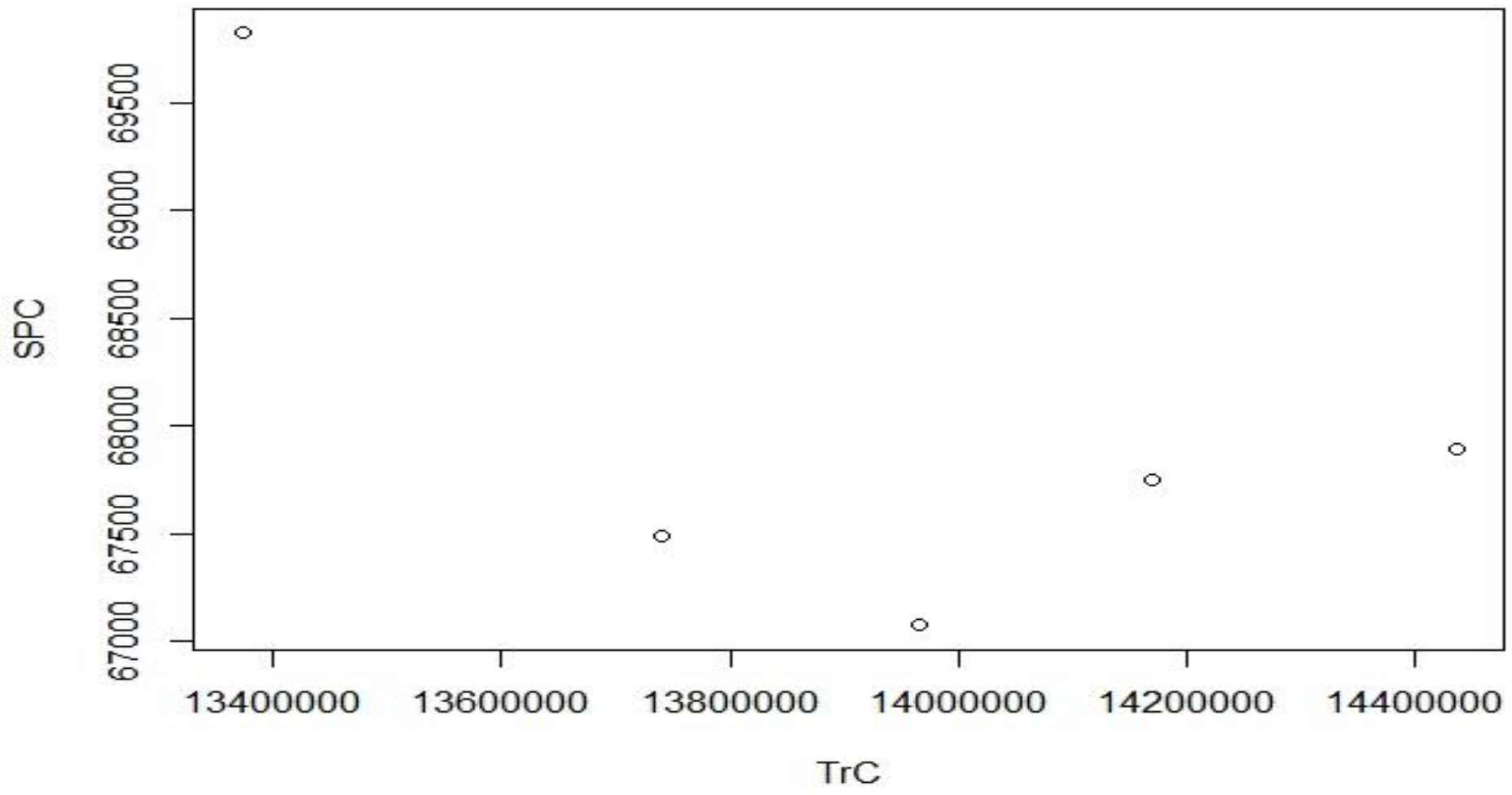

**Figure 7 Analysis of SPC 2**

Source: author's own calculations

The sample correlation between TrC and SPC is -0.6163201, with 95% confidence interval (-0.9707389  0.5829070). Because of the scarce data available, it is impossible to be rejected $H_0$, which says that there is no correlation between TrC and SPC. If we ignore the first year of the period, the trend is positive and probably there were some extraordinary expenditures incurred then.

**Empirical analysis-results interpretation:**

The results of **Equation 8.1**, the technical capacity of the processing plants in Bulgaria for the period 2016-2021 exceeds more than four times the amount of the „regenerated waste oil“, which fact is reported as an output of the activity of collective waste recovery systems.

According to **equation 8.2**, the demanded quantity of produced goods by "regenerated waste oil" for the period 2016-2021 exceeds more than ten times the amount of "regenerated waste oil" which fact is reported as an output of the activity of collective waste recovery systems. Of course, here we must also consider the amount of import of such final products, and on the other hand the needs of exports and the demand for such products in international markets with the prices there. The quantity of export offsets the quantity of import and it is even higher than it.

This means that even without the existence of collective waste recovery systems, the companies, processing used oil products and waste petroleum products would buy them from the companies applying the extended producer responsibility.

In this case, the social costs for the society will equal the expenditures of all these 5 collective waste recovery systems for processing of used oil products and waste petroleum products in Bulgaria. It can be measured through the opportunity costs of the sum, which commercial companies have paid for the services, provided by the collective waste recovery systems. The direct opportunity costs will be measured as loss for the society from the missed opportunity these companies have had to invest in new environmentally friendly technologies.

For the analysis of equation (9) there is not sufficient data to be conducted complete empirical testing with the respective conclusions. But nevertheless, it is obvious that the collective waste recovery systems for used oils and waste petroleum products are decapitalized, they do not enough staff and material assets and do not have sufficient liquidity. Most of these organizations operate at permanent economic loss, while the other companies in the field of hazardous waste processing operations, realize economic profit. These conclusions are also verified in some of the earlier publications of the author (Nozharov 2018b), and they are confirmed by the Registry Agency of Bulgaria.

In the first section of the third chapter, it was presented the theory for institutional inefficiency. The purpose of this presentation was the theory to serve as possible basis for establishing analytical model for the correlation between social costs and transaction costs. In the second section of third chapter, it was proposed a model to study the correlation between social and transaction costs in the context of circular economy. They were examined similar existing models in theoretical aspect in order to be distinguished the proposition for a new model from the existing ones and to be outlined its pros and cons. They were outlined the basic requirements that the components of a model for imitating market in the field of waste management with public intervention must meet. A partial statistical test was performed, based on the available statistics.

# CONCLUSION

The purpose of the present study was to verify whether the leading definition of the economic theory, according to which social costs equal private costs plus externalities, is a universal in nature (Pigou, 1954; Kapp, 1953; Berger, 2017) and can be applied to the EU circular economy concept.

The results of the study indicate that this definition can be contested, regarding the opportunity for its application in the context of circular economy and in this regard – its universal nature.

First, private costs in equation (1) do not exist in their neo-classical understanding in the context of circular economy. The waste recovery organizations do not operate with the goal of profit maximization, as they are legally prevented from generating an accounting profit and do not report such. Their main goal is to reduce the financial burden for companies, generating waste (based on the principle for extended producer responsibility, which accumulate the polluter pay principle) in accordance with the EU legal norms (COM/2015/0595). The function of waste recovery organizations is to cover the expenditures for the activity they perform on behalf of others by accumulating a shared financial resource, which leads to cost-covering funding. Next, these costs serve quasi-market subjects – created administratively by the power of coercion (normatively created and binding a certain category of companies) that do not operate in a free competitive market. These costs are reported to the state authority (Ministry of environment and waters, which report them in order to accept the fulfillment of waste recovery goals), but not exclusively to the shareholders of the companies.

In relation to the abovementioned, we cannot speak about “private costs” in the traditions of the neo-classical understanding. Given this, one cannot speak about “private costs” as a component of the equation for social costs (equation (1)) in the context of circular economy.

Second, the negative “externalities” in equation (1) represent uncompensated externalities, in which the actions of one economic agent (companies) affect the well-being of a third party (observer), who has not contested to them (Lichfield, 1996; Cheung, 1978). In their neo-classical dimension, they do not exist in the context of circular economy.

In this concept, collective waste recovery systems have a main goal to reduce the harmful impact of waste on the human life, as well as to turn waste into useful resources for the economy. In this way, the waste recovery organizations, which play the role of private hierarchies in the present analysis, and they are companies in the context of the neo-classical analysis in the circular economy hypothesis, do not generate externalities. The entire concept of circular economy is devoted to the elimination of externalities and its participants – the economic agents interact one to another precisely with this purpose. Waste recovery organizations and their members have normatively defined goals, they must follow and must report each year to the Ministry of environment and waters how much waste they have recycled, regenerated, and neutralized. They build, use or rent technical production capacity in this regard.

In relation to the above, we cannot speak about “externalities” in the context of their neo-classical understanding. Given this, we cannot speak of “externalities” as a component of the equation for social costs (equation (1)) in the context of circular economy.

Then all two components of the neo-classical equation (1) for social costs calculation do not exist in the context of circular economy. In this hypothesis, social costs cannot be measured as a sum of private costs and negative externalities.

In the hypothesis of circular economy, one cannot speak about the presence of perfect competition, which is the case in which equation (1) functions. Waste recovery organizations form administratively imitating and centralized quasi-market, which is created by law, and it functions in accordance with the normative rules for evaluation of the costs and benefits in case of prohibiting profit maximization. In this hypothesis one cannot speak about the presence or even about striving for perfect competition.

That is why, the present study challenges the universal nature of the neo-classical definition and measurement of social costs.

Third, in the context of circular economy, the analysis of social costs can be done through using the theory for institutional ineffectiveness. There exist four possible approaches in searching the reasons for institutional ineffectiveness: transaction costs approach, principal-agent approach, the game equilibrium approach, and the evolutionary approach. The present research has adopted this classification (it was examined in chapter three) and it is used as a basis for the social costs approach. The social costs approach in the context of circular economy is developed in the model, presented in the third chapter of the research, where these costs are a function of the transaction costs. They depend on the level of institutional ineffectiveness of waste recovery organizations and on the administrative imitating and centralized quasi-market, which is created by law and function on the normative rules, adopted by the EU authorities. Such a similar market is that of the greenhouse gas emissions trading permits, which is also created by the EU authorities, but at this market there is some level of competition amongst the participants and the maximization profit principle is not prohibited. Given this, a direct comparison between these two quasi-markets cannot be done, even though there are some similarities, especially in the criticism of poor functioning of both markets.

Fourth, the creation of an imitating market at which to be traded the externalities, leads to their elimination. The pollution will not become zero, but it will only be possibly Pareto optimal (Berta & Bertrand, 2014; Hurwicz, 1995; Dahlman, 1979). The analysis then focuses on the transaction costs. The present research accepts transaction costs in their systematic context, which constitutes a certain theoretical nuance of their basic definition. They are examined as:

- Costs for system management (coordination and motivation costs), which cover the system's institutional structure and effectiveness (Platteau, 2008);
- Costs, imposed by the government institutional constraints, representing the efforts made by the entrepreneurs and market participants as a whole in order to be in line with the official institutional framework – the complex regulatory framework and the regulatory instability (Marinescu, 2012);
- Costs, associated with the public goods management, representing the costs for establishing, maintaining and exchange of property rights over revenues and benefits, obtained as a result of the negative externalities mitigation between the participants and affecting parties (Libecap, 2014);

In the presented new model for measurement of the social costs in the context of circular economy in the third chapter, the highest level of transaction costs equals the lowest level of social public costs. When the left side of the equation – "social public costs" becomes zero because of the absence of system effectiveness (the presence of administrative imitating quasi-market), then they equal the transaction costs. They include opportunity costs for the society because of the ineffectiveness, accepted as:

- Maintenance costs of waste recovery organizations, which are ceded public resource (as far as this activity could legally be done by the government, based on the products sent to companies, which generate waste by the industry, managing the activities for protecting the environment at the ministry of environment and waters);
- Costs for inefficient functioning of the waste recovery organizations, which in the analysis of the present research, are decapitalized, they do not create own technical capacity, they do not have efficient human resources and they operate on permanent economic loss.;
- Coordination costs – for the wages of the public administration officers, occupied with the control of the waste recovery organizations activity (carried out by law on monthly and yearly basis);

- Costs for creating the system – administrative imitating quasi-market (political and other expenditures for permanent changes in the regulations for functioning of this market).

Fifth, the present study, based on an expanded literature review, has identified the weaknesses of the existing models for social costs measurement, including those in the field of waste management. For example, the double reporting of environmental costs in the measurement of the social costs, mixing the understanding for internalization of externalities with the measurement of social costs, mixing the understanding for the municipal system for household waste management without the use of collective waste recovery organizations with the circular economy concept and others. In this way, the existing models are inapplicable for measurement of the social costs in the context of circular economy.

Sixth, the present study does not repeat other publications, analyzing the circular economy through the institutional approach. For example, Grafström and Aasma, (2021) argue that the institutions are important for the development of the circular economy, but they only flag this issue. Their conclusions, concerning the role of institutions are general and point to:

- Well defined and enforceable private property rights (if household waste is only owned by the municipalities, then no entrepreneurs will be interested in them),
- Rule of law (the possibility to enter the market and freedom of doing business),

- And a “moral code of behavior, which legitimizes and recognizes these traditions”.

Grafström and Aasma, (2021) indicate that the main problem for the absence of efficient institutions is the problem with information.

Seventh, it is analyzed what need to be the main components of a model in the field of imitating market in the area of waste management, if it is created by public intervention, covering the concept of collective fulfillment of the extended produce responsibility from the viewpoint of the EU policy for circular economy. The analysis was done in the third chapter of the present study and consists of seven components.

Eighth, a critical view of the definition for transaction costs of the U.S. Environment Protection Agency, which is used as the basis of most modern scientific analyzes of waste management (EPA-USA, 2000a). According to the present research, transaction costs in the field of the environmental policy exist not only in the implementation of “incentive-based policies”, as EPA-USA believes (2000a). They are also present in deformations of the institutional environment, in which the environmental policy is developed, which is presented in the model in the third chapter.

Nineth, recommendations for future studies in this field, which could upgrade the present one:

- Extensive empirical testing of the proposed model in the third chapter as a result of gathering and accumulating more statistical data at the end of a ten-years period from the normative introduction of the concept of circular economy of the EU in 2015;

- Upgrading and alternative modelling of the recommendations for improvement of the effectiveness of the imitating market in the field of waste management, established with public intervention, covering the concept for collective fulfillment of the extended producer responsibility from the viewpoint of the EU policy for circular economy.

Bulgaria's high backwardness in the introduction of the circular economy, according to the results received by the descriptive analysis, must be overcome. Otherwise, Bulgaria will suffer not only from economic damages, but also from image damages internationally as an EU member-state, which presents resource-efficiency, close to that of developing countries. Authors, like Mazur-**Wierzbicka, (2021)** start talking about "Europe on two speeds", supporting this thesis precisely with the progress of the EU member states in building a circular economy, placing Bulgaria in the group of laggards. This thesis is supported also by **Giannakitsidou et al. (2020) , Marino and Pariso, (2020)** and others. One of the possible ways to be overcome this problem is by changes in the institutional effectiveness, regarding waste management.

The principle for extended producer responsibility is criticized by many authors after it was introduced in EU. For example, Sachs, (2006) comments that this principle should affect higher value products because otherwise the transaction costs may exceed the social benefits. According to Massarutto, (2014), the insufficiently detailed regulation of this principle at the Community level, which leaves freedom to individual Member States for its implementation, is a serious drawback. In this regard, individual EU member-states will compete how to downgrade the requirements and regulations in order to attract more investments. Moreover, companies are not encouraged to develop a circular design of their products. They do not pay fees according to the uniquely developed persistent characteristics of their products, but only according to their class. In addition, the principle for extended producer responsibility discourages consumers to actively participate in the circular resource management, although they are the actual generators of waste.

It is surprising that despite all criticism to the extended producer responsibility principle, since 1990 the European Union adopted it from the design of the recycling economy and replicated it to the circular economy concept. Even if there are some changes in the regulations, related to this principle in the policies for circular economy in EU (encouraging eco-design of products, transparent structures of the collective systems, full expenditures coverage), its effectiveness remains low (Friant, Vermeulen and Salomone, 2021).

The present study supplements the critics of this principle and despite its theoretical propositions, it is drawn from the real economy. Its purpose is to compare the dominating economic theory with the reality in order to be avoided the so called "blackboard economy" effect, referred to by the Nobel laureate **Coase (1988, p.19):**

> *„The policy under consideration is one which is implemented on the blackboard. All the information needed is assumed to be available and the teacher plays all the parts. He fixes prices, imposes taxes, and distributes subsidies (on the blackboard) to promote the general welfare. But there is no counterpart to the teacher within the real economic system."*.

# APPENDICES

## APPENDIX 1

### Statistical data in accordance with Order № RD-255/16.05.2016 of the Minister of environment and waters

<table>
<tr><td rowspan="2">Waste recovery organizations 2015</td><td rowspan="2">Released on the market (tons)</td><td>Regene-ration</td><td>Regeneration</td><td>Note:</td></tr>
<tr><td>Total tons</td><td></td><td>Utilized quantities of tons by installations with a complex license</td></tr>
<tr><td>„Oil recycling“ LtD</td><td>4297,418</td><td>1720,950</td><td>„Polychim CC“ LtD 1720,950 Regeneration, collected and recovered</td><td rowspan="4">„Polychim CC“ LtD Total regenerated 3131.95 Annual capacity with complex license № 440-H0/2012 Ministry of environment and waters (MEW) 25 000 tons per year<br>-------------<br>„Lubrika“ LtD Total regenerated 7427.88 Annual capacity</td></tr>
<tr><td>„Lubrika Environmental activities“ LtD</td><td>10994,849</td><td>4397,940</td><td>„Lubrika“ LtD 4397,940 Regeneration, collected and recovered</td></tr>
<tr><td>„Ecorivais oil“ JsC</td><td>3021,433</td><td>1211</td><td>„Polychim CC“ LtD 1211 tons Regenerated, collected and recovered</td></tr>
<tr><td>„National company for collection and recovery of used oils“ LtD</td><td>5996,816</td><td>2405,496</td><td>„Lubrika“ LtD 1483,450 regenerated „Verila Lubricants“ JsC 922,046 regenerated</td></tr>
</table>

| | | | | |
|---|---|---|---|---|
| **„Nord oils“ JsC** | 6913,290 | 2771,930 | „Lubrika“ LtD 1546,490 regenerated „Polichim CC“ LtD 200 Regenerated, collected and recovered<br><br>„Eco sand Sofia“ LtD 1025,440 – only recovered, without regenerated | with complex license № 352-НО-И0-А2/ 2013 MEW **30 000 tons per year** -------------- **„Verila Lubricants“ JsC** Total regenerated **922,046** *No data available about complex license in the MEW's registries* -------------- „Eco sand Sofia“ LtD 1025,440 – only recovered, without regenerated -------------- **Total annual capacity in with complex license** for regeneration of two of the stations: **55 000 tons per year** |
| | | | | |
| **TOTAL:** | **31223.806** | **12507.316** | **11281.876 - R** | **11481.876 - R** |
| | | | **12507.316 - R+R** | **12507.316 – R+R** |

## APPENDIX 2

### Statistical data in accordance with Order № RD-322/15.05.2017 of the Minister of environment and waters

<table>
<tr><th rowspan="2">Waste recovery organizations 2015</th><th rowspan="2">Released on the market (tons)</th><th>Regene-ration</th><th>Regeneration</th><th>Note:</th></tr>
<tr><th>Total tons</th><th></th><th>Utilized quantities of tons by installations with a complex license</th></tr>
<tr><td>„Oil recycling“ LtD</td><td>4607,415</td><td>1843,014</td><td>„Polychim CC“ LtD<br>1843,014<br>Regenerated, collected, and recovered</td><td rowspan="3">„Polichim CC“ LtD<br>Total regenerated<br>2906.014<br>Annual capacity with complex license № 440-H0/2012 Ministry of environment and waters (MEW)<br>25 000 tons per year<br>-------------</td></tr>
<tr><td>„Lubrika Environmental activities“ LtD</td><td>12247,413</td><td>4899</td><td>„Lubrika“ LtD<br>4899<br>Regenerated, collected, and recovered</td></tr>
<tr><td>„Ecorivais oil“ JsC</td><td>2655,665</td><td>1063</td><td>„Polychim CC“ LtD<br>1063 tons<br>Regenerated, collected, and recovered</td></tr>
<tr><td>„National company for collection and recovery of used oils“ LtD</td><td>4455,919</td><td>1790,262</td><td>„Lubrika“ LtD<br>1132,340<br>regenerated<br>„Verila Lubricants“ JsC<br>657,922<br>regenerated</td><td rowspan="2">„Lubrika“ LtD<br>Total regenerated<br>7109.525<br>Annual capacity with complex license<br>№ 352-HO-И0-A2/ 2013 MEW</td></tr>
<tr><td>„Nord oils“ JsC</td><td>7750,339</td><td>3119,205</td><td>„Lubrika“ LtD<br>1078,185<br>regenerated</td></tr>
</table>

| | | | | |
|---|---|---|---|---|
| | | | „Eco sand Sofia София“ LtD 2041,020 – only recovered, without regenerated | **30 000 tons per year** **---------------** **„Verila Lubricants“ JsC** Total regenerated **657,922** *No data available about complex license in the MEW's registries* **---------------** „Eco sand Sofia“ LtD 2041,020 – only recovered, without regenerated --------------- **Total annual capacity in with complex license** for regeneration of two of the stations: **55 000 per year** |
| **TOTAL:** | **31716.751** | **12714.481** | **10673.461 - R** | **10673.461 - R** |
| | | | **12714.481 – R+R** | **12714.481 – R+R** |

## APPENDIX 3

## Statistical data in accordance with Order № RD-278/15.05.2018 of the Minister of environment and waters

<table>
<tr><td rowspan="2">Waste recovery organizations 2015</td><td rowspan="2">Released on the market (tons)</td><td>Regeneration</td><td>Regeneration</td><td>Note:</td></tr>
<tr><td>Total tons</td><td></td><td>Utilized quantities of tons by installations with a complex license</td></tr>
<tr><td>„Oil recycling“ LtD</td><td>6514,993</td><td>2608,200</td><td>„Polychim CC“ LtD 2608,200 Regenerated, collected, and recovered</td><td rowspan="4">„Polichim CC“ LtD<br>Total regenerated 3503.200<br><br>Annual capacity with complex license № 440-H0/2012 Ministry of environment and waters (MEW) 25 000 per year<br>----------------<br>„Lubrika“ LtD<br>Total regenerated 6700.92<br>Annual capacity with complex license<br>№ 352-НО-И0-А2/ 2013 MEW<br><br>30 000 tons per year<br>----------------</td></tr>
<tr><td>„Lubrika Environmental activities“ LtD</td><td>11420,335</td><td>4568,235</td><td>„Lubrika“ LtD 4568,235 Regenerated, collected, and recovered</td></tr>
<tr><td>„Ecorivais oil“ JsC</td><td>2236,349</td><td>895</td><td>„Polychim CC“ LtD 895 tons Regenerated, collected, and recovered</td></tr>
<tr><td>„National company for collection and recovery of used oils“ LtD</td><td>5123,956</td><td>2062,094</td><td>„Lubrika“ LtD 1585,500 regenerated „Verila Lubricants“ JsC 476,594 regenerated</td></tr>
</table>

| | | | | |
|---|---|---|---|---|
| **„Nord oils“ JsC** | 7497,281 | 3007,640 | „Lubrika“ LtD 547,185 regenerated „Insa Oil“ LtD 180.00 Recovered - complex license. „Vel metal“ LtD 887,400 „Man trading“ LtD – 393 tons only recovered, without regenerated | **„Verila Lubricants” JsC** Total regenerated **476,594** *No data available about complex license in the MEW’s registries* **----------------** only recovered, no regeneration„“Insa Oil“ LtD 180.00 Annual capacity – complex license №73-Н2-ИО-АО/ 2013. **3350 tons per year** Recovered - complex license „Vel metal“ LtD 887,400 „Man trading“ LtD – 393 **----------------** **Total annual capacity in complex license** for regeneration of 3 of the installations: **58 350 tons per year** |
| **TOTAL:** | **32792.914** | **13141.169** | **10680.714 - R** | **10680.714 - R** |
| | | | **13141.169 – R+R** | **12141.114 – R+R** |

## APPENDIX 4

## Statistical data in accordance with Order № RD-390/15.05.2019 of the Minister of environment and waters

| Waste recovery organizations 2015 | Released on the market (tons) | Regeneration | Regeneration | Note: |
|---|---|---|---|---|
| | | Total tons | | Utilized quantities of tons by installations with a complex license |
| **„Oil recycling“ LtD** | 7081,179 | 2833,100 | Regeneration. No statistical data available in the order | No statistical data available in the order<br>-------------<br>**Total annual capacity in complex license** for regeneration of 3 of the installations: **58 350 tons per year** |
| **„Lubrika Environmental activities“ LtD** | 11590,965 | 4640,153 | Regeneration. No statistical data available in the order | |
| **„Ecorivais oil“ JsC** | 1923,253 | 770,000 | Regeneration. No statistical data available in the order | |
| **„National company for collection and recovery of used oils“ LtD** | 5385,923 | 2172,058 | Regeneration. No statistical data available in the order | |
| **„Nord oils“ JsC** | 6727,519 | 2794,960 | Regeneration. No statistical data available in the order | |
| **„Eco oil resource“ LtD** | 1,960 | 0,900 | Regeneration. No statistical data available in the order | |
| **TOTAL:** | **32710.799** | **13211.171** | - | |
| | | | - | |

## APPENDIX 5

### Statistical data in accordance with Order № RD-534/19.05.2021 of the Minister of environment and waters

**Note:** *for the year of 2020 no public data about issued orders by the Minister of environment and waters is available*

| Waste recovery organizations 2015 | Released on the market (tons) | Regeneration | Regeneration | Note: |
|---|---|---|---|---|
| | | Total tons | | Utilized quantities of tons by installations with a complex license |
| **„Oil recycling“ LtD** | 6725,839 | 2698,000 | Regeneration. No statistical data available in the order | No statistical data available in the order<br>-------------<br>**Total annual capacity in complex license** for regeneration of 3 of the installations: **58 350 tons per year** |
| **„Lubrika Environmental activities“ LtD** | 10811,264 | 4324,710 | Regeneration. No statistical data available in the order | |
| **„Ecorivais oil“ JsC** | 1626,179 | 651,000 | Regeneration. No statistical data available in the order | |
| **„National company for collection and recovery of used oils“ LtD** | 5136,383 | 2102,525 | Regeneration. No statistical data available in the order | |
| **„Eco oil resource“ LtD** | 5931,931 | 2397,880 | Regeneration. No statistical data available in the order | |
| TOTAL: | **30231.596** | **12174.115** | - | |
| | | | - | |

## APPENDIX 6

## Annual statistical data about equation (8) for the period 2016-2021

| Year | Released on the market<br>Total tons | Regeneration | Total capacity for processing *(minimum)* | Quality demanded on the market *(internal market + export)* | Average mean for equation (8.1) equation (8.2) |
|---|---|---|---|---|---|
| | | Total tons | Total tons | Total tons | |
| | - | **Equation (8)<br>Equation (8.1)<br>Equation (8.2)** | **Equation (8.1)** | **Equation (8.2)** | **Equation (8)** |
| | | **Parameter (A)** | | | **Parameter (B)** |
| **2016** | **31223.806** | **12507.316** | **55000** | **130 000** | **-79992.684** |
| **2017** | **31716.751** | **12714.481** | **55000** | **130 000** | **-79785.519** |
| **2018** | **32792.914** | **13141.169** | **58350** | **130 000** | **-81033.831** |
| **2019** | **32710.799** | **13211.171** | **58350** | **130 000** | **-80963.829** |
| **2021** | **30231.596** | **12174.115** | **58350** | **130 000** | **-82000.885** |
| **Average value for the period:** | ***31735.173*** | ***12749.650*** | ***57010*** | ***130 000*** | |

## APPENDIX 7

**Institutional basis for the circular economy in EU (regulations, directives, and communications)**

| | |
|---|---|
| **REGULATIONS** | |
| **1** | **Regulation (EU) 2019/1009** of the European Parliament and of the Council of 5 June 2019 laying down rules on the making available on the market of EU fertilising products and amending Regulations (EC) No 1069/2009 and (EC) No 1107/2009 and repealing Regulation (EC) No 2003/2003 (Text with EEA relevance). URL: https://eur-lex.europa.eu/legal-content/EN/TXT/?uri=celex%3A32019R1009 |
| **2** | **Regulation (EU) 2020/852** of the European Parliament and of the Council of 18 June 2020 on the establishment of a framework to facilitate sustainable investment, and amending Regulation (EU) 2019/2088 (Text with EEA relevance). PE/20/2020/INIT. OJ L 198, 22.6.2020, 13–43. URL: https://eur-lex.europa.eu/legal-content/EN/ALL/?uri=celex%3A32020R0852 |
| **3** | **Commission Regulation (EU) 2019/424** of 15 March 2019 laying down ecodesign requirements for servers and data storage products pursuant to Directive 2009/125/EC of the European Parliament and of the Council and amending Commission Regulation (EU) No 617/2013 (Text with EEA relevance.). URL: https://eur-lex.europa.eu/legal-content/EN/TXT/?uri=CELEX%3A32019R0424 |
| **4** | **Commission Regulation (EU) 2019/1784** of 1 October 2019 laying down ecodesign requirements for welding equipment pursuant to Directive 2009/125/EC of the EP and of the Council (Text with EEA relevance). URL: https://eur-lex.europa.eu/legal-content/EN/TXT/?uri=CELEX%3A32019R1784 |
| **5** | **Commission Regulation (EU) 2019/2021** of 1 October 2019 laying down ecodesign requirements for electronic displays pursuant to Directive 2009/125/EC of the European Parliament and of the Council, amending Commission Regulation (EC) No 1275/2008 and |
| | repealing Commission Regulation (EC) No 642/2009 (Text with EEA relevance.). URL: https://eur-lex.europa.eu/legal-content/EN/TXT/?uri=CELEX:32019R2021 |
| **6** | **Commission Regulation (EU) 2019/2023** of 1 October 2019 laying down ecodesign requirements for household washing machines and household washer-dryers pursuant to Directive 2009/125/EC of the European Parliament and of the Council, amending Commission Regulation (EC) No 1275/2008 and repealing Commission Regulation (EU) No 1015/2010 (Text with EEA relevance.). URL: https://eur-lex.europa.eu/legal-content/EN/TXT/?toc=OJ%3AL%3A2019%3A315%3ATOC&uri=uriserv%3AOJ.L_.2019.315.01.0285.01.ENG |
| **7** | **Commission Regulation (EU) 2019/2019** of 1 October 2019 laying down ecodesign requirements for refrigerating appliances pursuant to Directive 2009/125/EC of the European Parliament and of the Council and repealing Commission Regulation (EC) No 643/2009 (Text with EEA relevance.). URL: https://eur-lex.europa.eu/legal-content/en/TXT/?uri=CELEX%3A32019R2019 |
| **8** | **Commission Regulation (EU) 2019/2024** of 1 October 2019 laying down ecodesign requirements for refrigerating appliances with a direct sales function pursuant to Directive 2009/125/EC of the European Parliament and of the Council (Text with EEA relevance.). URL:https://eur-lex.europa.eu/legal-content/EN/TXT/?uri=uriserv:OJ.L_.2019.315.01.0313.01.ENG |
| **9** | **Commission Regulation (EU) 2019/2022** of 1 October 2019 laying down ecodesign requirements for household dishwashers pursuant to Directive 2009/125/EC of the European Parliament and of the Council amending Commission Regulation (EC) No 1275/2008 and repealing Commission Regulation (EU) No 1016/2010 (Text with EEA relevance.). URL: https://eur-lex.europa.eu/legal-content/EN/TXT/?uri=uriserv:OJ.L_.2019.315.01.0267.01.ENG&toc=OJ:L:2019:315:TOC |

| DIRECTIVES | |
|---|---|
| **1** | **Directive (EU) 2018/849** of the EP and of the Council of 30 May 2018 amending Directives 2000/53/EC on end-of-life vehicles, 2006/66/EC on batteries and accumulators and waste batteries and accumulators, and 2012/19/EU on waste electrical and electronic equipment (Text with EEA relevance). URL: https://eur-lex.europa.eu/legal-content/EN/TXT/?uri=CELEX%3A32018L0849 |
| **2** | **Directive (EU) 2018/850** of the EP and of the Council of 30 May 2018 amending Directive 1999/31/EC on the landfill of waste (Text with EEA relevance). URL: https://eur-lex.europa.eu/legal-content/EN/TXT/?uri=CELEX%3A32018L0850 |
| **3** | **Directive (EU) 2018/851** of the EP and of the Council of 30 May 2018 amending Directive 2008/98/EC on waste (Text with EEA relevance). URL: https://eur-lex.europa.eu/legal-content/EN/TXT/?uri=celex%3A32018L0851 |
| **4** | **Directive (EU) 2018/852** of the EP and of the Council of 30 May 2018 amending Directive 94/62/EC on packaging and packaging waste (Text with EEA relevance). URL: https://eur-lex.europa.eu/legal-content/EN/TXT/?uri=celex:32018L0852 |
| **5** | **Directive (EU) 2019/883** of the EP and of the Council of 17 April 2019 on port reception facilities for the delivery of waste from ships, amending Directive 2010/65/EU and repealing Directive 2000/59/EC (Text with EEA relevance). URL: https://eur-lex.europa.eu/legal-content/EN/TXT/?uri=CELEX%3A32019L0883 |
| **6** | **Directive (EU) 2019/904** of the EP and of the Council of 5 June 2019 on the reduction of the impact of certain plastic products on the environment (Text with EEA relevance). URL: https://eur-lex.europa.eu/eli/dir/2019/904/oj |
| **7** | **Directive (EU) 2019/771** of the EP and of the Council of 20 May 2019 on certain aspects concerning contracts for the sale of goods, amending Regulation (EU) 2017/2394 and Directive 2009/22/EC, and repealing Directive 1999/44/EC (Text with EEA relevance.). URL: https://eur-lex.europa.eu/legal-content/en/TXT/?uri=CELEX:32019L0771 |

| COMMUNICATIONS | |
|---|---|
| **1** | **COM/2011/0571 final.** Communication from the Commission to the European Parliament, the Council, the European Economic and Social Committee and the Committee of the Regions: Roadmap to a Resource Efficient Europe. URL: https://eur-lex.europa.eu/legal-content/EN/TXT/?uri=CELEX%3A52011DC0571 |
| **2** | **COM/2015/0614 final**. Communication from the Commission to the European Parliament, the Council, the European Economic and Social Committee and the Committee of the Regions Closing the loop - An EU action plan for the Circular Economy. URL: https://eur-lex.europa.eu/legal-content/EN/TXT/?uri=CELEX%3A52015DC0614 |
| **3** | **COM/2016/0773 final.** Communication from the Commission: Ecodesign Working Plan 2016-2019. URL: https://eur-lex.europa.eu/legal-content/EN/TXT/?uri=CELEX%3A52016DC0773 |
| **4** | **COM/2017/0479 final.** Communication from the Commission to the European Parliament, the European Council, the Council, the European Economic and Social Committee, the Committee of the Regions and the European Investment Bank: Investing in a smart, innovative and sustainable Industry A renewed EU Industrial Policy Strategy. URL: https://eur-lex.europa.eu/legal-content/en/TXT/?uri=CELEX%3A52017DC0479 |
| **5** | **COM/2017/033 final.** Report from the Commission to the European Parliament, the Council, the European Economic and Social Committee and the Committee of the Regions: on the implementation of the Circular Economy Action Plan. URL: https://eur-lex.europa.eu/legal-content/GA/TXT/?uri=CELEX:52017DC0033 |
| **6** | **COM/2018/028 final.** Communication from the Commission to the European Parliament, the Council, the European Economic and Social Committee and the Committee of the Regions: A European Strategy for Plastics in a Circular Economy. URL: https://eur-lex.europa.eu/legal-content/EN/TXT/?uri=COM%3A2018%3A28%3AFIN |
| **7** | **COM/2018/029 final.** Communication from the Commission to the European Parliament, the Council, the European Economic and Social Committee and the Committee of the Regions: on a monitoring framework for the circular economy. URL: https://eur-lex.europa.eu/legal-content/EN/TXT/?uri=COM%3A2018%3A29%3AFIN |
| **8** | **COM/2018/032 final.** Communication from the Commission to the European Parliament, the Council, the European Economic and Social Committee and the Committee of the Regions: on the implementation of the circular economy package: options to address the interface between chemical, product and waste legislation (Text with EEA relevance) options to address the interface between chemical, product and waste legislation. URL: https://eur-lex.europa.eu/legal-content/en/ALL/?uri=CELEX%3A52018DC0032 |
| **9** | **COM/2018/035 final.** Report from the Commission to the EP and the Council: on the impact of the use of oxo-degradable plastic, including oxo-degradable plastic carrier bags, on the environment. URL: https://eur-lex.europa.eu/legal-content/EN/TXT/?uri=CELEX%3A52018DC0035 |
| **10** | **COM/2019/22.** Reflection Paper Towards a Sustainable Europe by 2030. URL: https://ec.europa.eu/info/publications/reflection-paper-towards-sustainable-europe-2030_en |
| **11** | **COM/2019/190 final.** Report from the Commission to the EP, the Council, the European Economic and Social Committee and the Committee of the Regions: on the implementation of the circular economy action plan. URL: https://eur-lex.europa.eu/legal-content/EN/TXT/?uri=CELEX%3A52019DC0190 |
| **12** | **COM/2019/640 final.** Communication from the Commission to the EP, the European Council, the Council, the European Economic and Social Committee and the Committee of the Regions: The European Green Deal. URL: https://eur-lex.europa.eu/legal-content/EN/TXT/?uri=COM%3A2019%3A640%3AFIN |
| **13** | **COM/2020/98 final.** Communication from the Commission to the European Parliament, the Council, the European Economic and |
| | Social Committee and the Committee of the Regions: A new Circular Economy Action Plan For a cleaner and more competitive Europe. URL: https://eur-lex.europa.eu/legal-content/EN/TXT/?uri=COM:2020:98:FIN&WT.mc_id=Twitter |

* In developing the above mentioned appendices are also used materials from: **Friant, M. C., Vermeulen, W. J., & Salomone, R. (2021).** Analysing European Union circular economy policies: words versus actions. Sustainable Production and Consumption, 27, 337-353. https://doi.org/10.1016/j.spc.2020.11.001